\documentclass[12pt]{article}
\pdfoutput=1
\usepackage{array}
\usepackage{graphicx}
\usepackage{amssymb}
\usepackage{amsmath}
\usepackage{latexsym}
\usepackage[usenames]{color}
\usepackage{cite}
\usepackage{wasysym}
\usepackage{parskip}
\usepackage{wrapfig}
\usepackage{graphicx}
\usepackage[colorinlistoftodos]{todonotes}
\usepackage{tikz-cd}
\usepackage{lipsum}
\usepackage{adjustbox}
\usepackage{amssymb}
\usepackage{slashed}
\usepackage{cancel}
\usepackage{indentfirst}
\definecolor{darkblue}{cmyk}{0.9,0.9,0,0}
\definecolor{darkred}{rgb}{0.6,0,0.3}
\usepackage{graphicx} 
\usepackage[setpagesize=false,pagebackref=false, linktocpage, bookmarksopen=true, colorlinks=true, linkcolor=darkblue,citecolor=darkblue,urlcolor=darkblue]{hyperref}
\usepackage{hyperref}

\renewcommand{\thefootnote}{\arabic{footnote}}
\allowdisplaybreaks

\def\eqref#1{(\ref{#1})}

\newcommand{\beq}{\begin{equation}}
\newcommand{\eeq}{\end{equation}}
\hypersetup{pdfauthor={Name}}
\begin{document}
\thispagestyle{empty}

\renewcommand{\thefootnote}{\fnsymbol{footnote}}
\setcounter{page}{1}
\setcounter{footnote}{0}
\setcounter{figure}{0}
\begin{center}
$$$$
{\Large\textbf{\mathversion{bold}
$AdS_5 \times S^5$ supergravity vertex
operators}\par}

\vspace{1.3cm}

\textrm{Thiago Fleury $^{\lambda_L}$\footnote{tsi.fleury@gmail.com}, Lucas N. S. Martins $^{\lambda_R}$\footnote{lucas$\_$nmartins@hotmail.com}}
\\ \vspace{1.2cm}
\footnotesize{\textit{
$^{\lambda_L}$ 
International Institute of Physics, Federal University of Rio Grande do Norte, \\ Campus Universit\'ario,  Lagoa Nova, Natal, RN 59078-970, Brazil
\vspace{3mm} \\
$^{\lambda_R}$ 
Instituto de F\'isica Te\'orica, UNESP - Univ. Estadual Paulista, 
ICTP South American Institute for Fundamental Research,
Rua Dr. Bento Teobaldo Ferraz 271, 01140-070, S\~ao Paulo, SP, Brazil
}  
\vspace{4mm}
}

\par\vspace{1.5cm}

\textbf{Abstract}\vspace{2mm}
\end{center}
In any type II superstring 
background, the supergravity 
vertex operators in the pure spinor formalism are described 
by a gauge superfield.
In this paper, we obtain 
for the first time an explicit 
expression for this superfield 
in an $AdS_5 \times S^5$ background.
Previously, the vertex operators were only known close to the boundary of $AdS_5$ or in the minus eight picture. Our strategy for the computation was to apply eight picture raising operators in the minus
eight picture vertices.
In the process, a huge number  
of terms are generated and we have developed numerical techniques to 
perform intermediary simplifications. 
Alternatively, the same numerical techniques 
can be used to compute 
the vertices directly in the zero picture by constructing 
a basis of invariants and fitting for the coefficients. 
One motivation for 
constructing the vertex operators is the computation of $AdS_5 \times S^5$ string amplitudes. 

\noindent

\setcounter{page}{1}
\renewcommand{\thefootnote}{\arabic{footnote}}
\setcounter{footnote}{0}
\setcounter{tocdepth}{2}
\newpage
\tableofcontents

\parskip 5pt plus 1pt   \jot = 1.5ex

\section{Introduction\label{sec:intro}}

\indent Superstring theory in an 
$AdS_5 \times S^5$ background 
can be studied with both 
the pure spinor 
\cite{BerkovitsFirst}
and 
the Green-Scharwz formalisms 
\cite{GreenScharwz}, see \cite{ReviewPure,Foundations} 
for reviews\footnote{For recent developments 
using the Neveu-Scharwz-Ramond formalism and ambitwistor strings in AdS,
see \cite{NSR} and \cite{AmbiI,AmbiII} respectively.}.
In particular, the worldsheet action is known. 
Note that both the action and the vertex operators are BRST invariant in the pure spinor formalism and kappa symmetric 
in the Green-Scharwz formalism. 
Despite the many progresses 
in the study of superstrings in this background,
explicit superfield expressions for the vertex
operators are unknown. 
In this work, we made progress 
in this direction by 
finding expressions for all the half-BPS vertex operators in the pure spinor 
formalism.
These vertex operators were known 
before only close to the boundary 
of $AdS_5$
\cite{BerkovitsFleury} and they 
were used to compute open-closed
string amplitudes in 
\cite{Thales,ThalesII} reproducing  
the expected holographic results. 
Note that the vertices were also known 
in a different picture \cite{BerkovitsHalf}
and in the plane wave limit \cite{PlaneWaveVertex}\footnote{The $\beta$ deformed vertex is discussed in  \cite{BetaDeformation}.}.
In \cite{BerkovitsFleury}, 
it was used that the half-BPS operators 
in $\mathcal{N}=4$ super-Yang-Mills (SYM) 
can be written in terms of traces of the Sohnius superfield \cite{Sohnius} and  
harmonic variables
\cite{GalperinHarmonic,HoweWestHarmonic,
AndrianopoliHarmonic}. The duals 
to these operators 
were constructed using the
same variables in  \cite{Howe:1983sra,ChiralSuperfields}.
In this paper, we are going to focus on the 
bottom components of the half-BPS 
supermultiplets.  However, acting with 
supersymmetry generators, it is possible  
to generate all the states and we expect  that the 
result can be written in terms of the  superfields just mentioned. 
In this work, we use the pure spinor formalism but it should be possible 
to extend our results to the Green-Scharwz formalism and it 
will also be very interesting to make 
connection with the recent understanding 
of both formalisms as 
Chern-Simons theories \cite{Costello,CostelloII}.

As already mentioned, the main results 
of this paper is explicit expressions for the supergravity vertex operators in $AdS_5 \times S^5$. 
The vertices are states in the cohomology
of the BRST operator with ghost number 
two. They are annihilated by twenty four
supersymmetries and this implies that they depend only on eight worldsheet fermionic variables $\theta_+^a$, with $a=1,\ldots,8$. The vertices are schematically of the form (bottom state of the supermultiplet)
\begin{equation}
V_0(n) = e^{n(z+w)}    
\left[(\bar\lambda^2) + 
(\bar\lambda^2 \theta_+^2) P_1(n)
+(\bar\lambda^2 \theta_+^4) P_2(n)
+ (\bar\lambda^2 \theta_+^6) P_3 (n)+
(\bar\lambda^2 \theta_+^8) P_4(n)\right] \, ,
\label{schematically} 
\end{equation}
where both $z$ and $w$ are worldsheet variables and $\bar\lambda$ are pure
spinors. The subscript zero
means zero picture and its meaning will be explained in the
next section. 
The variable $z$ is related to the distance to the $AdS_5$ boundary and 
$w$ parametrizes an equator of $S^5$.  
The integer $n$ measures the dimension and $R$-charge of the state. For example,
the case $n=0$ corresponds to the dilaton vertex operator of \cite{BerkovitsPrescription}. 
The $P_i(n)$ are complicated polynomials but with maximum degree four in $n$ and this truncation to a very low order might seem surprising. 
The vertices are expected to have a nonsingular flat space limit and in fact this implies the observed truncation of the polynomials.   
 The flat space limit consists in rescaling the generators of $\mathfrak{psu}(2,2|4)$
 and deforming this algebra continuously  
 to the ten dimensional super-Poincaré algebra. 
 The parameter $n$ and the 
 coordinates $\theta^a_+$ are also rescaled in the process. The only terms 
 that survive this limit are the ones
 of the form $n^k \theta_+^{2k}$ and any term of the form $n^{k^{\prime}} \theta_+^{2k}$ with $k^{\prime}>2k$ has a singular limit. Notice that the vertices do not depend on the other 
 bosonic worldsheet variables. This follows because we are considering a particular polarization and a specific spacetime position. It is possible to perform $PSU(2,2|4)$ transformations 
 and get more general expressions.



The vertices in (\ref{schematically}) 
 are in a very particular gauge and 
they do not depend on all the pure spinor variables. In order to understand this better recall that in any 
type II superstring background the supergravity multiplet is described in the pure spinor formalism by the unintegrated vertex operator 
\cite{BerkovitsChandia}
\begin{equation}
V=\lambda_{L}^{\underline\alpha}\lambda_{R}^{\hat{\underline\beta}}A_{\underline\alpha\hat{\underline\beta}}(x^{M},\theta_{L}^{\underline\alpha},\theta_{R}^{\hat{\underline\alpha}}) \, ,  
\label{vertextypeII}
\end{equation}
where $A_{\underline\alpha\hat{\underline\beta}}$ is a bispinor superfield depending on the appropriate type II superspace variables $(x^{M},\theta_{L}^{\underline\alpha},\theta_{R}^{\hat{\underline\alpha}})$,
with $M=1,\ldots,10$ being space-time 
indices and $\underline\alpha,\hat{\underline\alpha}=1,\dots,16$ 
Weyl spinor indices. The    $(\lambda_{L}^{\underline\alpha},\lambda_{R}^{\hat{\underline\alpha}})$ are the bosonic pure spinor variables satisfying $\lambda_{L}^{\underline\alpha} \gamma_{\underline\alpha \underline\beta}^{M}\lambda_{L}^{\underline\beta}=\lambda_{R}^{\hat{\underline\alpha}} \gamma_{\hat{\underline\alpha}\hat{\underline\beta}}^{M}\lambda_{R}^{\hat{\underline\beta}}=0$
with $\gamma^M_{\underline\alpha \underline\beta}$ the $SO(10)$ Pauli matrices.
Note that the superfield $A_{\underline\alpha\hat{\underline\beta}}$ appearing in
(\ref{vertextypeII}) is a supergravity gauge field.  
The BRST operator $Q$ has a complicated action on general states but it simplifies a lot when acting on the supergravity vertices. In this case, the operator is given by 
\begin{equation}
Q=\lambda_{L}^{\underline\alpha}\nabla_{L
\underline\alpha}+\lambda_{R}^{\hat{\underline\alpha}}\nabla_{R\hat{\underline\alpha}} \, , 
\label{BRSToperator}
\end{equation}
where $\nabla_{L\underline\alpha}$ and $\nabla_{R\hat{\underline\alpha}}$ are the super-covariant derivatives. A necessary condition 
for the vertex operator (\ref{vertextypeII}) to be a physical state 
is $Q \cdot V = 0$  and this implies 
\begin{equation}
\gamma_{MNPQR}^{\underline\alpha \underline\beta}\nabla_{L \underline\alpha}A_{\underline\beta \hat{\underline\gamma}}= \gamma_{MNPQR}^{\hat{\underline\alpha}\hat{\underline\beta}}\nabla_{R\hat{\underline \alpha}}A_{\underline\gamma\hat{\underline\beta}}=0 \, . 
\label{equationsforA} 
\end{equation}
In addition, if $V$ of (\ref{vertextypeII}) describes a 
physical state then it cannot be written 
as $V = Q \, \cdot \, \Omega$ for any $\Omega$. 
Note that the superfield $A_{\underline\alpha \hat{\underline\beta}}$ is
defined up to the gauge transformations 
$(V \sim V \; + \; Q \cdot \Omega)$
\begin{equation}
\delta A_{\underline\alpha \hat{\underline\beta}}
= \nabla_{L\underline\alpha} \Omega_{R \hat{\underline\beta}} + 
\nabla_{R\hat{\underline\beta}} \Omega_{L \underline\alpha}\, , 
\label{gaugetransform}
\end{equation}
with
\begin{equation}
\gamma_{MNPQR}^{\underline\alpha \underline\beta}\nabla_{L \underline\alpha}\Omega_{L \underline\beta} =\gamma_{MNPQR}^{\hat{\underline\alpha} \hat{\underline\beta}}\nabla_{R \hat{\underline\alpha}}\Omega_{R \hat{\underline\beta}}=0 \, . 
\label{OmegaConsistency}
\end{equation}

In some backgrounds such as 
flat and 
$AdS_5 \times S^5$, it is possible to 
choose a very convenient gauge for the vertex operators. We will demonstrate this for a flat background below. 
The $AdS_5 \times S^5$ case will be explained in section \ref{halfBPSsection}.
Note that the existence of this gauge is important for defining the vertex operators in different pictures. 
In a general type II background 
the covariant derivatives satisfy the following algebra
\begin{equation}
[\nabla_{L \underline\alpha},\nabla_{L \underline\beta}]=2\gamma^{M}_{\underline\alpha \underline\beta}\nabla_{M},\qquad [\nabla_{R\hat{\underline\alpha}},\nabla_{R\hat{\underline\beta}}]=2\gamma^{M}_{\hat{\underline\alpha} \hat{\underline\beta}}\nabla_{M},\qquad [\nabla_{L \underline\alpha},\nabla_{R\hat{\underline\alpha}}]=F_{\underline\alpha\hat{\underline\alpha}} \, ,     
\end{equation}
where 
$F_{\underline\alpha\hat{\underline\alpha}}$ is a bi-spinor that describes the superspace curvature  
\cite{BerkovitsHowe}.
In what follows, 
we will organize the covariant derivatives with $SO(1,9)$ spinorial indices into chiral and anti-chiral $SO(8)$ spinors $(\nabla_{a},\nabla_{\dot a})$ with $a,\dot{a}=1,\ldots,8$.
In a flat background, we can go to  
a frame where the only non-vanishing component of the momentum is $k_+ = k_0 + k_9$. It is possible to show that in this frame 
the covariant derivatives when acting on a massless state obey the following algebra \cite{BerkovitsHalf}
\begin{equation}
[\nabla_{La},\nabla_{Lb}]=2k_+\delta_{ab} \, ,\qquad [\nabla_{Ra},\nabla_{Rb}]=2k_+\delta_{ab} \, , \qquad  
[\nabla_{L},\nabla_{R}]=0
\label{flatcommutators} \, , 
\end{equation}
and the other commutators are not going to be used.  
The commutators above imply that we 
can invert both $\nabla_{La}$ and $\nabla_{Ra}$ and using this property 
we can go 
to a gauge where the supergravity gauge fields have the components 
\begin{equation}
A_{ab}=A_{\dot a b}=A_{a\dot b}=0 \, . 
\label{allconditions}
\end{equation}
The proof is as follows. Let's do a
gauge transformation with 
parameters $\Omega_{R \hat\beta}$. The equation 
(\ref{gaugetransform}) implies
\begin{equation}
\delta A_{a  \dot a}
= \nabla_{L a} \Omega_{R \dot a} \, , \quad  \delta A_{a  b}
= \nabla_{L a} \Omega_{R b} \, . 
\end{equation}
We want $\delta A_{a  \dot a}=- A_{a  \dot a}$ and 
$\delta A_{a  b}=- A_{a b}$.
By replacing these variations above, we can solve the equations and get  $
\Omega_{R \hat\beta} \propto \nabla_{L a} A_{a \hat{\beta}}$. 
Moreover, it is easy to see
that this $\Omega_{R \hat\beta}$ solves the necessary conditions (\ref{OmegaConsistency}).
This follows from 
equation
(\ref{equationsforA}) and 
the last commutators of (\ref{flatcommutators}).
After this gauge transformation, we have 
$A_{a  \dot a}= A_{a b}=0$.
Now we can perform a new gauge transformation 
with parameter $\Omega_{L \dot{a}}$ and using the same reasoning we arrive at
(\ref{allconditions}). 
We believe that it is possible to adapt the argument just given to $AdS_5 
\times S^5$.
However, one needs to carefully 
deal with the transformations of the pure spinors variables (they are not BRST invariant any more).
In addition the commutation relations of the covariant derivatives are more complicated 
and it is necessary to fix the 
$USp(2,2) \times USp(4)$ gauge transformations of the pure spinors variables. We will argue that the gauge \eqref{allconditions} is reachable for this background in section \ref{halfBPSsection}\footnote{The arguments are only valid for the bottom state of the multiplet. It is possible that there are subtleties for the other states and 
it deserves further investigation. We thank Nathan Berkovits for pointing this out to us.}. 

The strategy used in this paper for 
obtaining the vertex operators was 
to start with their expressions in
the minus eight picture proposed 
in \cite{BerkovitsHalf} and higher 
their picture up to zero. This strategy was proposed recently.  
Previously, in a flat background, 
the closed superstring vertex operators were
constructed by taking the left-right product of the open superstring vertex operators, see 
\cite{ICTPlectures} for a review. Note that any method based on  
holomorphicity cannot work in $AdS_5$ as 
the current
algebra computed in \cite{Current1,Current2,Current3} is not
holomorphic already at leading order 
in $\alpha^{\prime}$. 
The picture raising procedure is 
best understood by  
bosonizing some of the variables and 
this will be reviewed in the next section. In a flat background, it is possible to 
change the picture of a vertex operator  
analytically and write a close expression for the vertex in any picture, see \cite{LucasFlat,MikhailovZavaleta}. 
In $AdS_5 \times S^5$, the BRST transformations of the worldsheet variables are very complicated, in
particular, the pure spinor variables 
transform. It is more complicated to do any 
analytic calculation in $AdS_5$ as a huge number of terms are generated every time we act with a new picture raising operator. 
Many simplifications are possible but
they are hard to implement as 
the pure spinor constraints are quadratic. 
In this work, the majority of the simplifications were done numerically.
Notice that the same numerical techniques can be used to compute 
the cohomology of the BRST operator directly 
at picture zero. This is 
an alternative way of obtaining the vertex operators. 
The procedure consists in writing the vertices using a complete basis of invariants respecting all the symmetries and depending 
on two pure spinors 
and fitting for the coefficients by imposing
that the vertex is BRST closed and not BRST exact. 
In this paper, we chose a particular 
parametrization of the $AdS_5 \times S^5$ superspace. 
If one wants to compute the vertices using a different 
parametrization, we noted that the fitting procedure just described is much faster in terms 
of computer time. 

One very strong motivation for 
computing the vertex operators is 
the calculation of string amplitudes 
\cite{FleuryMartinsII}. 
Apart from the open-closed amplitude mentioned before, no other amplitude has been 
computed using the pure spinor formalism in $AdS_5 \times S^5$ 
and one of the reasons 
was the lack of explicit expressions for the vertices. The other ingredient 
needed for amplitude computations is a prescription
for dealing with the zero modes of the 
worldsheet variables including the 
pure spinor ghosts. For example, 
in a flat background to obtain the  tree-level three open strings amplitude,
the prescription consist in multiplying the three vertices and 
reading the term proportional 
to a combination of three pure spinors 
and five thetas, see 
\cite{MafraThesis} for more details.  
In \cite{BerkovitsPrescription}, a $PSU(2,2|4)$ invariant prescription for the $AdS_5 \times S^5$ amplitude was proposed and it involves an integration over all the 
thirty two $\theta$'s. However, this prescription gives a vanishing result for any three point amplitude involving massless states since, in the light cone gauge, each massless state can contribute with at maximum of eight $\theta$'s\footnote{
This fact was explained to us by Nathan Berkovits (private communication).}.
Having the explicit expressions for the vertex operators, it is possible to
test any proposal by comparing 
with existing results in the literature.
For example, it is known that the three point 
amplitude of half-BPS operators is completely fixed by supersymmetry and the result
is just a normalization constant times
kinematical factors 
\cite{ThreePoint1,ThreePoint2,ThreePoint3}.
In order to compute higher than three point amplitudes,
it will be necessary to compute the integrated
vertex operators as well.
In this work, we only consider the unintegrated
one. However, it is possible to adapt several existing 
results in the literature \cite{IntegratedVertexOperators,IntegratedVertexOperatorsII}. 
Another important element for multiloop
calculations is the $b$ ghost which
is a composite operator in the pure spinor formalism. The $b$ ghost is known in this background and it was 
computed in \cite{bghost}.

This paper is organized as follows. In section 
\ref{flatspace}, we review 
the flat space vertex operators and their recent minus eight picture realizations. 
The section \ref{purespinorformalism} 
contains our 
conventions for the 
$AdS_5 \times S^5$ pure spinor superstring.
In particular, we define 
our worldsheet variables 
and the BRST operator. 
Our main results, the expressions for the vertex 
operators, is presented in 
section \ref{halfBPSsection}.
In the same section, we argue that the picture raising procedure is well defined in $AdS_5 \times S^5$, 
that our results reduce 
correctly to the well known 
dilaton vertex operator in an appropriate limit and 
that the vertices have the correct 
flat space limit. 
In section \ref{theboundary},
we find how the boundary of $AdS_5$ is described in our coordinates and explain
how to obtain a vertex operator with a general polarization and for an arbitrary spacetime position.
Finally, section 
\ref{Conclusions}
has our conclusions. 
The Appendices have 
our conventions for the 
$\mathfrak{psu}(2,2|4)$ algebra, explicit expressions
for the BRST transformations 
and the vertex operators written in a compact $SO(8)$
notation. 

\section{The Flat Space Vertex Operators}\label{flatspace}

In this section, we review the known massless vertex operators in a flat background. 
In particular, we explain their recent minus eight picture realizations. 
Recall that in \cite{Howe:1983sra}, the ten dimensional type IIB superspace was parametrized in terms of holomorphic 
$\theta^{\underline\alpha}_{+}$
and anti-holomorphic 
$\theta^{\underline\alpha}_{-}$ combinations of 
fermionic coordinates. These variables are given by
\begin{equation}
\theta^{\underline\alpha}_{+}=\theta_{L}^{\underline\alpha}+ i\theta_{R}^{\hat{\underline\alpha}} \, ,\qquad \theta^{\underline\alpha}_{-}=\theta_{L}^{\underline\alpha}- i\theta_{R}^{\hat{\underline\alpha}} \, ,   
\label{thetaplusandminus}
\end{equation}
where $\theta_{L}^{\underline\alpha}$ and $\theta_{R}^{\hat{\underline\alpha}}$ were defined below 
(\ref{vertextypeII}).
The complex conjugate 
of these variables can be deduced from $(\theta_{L}^{\underline\alpha})^{*}=\theta_{L}^{\underline\alpha}$ and $(\theta_{R}^{\hat{\underline\alpha}})^{*}=\theta_{R}^{\hat{\underline\alpha}}$, which implies that $(\theta_{\pm}^{\underline\alpha})^{*}=\theta_{\mp}^{\underline\alpha}$. Note that the signs $\pm$ denotes the $R$-charge. 
In addition, it was shown that the type IIB onshell supergravity multiplet can be described in terms of an analytic superfield $\Phi$ obeying a certain reality condition, which is
\begin{equation}
(\nabla_{-})_{\underline\alpha}\Phi=0,\qquad (\nabla_{+})^{4}_{\underline\alpha \underline\beta \underline\gamma \underline\delta}\Phi = (\nabla_{-})^{4}_{\underline\alpha \underline\beta \underline\gamma \underline\delta}\Phi^{*} \, .  
\label{eq:superA}
\end{equation}
It is possible to work 
in coordinates 
where 
\begin{equation}
(\nabla_-)_{\underline\alpha} = \frac{\partial}{\partial \theta_+^{\underline\alpha}} \, , 
\end{equation}
and the constraint of analyticity is easily solved
in these coordinates by requiring that the fermionic variables that enters in $\Phi$ is only $\theta_-^{\alpha}$. Expanding in components the superfield $\Phi$, we have 
\begin{equation}
\Phi(y,\theta_-)=\phi(y)+\psi_{\underline\alpha}(y)\theta_-^{\underline\alpha}+F_{\underline\alpha \underline\beta}(y)\theta_{-}^{\underline\alpha}\theta_-^{\underline\beta}+\psi_{\underline\alpha \underline\beta \underline\gamma}(y)\theta_{-}^{\underline\alpha}\theta_-^{\underline\beta}\theta_-^{\underline\gamma}+F_{\underline\alpha \underline\beta \underline\gamma \underline\delta}(y)\theta_{-}^{\underline\alpha}\theta_-^{\underline\beta}\theta_-^{\underline\gamma}\theta_-^{\underline\delta}+\dots \, ,
\label{superfieldexpansion}
\end{equation}
where $y$ are the bosonic coordinates and the terms with higher orders in $\theta^{\underline\alpha}_{-}$ are not independent but are derivatives of the fields showed above.

In the expansion (\ref{superfieldexpansion}), the $\phi(y)$ is 
a combination of the dilaton and the axion. The $\psi_{\underline\alpha}$ contains the two dilatinos. The $F_{\underline\alpha \underline\beta}$ contains the NSNS and the RR 3-form field strengths. The $\psi_{\underline\alpha \underline\beta 
\underline\gamma}$ contains the two gravitinos. 
Finally, the $F_{\underline\alpha \underline\beta \underline\gamma 
\underline\delta}$ combines the Weyl tensor and the self-dual RR 5-form field strength.

In what follows, we are going
to work in the light-cone frame where the only non-zero component of the momentum 
is $k_+=k_0+k_9$.
Going to that frame breaks Lorentz symmetry $SO(1,9)$ down to $SO(1,1)\times SO(8)$. 
Thus the ten dimensional Weyl spinors will split into a pair of Weyl and anti-Weyl $SO(8)$ spinors
\begin{equation}
\theta^{\underline\alpha}_{\pm}=(\theta_{\pm}^{a},\bar\theta_{\pm}^{\dot a}) \, ,    
\end{equation}
with $a,\dot a=1,\dots 8$ being the Weyl and anti-Weyl indices respectively.

The superfield $\Phi$ in light-cone frame becomes the superfield of \cite{Green:1983hw} where the ten dimensional reality condition and analyticity becomes
\begin{equation}
\nabla_{-}^{a}\Phi=0,\qquad\bar\nabla_{\pm}^{\dot a}\Phi=0,\qquad (\nabla_{+})^{4}_{abcd}\Phi=\frac{1}{4!}\varepsilon_{abcd}\,^{efgh}(\nabla_{-})_{efgh}^{4}\Phi^{*} \, . 
\label{eq:superphi}
\end{equation}
Note that generically  it is possible to choose coordinates 
such that the superfield $\Phi$ depends only on sixteen thetas instead of thirty two as in 
(\ref{superfieldexpansion}).
However, in light-cone frame, it is possible to reduce the number of fermionic coordinates even further down to eight.

Recently in \cite{LucasFlat,BerkovitsHalf,MikhailovZavaleta}, a closed string vertex operator was constructed in terms of the above light-cone superfield. 
It was also shown that 
the vertex have 
a very compact expression 
(just one term) 
when written in the minus eight picture. 
In this picture the vertex becomes
\begin{equation}
V^{\rm{flat}}_{-8}=(\bar \lambda^{\dot a}_{L}\bar\lambda_{R}^{\dot a})\left(\prod_{a=1}^{8}\delta(\lambda_+^{a})\right) \Phi^{*} =(\bar \lambda^{\dot a}_{L}\bar\lambda_{R}^{\dot a})\left(\prod_{a=1}^{8}\delta(\lambda_{+}^{a})\nabla_{+}^{a}\right)\Phi \, . 
\label{vertexflat}
\end{equation}
Several comments are in order. 
First, the variables $\lambda_+^a$ appearing 
above and their 
cousin
variables
$\lambda_-^a$ are defined 
similarly to the $\theta_{\pm}$ in 
(\ref{thetaplusandminus})
as
\begin{equation}
\lambda^{\underline\alpha}_{+}=\lambda_{L}^{\underline\alpha}+ i\lambda_{R}^{\hat{\underline\alpha}} \, ,\qquad \lambda^{\underline\alpha}_{-}=\lambda_{L}^{\underline\alpha}- i\lambda_{R}^{\hat{\underline\alpha}} \, .   
\label{lambdaplusandminus}
\end{equation}
Notice that it is possible to take $\lambda_+^a$ as an unconstrained variable 
and the product of eight delta functions is well defined. The proof 
was given in \cite{LucasFlat} 
and it consists in showing 
that $\lambda_-^a$ 
can be expresssed as a function of the other $\lambda$'s.  
We are going to repeat the proof
here for the readers convenience. 
The pure spinor constraints $(\lambda_L \gamma^M \lambda_L) =(\lambda_R \gamma^M \lambda_R) =0$ become in $SO(8)$
notation
\begin{equation}
\lambda_L^a \lambda_L^a =
\lambda_R^a \lambda_R^a =
\bar\lambda_L^{\dot{a}}
\bar\lambda_L^{\dot{a}}=
\bar\lambda_R^{\dot{a}}
\bar\lambda_R^{\dot{a}}=0 \, , \quad \lambda_L^a \sigma^{\underline{m}}_{a
\dot{a}} \bar\lambda^{\dot{a}}_L =
\lambda_R^a \sigma^{\underline{m}}_{a
\dot{a}} \bar\lambda^{\dot{a}}_R=0 \, ,
\end{equation}
or, equivalently, using 
(\ref{lambdaplusandminus}) 
the constraints are
\begin{equation}
\begin{aligned}
&\lambda^a_+ \lambda^a_+ + \lambda^a_- \lambda^a_- =0 \, , \quad \bar\lambda^{\dot{a}}_+ \bar\lambda^{\dot{a}}_+ +
\bar\lambda^{\dot{a}}_- \bar\lambda^{\dot{a}}_- =0 \, , \quad \lambda^a_+ \lambda_-^a =0 \, , \quad  
\bar\lambda^{\dot{a}}_+ \bar\lambda_-^{\dot{a}} =0 \, ,\\
&\lambda^a_+ \sigma^{\underline{m}}_{a \dot{a}} \bar{\lambda}^{\dot{a}}_+ + \lambda^a_- \sigma^{\underline{m}}_{a \dot{a}} \bar{\lambda}^{\dot{a}}_-=0 \, , \quad
\lambda^a_+ \sigma^{\underline{m}}_{a \dot{a}} \bar{\lambda}^{\dot{a}}_- + \lambda^a_- \sigma^{\underline{m}}_{a \dot{a}} \bar{\lambda}^{\dot{a}}_+=0 \, ,
\end{aligned} 
\label{eq:so8pure} 
\end{equation}
and we have used the $SO(8)$ Pauli Matrices
\begin{equation}
\sigma^{\underline{m}}_{a \dot{a}} \, , \quad
{\rm{and}} \quad \sigma^{\underline{m}}_{\dot{a} a} \, , 
\end{equation}
with $\underline{m}=1, \ldots,8$. 
Using the Fierz identity 
\begin{equation}
\delta_{a b} \delta_{\dot{a} \dot{b}} = \sigma^{\underline{m}}_{a \dot{b}} \sigma^{\underline{m}}_{\dot{a} b} -\frac{1}{4}
\sigma^{\underline{m} \, \underline{n}}_{ab} \sigma^{\underline{m} \, \underline{n}}_{\dot{a} \dot{b}} \, , 
\end{equation} 
one has as desired, 
\begin{equation}
\begin{aligned}
\lambda^a_- & = \frac{\bar{\lambda}_+^{\dot{a}} \bar{\lambda}_+^{\dot{a}}}{(\bar{\lambda}_+ \bar{\lambda}_+)} \lambda^a_- = 
\frac{\bar{\lambda}_+^{\dot{a}}}{(\bar{\lambda}_+ \bar{\lambda}_+)} \left((\sigma^{\underline{m}} \bar{\lambda}_+)^a (\sigma^{\underline{m}} \bar{\lambda}_-)^{\dot{a}}
- \frac{1}{4} (\sigma^{\underline{m} \, \underline{m}} \bar{\lambda}_+)^{\dot{a}} (\sigma^{\underline{m} \, \underline{m}} \bar{\lambda}_-)^{a}   \right) \\
&= (\sigma^{\underline{m}} \bar{\lambda}_+)^a \frac{(\bar{\lambda}_+ \sigma^{\underline{m}} \lambda_-)}{(\bar{\lambda}_+ \bar{\lambda}_+)}
= - (\sigma^{\underline{m}} \bar{\lambda}_+)^a \frac{(\lambda_+ \sigma^{\underline{m}} \bar\lambda_-)}{(\bar{\lambda}_+ \bar{\lambda}_+)}
= \frac{1}{4} \frac{(\bar{\lambda}_+ \sigma^{\underline{m} \, \underline{n}} \bar{\lambda}_-)}{(\bar{\lambda}_+ \bar{\lambda}_+)}
(\sigma^{\underline{m} \, \underline{n}} \lambda_+)^{a} \, . 
\end{aligned}
\label{eq:minusintermsofplus}
\end{equation}
Note that the expression above is only defined for $(\bar\lambda_{+}\bar\lambda_{+})\neq 0$. This means that we solve partially the constraint such that $\lambda_{-}^{a}$ is completely fixed by the other components while $\lambda_{+}^{a}$ is completely unconstrained.

Then by taking $\lambda_+^a$
as independent variables, 
one can formally bosonizes them together with their conjugate 
momenta $w_a^+$ as \cite{FMS}
\begin{equation}
\lambda^a_+ \cong \eta^{(a)} e^{\phi^{(a)}} \, , \quad w^+_{a} \cong - e^{- \phi^{(a)}} \partial\xi_{(a)}  \, , \quad \delta(\lambda^a_+) \cong - e^{- \phi^{(a)}} \, . 
\label{bosonization}
\end{equation}
The number of delta functions determines the picture of the vertex operators. Notice that the delta functions are fermionic 
objects. In \cite{BerkovitsMaldacenaConjecture,BerkovitsHalf}, it was postulate how to change the picture of
a given vertex operator. For example, the picture minus seven vertex operator is obtained as follows
\begin{equation}
V^{(b)}_{- 7} = Q \cdot \xi_{(b)} \cdot V_{- 8} \, , 
\end{equation}
and 
\begin{equation}
\xi_{(a)} \cdot \delta(\lambda_+^a) = \frac{1}{\lambda_+^a} \, . 
\label{xioriginal}
\end{equation}
Apparently, due to the rule above the $V^{(b)}_{-7}$ vertex would have $\lambda$'s in the denominator and thus a problematic pole. 
However, it was argued in \cite{BerkovitsMaldacenaConjecture,BerkovitsHalf} that 
the denominators always disappear in a flat background. The same is true in $AdS$ and this will be proven in section 
\ref{halfBPSsection}. 
The flat space argument is as follows, if $V\delta(\lambda)$ is a BRST invariant operator then
necessarily 
$QV=\lambda(\dots)$ 
with the same $\lambda$
appearing in the argument of the delta function, 
thus $Q(\xi V\delta(\lambda))=\frac{\lambda}{\lambda}(\dots)$, i.e.
the denominator always cancel. 
The absence of denominators  
is used as a consistency check in the calculations.

It is not hard to see 
that $V^{\rm{flat}}_{-8}$ of
(\ref{vertexflat}) is BRST invariant as required, see 
\cite{LucasFlat,BerkovitsHalf,MikhailovZavaleta} for more details. In a flat background the pure spinors are BRST invariant 
and both the prefactor 
in (\ref{vertexflat}) 
and the delta functions are invariant. Moreover, expressing the BRST operator 
of (\ref{BRSToperator}) 
in terms of the covariant 
derivatives $\nabla_{\pm}$
and the pure spinors 
$\lambda_{\pm}$
and using the commutation relations 
\begin{equation}
[ \nabla_+ , \nabla_-] \propto k_+ \, , 
\end{equation}
and the
properties of the superfield $\Phi$ given in 
(\ref{eq:superphi}), one can show that only terms proportional to  $\lambda^a_{-}$ are left when acting with the BRST operator on the vertex. However, we have
\begin{equation}
\lambda_{-}^{a}\left(\prod_{a=1}^{8}\delta(\lambda_+^{a})\right)=0 \, . 
\end{equation}
This follows because of \eqref{eq:minusintermsofplus}, i.e. $\lambda^a_{-}$ can be expressed as a sum of terms 
and each term is proportional to $\lambda^{a}_+$
.
So, the vertex is BRST invariant. 

To go from the minus eight picture to the usual zero picture, we act with eight picture raising operators $Q 
\cdot \xi_{(a)}$. 
In this way, the zero picture vertex operator $V_0$ is 
written in terms of the minus eight picture vertex $V_{-8}$ as
\begin{equation}
V_0= Q\left(\xi_{(8)}\dots Q\left(\xi_{(1)}\cdot V_{-8}\right)\dots\right) \, . 
\end{equation} 
In \cite{LucasFlat,MikhailovZavaleta}, it was shown that $V^{\rm{flat}}_0$ 
obtained from the procedure above takes the form  
\begin{equation}
V^{\rm{flat}}_{0}=V^{\rm{flat}}_{0,0}+V^{\rm{flat}}_{0,2}+V^{\rm{flat}}_{0,4}+V^{\rm{flat}}_{0,6}+V^{\rm{flat}}_{0,8} \, . 
\label{Vflat}
\end{equation}
where $V_{0,n}$ 
preserves $SO(8)$ invariance and contains $n$ derivatives
$\nabla_+$ acting on the superfield $\Phi$.
In the section 
\ref{halfBPSsection},
we are going to take the flat space limit of our $AdS$ vertex operators and compare the results with the flat space ones. 
In $AdS$ we have only computed the bottom state
of the supermultiplet so we are going to show explicit 
expressions for 
$V^{\rm{flat}}_{0}$ 
only for  
$\Phi \sim e^{(i k_+ y)}$. 
Notice that the conventions in this paper differ from the ones 
in \cite{LucasFlat,BerkovitsHalf} by some numerical factors. The relevant flat space 
BRST transformations in this paper are 
\begin{equation}
Q \cdot \theta^a_+  
= \lambda^a_+ \, , \quad \quad 
Q \cdot e^{(i k_+ y)}
=  i k_+ \lambda^a_- \theta^a_+ \, e^{(i k_+ y)} \, . 
\label{flatspaceBRST2}
\end{equation}
The term $V^{\rm{flat}}_{0,0}$ 
in the vertex 
is given by 
\begin{equation}
V^{\rm{flat}}_{0,0} 
= 2 i (\bar{\lambda}_L^{\dot{a}} \bar{\lambda}_R^{\dot{a}} ) e^{(i k_+ y)} 
=( \bar\lambda^{\dot a}_{+} \bar\lambda^{\dot a}_+) e^{(i k_+ y)} \, , 
\end{equation}
where we have used 
the constraints 
(\ref{eq:so8pure}).  
In addition, 
\begin{equation}
V^{\rm{flat}}_{0,2}   
= - \frac{k_+}{4} 
( \bar{\lambda}_L \sigma^{ij} \bar{\lambda}_R)
(\theta_+ \sigma_{ij} \theta_+) e^{(i k_+ y)} \, , \end{equation}
and 
\begin{equation}
V^{\rm{flat}}_{0,4}  
= ( i k_+)^2 \frac{i}{192}
(\bar{\lambda}_L \sigma^{ijkl} \bar{\lambda}_R) (\theta_+ \sigma_{ij} \theta_+)
(\theta_+ \sigma_{kl} \theta_+)  e^{(i k_+ y)} \, .
\end{equation}
The remaining terms are 
\begin{equation}
V^{\rm{flat}}_{0,6}  
= (i k_+)^3 \frac{i}{11 520}
(\bar{\lambda}_L \sigma^{ijklmn} \bar{\lambda}_R) (\theta_+ \sigma_{ij} \theta_+)
(\theta_+ \sigma_{kl} \theta_+)
(\theta_+ \sigma_{mn} \theta_+)e^{(i k_+ y)} \, ,
\end{equation}
and 
\begin{equation}
 V^{\rm{flat}}_{0,8}  
=   2 i 
(i k_+)^4
(\bar{\lambda}^{\dot{a}}_L \bar{\lambda}^{\dot{a}}_R ) \, 
\theta_+^1 
\theta_+^2
\theta_+^3
\theta_+^4
\theta_+^5
\theta_+^6
\theta_+^7
\theta_+^8 \, . 
\end{equation}
In order to derive 
the expressions for $V^{\rm{flat}}_{0,n}$ given above,
in addition to
\eqref{eq:minusintermsofplus}
the following relations are useful (see  
\cite{LucasFlat,BerkovitsHalf,MikhailovZavaleta} for more details) 
\begin{equation}
\begin{aligned}
4! (\bar{\lambda}_+ \sigma^{[ij} \bar{\lambda}_-)     (\bar{\lambda}_+ \sigma^{kl]} \bar{\lambda}_-) 
= 4 (\bar{\lambda}^{\dot{a}}_+ \bar{\lambda}^{\dot{a}}_+) (\bar{\lambda}_+ \sigma^{ijkl} \bar{\lambda}_+ -
\bar{\lambda}_- \sigma^{ijkl} \bar{\lambda}_-) \, ,  \\
6! (\bar\lambda_+ \sigma^{[ij}
\bar\lambda_-)
(\bar\lambda_+ \sigma^{klmn]} \bar\lambda_+ - 
\bar\lambda_- \sigma^{klmn]} \bar\lambda_-)=
288 (\bar\lambda_+^{\dot{a}}
\bar\lambda_+^{\dot{a}})
(\bar\lambda_+ \sigma^{ijklmn} \bar\lambda_-) \, , 
\end{aligned}
\end{equation}
and
\begin{equation}
\begin{aligned}
\\
(\bar\lambda_+ \sigma^{ij}
\bar\lambda_-)
(\bar\lambda_+ \sigma^{klmn} \bar\lambda_+ - 
\bar\lambda_- \sigma^{klmn} \bar\lambda_-)
(\lambda_+ \sigma_{ij} \theta_+) 
(\theta_+ \sigma_{kl} \theta_+) 
(\theta_+ \sigma_{mn} \theta_+) 
= \\
2 (\bar\lambda_+ \sigma^{[ij}
\bar\lambda_-)
(\bar\lambda_+ \sigma^{klmn]} \bar\lambda_+ - 
\bar\lambda_- \sigma^{klmn]} \bar\lambda_-)
(\lambda_+ \sigma_{ij} \theta_+) 
(\theta_+ \sigma_{kl} \theta_+) 
(\theta_+ \sigma_{mn} \theta_+) \, . 
\end{aligned}
\end{equation}
Note that the vertex operator
(\ref{Vflat}) is in the gauge \begin{equation}
V_0^{\rm{flat}}=\bar\lambda^{\dot a}_{L}\bar\lambda^{\dot b}_{R}A_{\dot a\dot b} \, .  
\end{equation}

Naively, it seems that all the vertex operators obtained by acting with picture raising operators in lower picture vertices are BRST exact. 
In fact, for example, we have by construction that
\begin{equation}
V_0=Q(\xi_{(8)} \cdot V_{-1})=Q\left(\frac{\tilde V_{-1}}{\lambda_{+}^{8}}\right),\qquad {\rm{with}} \qquad V_{-1}=\tilde V_{-1} \delta(\lambda_{+}^{8}) \, . \end{equation}
It turns out that we must be careful with the inverse powers of $\lambda_{+}^{a}$. Operators that involve inverse powers of $\lambda_{+}^{a}$ are not globally defined in the pure spinor space, but only in a patch where $\lambda_{+}^{a}\neq 0$. If we include states that are not globally defined in the pure spinor space the BRST cohomology trivializes since $Q(\theta^{1}/\lambda^{1})=1$, so any BRST closed state would also be exact. This force us to consider only states that are globally defined in the pure spinor space and this implies 
that the vertices in  this construction are not necessarily BRST trivial.  

In the construction of the flat space vertex operators revised above, we see that both the vertex in the minus eight picture (\ref{vertexflat}) and the vertex in the zero picture 
(\ref{Vflat}) are in the gauge where only the component 
$A_{\dot{a} \dot{b}}$ of the gauge superfield is non-zero. This is not a coincidence and in fact the vertex in all the intermediate pictures between minus eight and zero are also in this gauge in flat space, see below.
Before ending this section, we are going to describe the possibility of redefining $\xi_{(a)}$ and changing the argument of the delta functions to $Q(\theta_{+}^{a})$ instead
of $\lambda^a_+$. 
This redefinition will not be used in the next sections, but it may be useful for some readers.   
In a flat background, these two quantities are equal but this is not true in general.
The delta functions 
$\delta(Q(\theta_{+}^{a}))$ 
are well defined 
if the 
$Q(\theta_{+}^{a})$'s are 
independent of each other and one way of verifying this is by computing the 
Jacobian to change the delta functions to 
$\delta(\lambda_{+}^{a})$ and verifying that it is non-singular. 
Suppose we are considering 
a supergravity background 
where some of its vertex 
operators have a minus eight picture description. 
These vertices will have the following general form 
\begin{equation}
V_{-8}=\frac{1}{8!} \varepsilon_{a_1\dots a_{8}}(\tilde V_{-8})^{a_1\dots a_8} \prod_{a=1}^{8}\delta(Q(\theta_{+}^{a})) \, . 
\label{generalminuseight}
\end{equation}
In
the formula above, 
$(\tilde V_{-8})^{a_1\dots a_{8}}$ is anti-symmetric on the indices $a_{i}=1,\dots,8$. The BRST invariance of $V_{-8}$ implies that
\begin{equation}
Q (\tilde V_{-8})^{a_1\dots a_8}=Q(\theta_+^{[a_1})(\tilde V_{-7})^{a_2\dots a_8]} \, , \end{equation}
with $(\tilde V_{-7})^{a_{1}\dots a_{7}}$ also anti-symmetric in all its indices. This follows 
because the BRST operator is nilpotent and it trivially annihilates the delta functions. We can use the $(\tilde V_{-7})$'s to construct a vertex operator in the minus seven picture as 
\begin{equation}
V_{-7}^{(a)}=\frac{1}{7!}\varepsilon^a_{ \; \,  a_1\dots a_7}(\tilde V_{-7})^{a_1\dots a_7} \prod_{b\neq a}\delta (Q(\theta^{b}_+)) \, .    
\end{equation}
Note that the minus eight picture vertex operator and the minus seven picture vertex operator constructed above are related by the generalized picture raising procedure given by 
\begin{equation}
Q(\xi_{(a)} \cdot V_{-8})  = V_{-7}^{(a)},\qquad {\rm{with}} \qquad \xi_{(a)}\delta (Q(\theta^{a}_+))=\frac{1}{Q(\theta^{a}_+)} \, . 
\end{equation}
Similar relations are obtained for all pictures. BRST invariance implies 
\begin{equation}
Q(\tilde V_{-n})^{a_1\dots a_{a_n}}= Q(\theta_+^{[a_1})(\tilde V_{-n+1})^{a_{2}\dots a_{n}]} \, , 
\end{equation}
and the $-n$ picture vertex operator is given by
\begin{equation}
V_{-n}^{(a_{1}\dots a_{(8-n)})}= \varepsilon_{a_1\dots a_{(8-n)}b_1\dots b_{n}}(\tilde V_{-n})^{b_1\dots b_{n}}\prod_{c\neq (a_{1}\dots a_{8-n})}\delta(Q(\theta_{+}^{c})) \, .     
\end{equation}
Notice that following the  
manipulations above, 
one can see that in a flat background if 
$\tilde V_{-8}$ only contains $\bar\lambda$ this will also be true for all the $\tilde V_{-n}$ until we reach $V_0$.
So the gauge condition 
is maintained by the picture raising procedure in the flat case. 

In the section 
\ref{halfBPSsection}, we are going to present 
the supergravity vertex 
operators in an 
$AdS_5\times S^{5}$ background. 
In that section, we are going to focus on a particular state instead of the full supermultiplet. 
However, 
it was argued in 
\cite{ChiralSuperfields} that first order on-shell  fluctuations 
about $AdS_5\times S^{5}$ can also be described by a chiral superfield $\Phi$ obeying similar constraints as the ones in 
(\ref{eq:superA}). 
Thus we expect that our vertex operators can be covariantizied similarly to what was done in flat space 
but this time using the $AdS_5 \times S^5$ algebra of
covariant derivatives.

\section{The AdS$_5 \times$ S$^5$ Pure Spinor Formalism} 
\label{purespinorformalism}

The $AdS_5 \times S^5$ pure spinor formalism is based in the superspace given by the following supercoset
\begin{equation}
\frac{PSU(2,2|4)}{USp(2,2) \times USp(4)} \, .
\label{supercoset}
\end{equation}
The bosonic subgroups $USp(2,2)$ and $USp(4)$ appearing in the denominator are the spin groups associated to the $SO(4,1)\subset SO(4,2)$ and $SO(5)\subset SO(6)$ isometries of $AdS_5\times S^{5}$. 
These subgroups are usually called
the isotropy groups. Note that the bosonic part of the numerator is given by $SU(2,2)\times SU(4)$ which are the spin groups respective to $SO(4,2)$ and $SO(6)$. This coset can be parametrized by ten bosonic coordinates  $x^{\alpha}_{\dot\alpha}, z, y^{\bar{i}}_i, w$ 
with $\alpha$, $\dot\alpha$, $i$, $\bar{i} =1,2$ 
and thirty two fermionic
coordinates $\theta^I_{\hat{I}}$ and $\theta^{\hat{I}}_I$, and the indices decompose as $I = \{ i, \bar{i} \}$ 
and $\hat{I}= \{ \alpha, \dot\alpha \}$. Notice that the fermionic variables are distinguished by the positions of their indices so
we must be careful if we want to lower and raise them using the $\epsilon$ symbols.  
Our 
coset parametrization will be given by
\begin{equation}
g = g_+ \bar{g}_+ \bar{g}_- g_- g_0 =e^{(\theta^{\alpha}_i q^{i}_{\alpha} + \theta^{\bar{i}}_{\dot\alpha} q^{\dot\alpha}_{\bar{i}} + x^{\alpha}_{\dot{\alpha}} K^{\dot{\alpha}}_{\alpha}+ y^{\bar{i}}_i K^{i}_{\bar{i}})}e^{(\theta^{\dot{\alpha}}_i q^i_{\dot\alpha}+ \theta^{\bar{i}}_{\alpha} q^{\alpha}_{\bar{i}})}
e^{(\theta^{i}_{\dot\alpha} q^{\dot\alpha}_{i}+ \theta^{\alpha}_{\bar{i}} q^{\bar{i}}_{\alpha})} e^{(\theta^{i}_{\alpha} q^{\alpha}_{i}+ \theta^{\dot\alpha}_{\bar{i}} q^{\bar{i}}_{\dot\alpha})} e^{ z \Delta + w J } \, ,
\label{eq:cosetparametrization} 
\end{equation}
where the $q$'s are the supersymmetry generators\footnote{Usually the fermionic generators 
of $\mathfrak{psu}(2,2|4)$ are divided into supercharges $q$ and special conformal
supercharges $s$. 
In our notation, the sixteen generators $s$ are 
$\{q^i_{\alpha}, q^{\bar{i}}_{\alpha},q^{\dot{\alpha}}_{i},q^{\dot{\alpha}}_{\bar{i}} \} $.}, the $K$'s are the conformal boosts, the $\Delta$ is the dilatation 
generator and $J$ is its $R$-charge analogous. The additional generators of the $\mathfrak{psu}(2,2|4)$ algebra are the translations $P$'s and the rotations
$\hat{M}$'s. Our conventions for the  algebra are given in the Appendix \ref{SectionAlgebra}. 
Choosing a coset parametrization as the one given in (\ref{eq:cosetparametrization}) is equivalent to fix the $USp(2,2)\times USp(4)$ gauge symmetries and the coordinates appearing in the coset are all gauge invariant by construction. 
Recall
that the coset transforms under an infinitesimal $USp(2,2)\times USp(4)$ gauge transformation with parameters $\Sigma^{\prime}$'s as 
\begin{equation}
\delta g = g \left( \hat{M}^{\alpha}_{\beta} \Sigma^{\prime \beta}_{\alpha} + \hat{M}^{\dot\alpha}_{\dot\beta} \Sigma^{\prime \dot\beta}_{\dot\alpha} 
+\hat{M}^{i}_{j} \Sigma^{\prime j}_{i} + \hat{M}^{\bar{i}}_{\bar{j}} \Sigma^{\prime \bar{j}}_{\bar{i}}+
(P^{\alpha}_{\dot{\alpha}} + \epsilon_{\dot\alpha \dot\beta} \epsilon^{\alpha \beta} K_{\beta}^{\dot\beta}) \Sigma_{\alpha}^{\prime \dot\alpha} +
(P^{\bar{i}}_i + \epsilon_{ij} \epsilon^{\bar{i} \bar{j}} K_{\bar{j}}^j) \Sigma^{\prime i}_{\bar{i}}  \right) \, .
\label{gaugetransformation}
\end{equation}

The coset choice (\ref{eq:cosetparametrization}) was not arbitrary. We have chosen a parametrization where all the generators that annihilate our
vertex operators (see the next section) are located at the left. This choice greatly simplifies the calculations. In particular, the vertex operators  
will not depend on $\theta^{\alpha}_i, \theta^{\bar{i}}_{\dot\alpha}, \theta^{\bar{i}}_{\alpha},
\theta^{\dot\alpha}_{i},
\theta_{\bar{i}}^{\alpha},
\theta_{\dot\alpha}^{i},
x^{\alpha}_{\dot{\alpha}}$ and  $y^{\bar{i}}_i$.

Under a BRST transformation generated by $Q$ the coset transforms as 
\begin{equation}
\begin{aligned}
Q \cdot g & = g \, (\lambda_+ q_- + \lambda_-q_+ + \bar{\lambda}_+ \bar{q}_- + \bar{\lambda}_- \bar{q}_+) \\ 
& + g \left( \hat{M}^{\alpha}_{\beta} \Sigma^{\beta}_{\alpha} + \hat{M}^{\dot\alpha}_{\dot\beta} \Sigma^{\dot\beta}_{\dot\alpha} 
+\hat{M}^{i}_{j} \Sigma^{j}_{i} + \hat{M}^{\bar{i}}_{\bar{j}} \Sigma^{\bar{j}}_{\bar{i}}+
(P^{\alpha}_{\dot{\alpha}} + \epsilon_{\dot\alpha \dot\beta} \epsilon^{\alpha \beta} K_{\beta}^{\dot\beta}) \Sigma_{\alpha}^{\dot\alpha} +
(P^{\bar{i}}_i + \epsilon_{ij} \epsilon^{\bar{i} \bar{j}} K_{\bar{j}}^j) \Sigma^i_{\bar{i}}  \right) \, , 
\end{aligned}
\label{eq:BRST}
\end{equation}
where the $\lambda$'s are the bosonic pure spinor variables and we have used the definitions 
\begin{equation}
\lambda^a_+ = (\lambda^i_{\alpha}, \lambda^{\dot\alpha}_{\bar{i}}) \, , \quad \lambda^a_- =(\lambda^{\bar{i}}_{\dot\alpha}, \lambda^{\alpha}_i) \, , \quad
\bar\lambda^{\dot{a}}_+ = (\lambda^i_{\dot{\alpha}} , \lambda^{\alpha}_{\bar{i}})  \, , \quad \bar{\lambda}^{\dot{a}}_- = (\lambda^{\bar{i}}_{\alpha}, \lambda^{\dot{\alpha}}_i) 
\, ,
\label{eq:so8lambda}
\end{equation}
and 
\begin{equation}
q^a_+ = (q^i_{\alpha}, q^{\dot\alpha}_{\bar{i}}) \, , \quad q^a_- =(q^{\bar{i}}_{\dot\alpha}, q^{\alpha}_i) \, , \quad 
\bar{q}^{\dot{a}}_+ = (q^{\alpha}_{\bar{i}}, q^i_{\dot{\alpha}} ) \, , \quad \bar{q}^{\dot{a}}_- = (q^{\bar{i}}_{\alpha}, q^{\dot\alpha}_i) \, , \quad
\label{eq:qinso8} 
\end{equation} 
where $a$ and $\dot{a}$ are $SO(8)$ spinor indices. The subscripts in the definitions above indicate the charge under the generator $J$. 
The contraction of the indices in  (\ref{eq:BRST}) is the obvious one, for example, 
\begin{equation}
\lambda_+ q_- = \lambda^i_{\alpha} q^{\alpha}_i + \lambda^{\dot\alpha}_{\bar{i}} q^{\bar{i}}_{\dot\alpha} \, . 
\end{equation}

In (\ref{eq:BRST}), the second line contains a restoring gauge transformation parametrized by the $\Sigma$'s. This is always the case when we want to preserve a coset parametrization. 
In $AdS$,
the $\lambda$'s are not BRST invariant anymore and they transform as  
\begin{equation}
Q \cdot \lambda^{I}_{\hat{I}} = \Sigma^I_J \lambda^J_{\hat{I}} - \Sigma^{\hat{J}}_{\hat{I}} \lambda_{\hat{J}}^I \, ,
\label{brstlambda} 
\end{equation}
and similarly for $\hat{I} \leftrightarrow I$. 
Notice that a different approach is possible, one example being 
\cite{BerkovitsChandia}, where the $USp(2,2)\times USp(4)$ gauge transformations are not fixed. In this case, the BRST transformations are defined up to a gauge transformation. Here we are fixing the gauge and fixing a particular BRST transformation that preserves our coset parametrization.

In order for the BRST operator to be nilpotent\footnote{If the gauge symmetry is not fixed the BRST operator can be nilpotent up to gauge transformations, but in this work, since we gauge fix, the BRST operator must be nilpotent.}, and the theory well defined,
the variables $\lambda$'s have to satisfy several quadratic constraints,
the so called pure spinor constraints. One way to derive these constraints is by noticing that (with a compact notation) 
\begin{equation}
Q^2 \cdot g = g (\lambda q + \Sigma)(\lambda q + \Sigma) + g ((Q \cdot \lambda) q + (Q \cdot \Sigma)) \, , 
\end{equation}
and using the BRST transformation of $\lambda$ given in  (\ref{brstlambda}), one concludes that the necessary condition for $Q$ to be nilpotent is 
\begin{equation}
\{ \lambda q , \lambda q \} \, \in \,
\mathfrak{usp}(2,2) \times \mathfrak{usp}(4)  \, . 
\end{equation}
To see the implications of the above constraints, we need the following definitions    
\begin{equation}
\begin{aligned}
& \tilde{\lambda} \equiv \{ \lambda^{\alpha}_i, \lambda_{\bar{i}}^{\dot{\alpha}}, \lambda^{\alpha}_{\bar{i}} , \lambda_i^{\dot\alpha} \} \, , \quad \tilde{q} 
\equiv  \{ q^i_{\alpha}, q^{\bar{i}}_{\dot{\alpha}},  q^{\bar{i}}_{\alpha}, q^i_{\dot\alpha} \} \, , \\
& 
\tilde{\bar{\lambda}} \equiv
 \{  -\epsilon_{i j} \epsilon^{\alpha \beta} \lambda_{\beta}^{j}, -\epsilon_{\bar{i} \bar{j}} \epsilon^{\dot\alpha \dot\beta} \lambda_{\dot\beta}^{\bar{j}} ,
\epsilon_{\bar{i} \bar{j}} \epsilon^{\alpha \beta} \lambda_{\beta}^{\bar{j}} , \epsilon_{i j} \epsilon^{\dot\alpha \dot\beta} \lambda_{\dot\beta}^{j}   \} \, ,  \\
& 
\tilde{\bar{q}} \equiv \{  -\epsilon^{i j} \epsilon_{\alpha \beta} q^{\beta}_{j}, -\epsilon^{\bar{i} \bar{j}} \epsilon_{\dot\alpha \dot\beta} q^{\dot\beta}_{\bar{j}} ,
\epsilon^{\bar{i} \bar{j}} \epsilon_{\alpha \beta} q^{\beta}_{\bar{j}} , \epsilon^{i j} \epsilon_{\dot\alpha \dot\beta} q^{\dot\beta}_{j}   \} \, . 
\label{Definitionsoftildespinors}
\end{aligned} 
\end{equation} 
The last two definitions can be written compactly as 
\begin{equation} 
\tilde{\bar{\lambda}}^{\hat{I}}_I = \sigma_6^{\hat{I} \hat{J}} (\sigma_{6})_{I J} \lambda^{J}_{\hat{J}}  \, , \quad \quad 
\tilde{\bar{q}}^{I}_{\hat{I}} = \sigma_6^{I J} (\sigma_{6})_{\hat{I} \hat{J}} q^{\hat{J}}_{J} \, , 
\end{equation}
where $\sigma_n^{IJ}$ are $SO(6)$ Pauli matrices, with $n=1, \ldots, 6$. They obey
\begin{equation}
\sigma_n^{IJ} = \frac{1}{2} \epsilon^{IJKL} (\sigma_n)_{KL} \, , 
\label{loweringindices}
\end{equation} 
and in our conventions 
\begin{equation}
\sigma_6^{IJ} = \left(
\begin{tabular}{cc}
 $\epsilon^{i j}$ & 0  \\
0& $-\epsilon^{\bar{i} \bar{j}}$ 
\end{tabular} \right) \, . 
\end{equation}
This motivates the definitions 
\begin{equation}
\epsilon^{IJ}=\sigma_{6}^{IJ},\quad \epsilon^{\hat I\hat J}=\sigma_{6}^{\hat I\hat J},\quad \epsilon_{IJ}=\sigma^{6}_{IJ},\quad \epsilon_{\hat I\hat J}=\sigma^{6}_{\hat I\hat J} \, , \label{sigma6}
\end{equation}
which enables us to raise and lower indices that transform under $USp(2,2)$ and $USp(4)$. Taking the following linear combination of the generators \begin{equation}
i q_R \equiv \tilde{q} + \tilde{\bar{q}} \, , \quad q_L \equiv \tilde{q} - \tilde{\bar{q}} \, ,
\end{equation}
it is possible to show using the commutation relations of the Appendix \ref{SectionAlgebra} that 
\begin{equation}
\{ q_R , q_L \}  \; \in \; 
\mathfrak{usp}(2,2) \times \mathfrak{usp}(4)\, , \quad {\rm{and}} \quad 
\{ q_R , q_ R \} , \; \{ q_L , q_L\} \, \not\in \, 
\mathfrak{usp}(2,2) \times \mathfrak{usp}(4) \, . 
\end{equation}
The next step is to write the combination of $\lambda$'s and $q$'s appearing in the BRST transformation 
(\ref{eq:BRST}) in the new notation
\begin{equation}
(\lambda_+ q_- + \lambda_-q_+ + \bar{\lambda}_+ \bar{q}_- + \bar{\lambda}_- \bar{q}_+) = \tilde{\lambda} \tilde{q} + 
\tilde{\bar{\lambda}} \tilde{\bar{q}} = \frac{1}{2 i}(\tilde{\lambda}+\tilde{\bar{\lambda}}) q_R + \frac{1}{2}(\tilde{\lambda}-\tilde{\bar{\lambda}}) 
q_L \equiv \frac{1}{2} \lambda_R  \, q_R + \frac{1}{2} \lambda_L \, q_L \, .
\label{definitionLR}
\end{equation}
Thus the pure spinor constraints become
\begin{equation}
\{ \lambda_R \, q_R, \lambda_R  \, q_R \} =0 \, , \quad \quad \{ \lambda_L \, 
q_L, \lambda_L \, 
q_L \} =0 \, , 
\end{equation} 
or equivalently 
\begin{equation}
[ ( \tilde{\lambda} \tilde{q}) , (\tilde{\lambda} \tilde{\bar{q}})] + 
[ ( \tilde{\bar{\lambda}} \tilde{q}) , (\tilde{\bar{\lambda}} \tilde{\bar{q}})] =0 \, , \quad \quad 
[ ( \tilde{\lambda} \tilde{q}) , (\tilde{\bar\lambda} \tilde{\bar{q}})] + 
[ ( \tilde{\bar{\lambda}} \tilde{q}) , (\tilde{\lambda} \tilde{\bar{q}})] =0 \, . \end{equation}
Evaluating the commutators, the constraints become in components 
\begin{equation}
\begin{aligned} 
& \hat\lambda^I_{\hat{I}} \hat\lambda^{\hat{I}}_J + \epsilon^{I K} \epsilon_{J L} \hat{\lambda}^L_{\hat{I}} \hat{\lambda}^{\hat{I}}_K = \frac{1}{2} \delta^I_J 
(\hat\lambda^K_{\hat{I}}  \hat\lambda^{\hat{I}}_K) \, , \\
& \epsilon^{\hat{I} \hat{J}} \hat\lambda^I_{\hat{I}} \hat\lambda^J_{\hat{J}} + \epsilon^{I K} \epsilon^{J L} \epsilon_{\hat{I} \hat{J}} \hat\lambda^{\hat{I}}_K \hat\lambda^{\hat{J}}_L 
=  \frac{1}{4} \epsilon^{I J} (\epsilon_{\hat{K} \hat{L}} \epsilon^{K L} \hat\lambda^{\hat{K}}_K \hat\lambda^{\hat{L}}_L+
\epsilon^{\hat{K} \hat{L}} \epsilon_{K L} \hat\lambda_{\hat{K}}^K \hat\lambda_{\hat{L}}^L) \, , 
\end{aligned} 
\label{eq:so6pure}
\end{equation}
and similarly for $I \leftrightarrow \hat{I}$. In the expression above, we have used the definitions 
\begin{equation}
\hat\lambda^{\bar{i}}_{\alpha} = - \lambda^{\bar{i}}_{\alpha} \, , \quad \hat\lambda^{\dot\alpha}_{i} = - \lambda^{\dot\alpha}_{i} \, , \quad
{\rm{and}} \quad \hat\lambda^I_{\hat{J}} = \lambda^I_{\hat{J}} \, , \quad \hat\lambda^{\hat{I}}_J = \lambda^{\hat{I}}_J \quad {\rm{otherwise}}. 
\end{equation}
Note that the pure spinor constraints above are given in $SO(8)$ notation in  
(\ref{eq:so8pure}).

In the next section, we will define and compute the half-BPS vertex operators in two different pictures.  
The procedure for moving between pictures was defined in \cite{BerkovitsMaldacenaConjecture,BerkovitsHalf} and involves BRST
transformations. 
In flat space, the BRST transformation of the variables is simpler and one can change the picture analytically  \cite{LucasFlat,MikhailovZavaleta}. In $AdS$, the calculation is complicated by the fact 
that the $\lambda$'s are not BRST invariant, see (\ref{brstlambda}), and the transformations of the remaining 
worldsheet variables have many terms. The full set of BRST transformations solving (\ref{eq:BRST}) are given in the
Appendix \ref{appendixcoset}. Thus in this work, all the $AdS$ calculations were performed using a computer. 
Our strategy was to replace every $\lambda_-^a$ by the other $\lambda$'s by using (\ref{eq:minusintermsofplus})
and then using the additional constraints \begin{equation}
\quad \bar{\lambda}_+ \bar{\lambda}_+ + \bar{\lambda}_- \bar{\lambda}_- =0 \, , \quad \bar{\lambda}_+ \bar{\lambda}_ -=0 \, . 
\label{barlambdapure}
\end{equation}
The remaining simplifications using the pure spinor constraints where done numerically. The solution of the 
pure spinor constraints using an $U(5)$ notation is well known and it can be found in many places in the literature, see \cite{MafraThesis} for example. 
In particular, the solution is parametrized by eleven numbers which is the number of independent components
of a pure spinor in ten dimensions. It is not difficult to map numerical solutions in $U(5)$ notation to solutions of 
the constraints in $SO(8)$ or $SO(6)$ notation given in  (\ref{eq:so8pure})  and (\ref{eq:so6pure}) respectively. 

The pure spinor action in $AdS_5 \times S^5$ is known and it is written in terms of currents constructed
from the coset $g$. The action will not be needed in this work and we refer the reader to the literature 
for its expression. However, the action will be important for the computation of amplitudes because we will
need to know the OPE's of various worldsheet fields. In this work, only the BRST transformations are needed 
and the knowledge that physical states are states in the cohomology of the BRST operator with ghost
number two. 

We have chosen to work with the coset parametrization (\ref{eq:cosetparametrization}) where the worldsheet coordinates 
are gauge invariant. As mentioned before, in this case the $\lambda$'s are not BRST invariant and transform under restoring
gauge transformations. 
Alternatively, it is possible to use different cosets and define gauge invariant $\lambda$'s. In the Appendix \ref{appendixcoset}, we give an example of such
a coset. The main disadvantage of such cosets is that the pure spinor constraints are more involved and 
working with them even numerically 
is quite complicated. 
Nevertheless, we believe that the easiest way of obtaining the vertex operators for any alternative cosets
is to generate a basis of invariants and solving the condition of BRST closeness by numerically fitting for the coefficients of the basis. It is also possible to write a second basis and verify that the obtained vertex is not BRST exact.    
 
\section{The Half-BPS Vertex operators}\label{halfBPSsection} 

In this section, we are going to compute the vertex operator for any half-BPS state in the zero picture. 
This is the main result of the paper. As mentioned in the Introduction, it is known that 
these vertex operators can be described by just one superfield \cite{BerkovitsChandia}. However,
the explicit expression for this superfield was not known. 
Our strategy for the calculation was to start with the vertices in the minus eight picture 
derived in \cite{BerkovitsMaldacenaConjecture,BerkovitsHalf}  and change the picture step by step 
until we reach the zero picture vertices. 
In this section, we are going to redefine the $\theta$'s in order to absorb some of the exponential factors  
and have more compact expressions, we will use 
\begin{equation}
\tilde{\theta}_{\pm} \equiv e^{\pm \frac{w-z}{2}} \theta_{\pm} \, , 
\quad 
\tilde{\bar{\theta}}_{\pm} \equiv e^{\pm \frac{w+z}{2}} \bar{\theta} _{\pm} \, ,
\label{eq:thetatilde} 
\end{equation}
where 
\begin{equation}
\theta^a_+ = (\theta^i_{\alpha}, \theta^{\dot\alpha}_{\bar{i}}) \, , \quad \theta^a_- =(\theta^{\bar{i}}_{\dot\alpha}, \theta^{\alpha}_i) \, , \quad
\bar\theta^{\dot{a}}_+ = (\theta^i_{\dot{\alpha}} , \theta^{\alpha}_{\bar{i}})  \, , \quad \bar{\theta}^{\dot{a}}_- = (\theta^{\bar{i}}_{\alpha}, \theta^{\dot{\alpha}}_i) \, .  
\label{eq:deftheta}
\end{equation} 
 The BRST transformations of the $\tilde{\theta}$'s are easily deduced
\begin{equation}
\begin{aligned}
Q(\tilde{\theta}_{\pm}) = e^{\pm \frac{w-z}{2}} Q(\theta_{\pm})  \pm \frac{1}{2} Q(w-z) \tilde{\theta}_{\pm} \, , \quad
Q(\tilde{\bar{\theta}}_{\pm}) = e^{\pm \frac{w+z}{2}} Q(\bar{\theta}_{\pm})  \pm \frac{1}{2} Q(w+z) \tilde{\bar{\theta}}_{\pm} \, .
\end{aligned}
\end{equation}
Notice that the $\tilde{\theta}$ are chargerless under
$\Delta$ and $J$, see the Appendix \ref{SectionAlgebra} for our conventions.  
In what follows, we are going to suppress the tilde from the $\theta$'s to avoid cluttering 
and we hope that this does not cause any confusion.

Let's start by reviewing the properties of half-BPS operators. 
It is well known that the $AdS_5 \times S^5$ superstring theory is dual to $\mathcal{N}=4$ Super-Yang-Mills (SYM). 
All the single trace gauge invariant half-BPS operators in SYM have been classified and they take the following form
\begin{equation}
\mathcal{O}_L (x, y) =  {\rm{Tr}} \, ((\bar{y} \cdot \hat\Phi)^L)(\tilde{x}) \, ,  
\end{equation}
where the trace is over the gauge group indices, $\bar{y}$ is a null six dimensional vector $\bar{y} \cdot \bar{y} =0$ called the polarization vector,
$L$ is the length of the operator, $\tilde{x}$ is its spacetime position 
(the relation between the variables $\tilde{x}^{\mu}$ and the variables $x^{\alpha}_{\dot\alpha}$
appearing in the coset parametrization (\ref{eq:cosetparametrization})
will be discussed in the next section)
and finally $\hat\Phi$'s 
are the six real scalars of the theory.  
We are going to consider these operators at the spacetime position $\tilde{x}^{\mu}=0$ and select a specific polarization vector 
such as the operator has charge $L$ under the $U(1)$ generator $J$. One possible choice of polarization is 
\begin{equation}
\bar{y} = \{ 1, i, 0,0,0,0 \} \, .
\end{equation}
Starting with an operator with this properties, it is possible to get more general ones by applying spacetime translations 
and $SO(6)$ rotations. This specific set of operators have $\Delta = J = L$. 
Being a half-BPS operator, it is annihilated by twenty four supercharges. The operators are superconformal primaries and thus any susy generator that lower
its dimension has to annihilated it. The operators are also $SO(6)$ highest weight operators so any susy generator
that raises its $J$ charge must annihilate it. The conclusion is that all the susy's generators with $J-\Delta \geq 0$ annihilate it.
The charge of all the generators can be found in the Appendix \ref{SectionAlgebra}. The conclusion is that the supermultiplet
that our half-BPS operators belong only depends on $\theta^a_+$ defined in (\ref{eq:deftheta}). In what follows,
we will look for vertex operators with the same properties. In the minus eight picture they are given by 
\begin{equation}
V_{- 8}(n)  = (\bar{\lambda}_+ \bar{\lambda}_+)  \, e^{n(z+w)} \prod_{\dot\alpha, \bar{i} =1,2} \theta^{\dot\alpha}_{\bar{i}} \prod_{ \alpha,i =1,2} 
\theta^i_{\alpha} \prod_{\dot\alpha, \bar{i} =1,2} \delta(\lambda^{\dot{\alpha}}_{\bar{i}}) \prod_{ \alpha,i =1,2} \delta(\lambda^i_{\alpha}) \, , 
\label{eq:minus8vertex} 
\end{equation}
where $n$ is equal to $-L$ and that comes from how the vertex operators transform under $PSU(2,2|4)$. The prefactor is 
\begin{equation}
(\bar{\lambda}_+ \bar{\lambda}_+) = \lambda^{\alpha}_{\bar{i}} \lambda^{\beta}_{\bar{j}} 
\epsilon_{\alpha \beta} \epsilon^{\bar{i} \bar{j}} - \lambda^{i}_{\dot{\alpha}} \lambda^{j}_{\dot{\beta}} \epsilon_{ij} \epsilon^{\dot\alpha \dot\beta} \, . 
\label{eq:lambdabarplus}   
\end{equation}
Notice that using the more usual scalar prefactor $(\lambda_L \lambda_R)$ where 
both $\lambda_L$ and $\lambda_R$ were defined in  (\ref{definitionLR}) is equivalent. The term $(\lambda_L \lambda_R)$
depends on all the $\lambda$'s but the dependency in $\lambda_+$ and $\lambda_-$ are killed by the delta functions in $V_{-8}$
and the dependency in $\bar\lambda_-$ can be eliminated due to the first pure spinor constraint given in 
(\ref{barlambdapure}). It is possible to show that all $V_{-8}(n)$ are annihilated by the same set 
of generators that annihilate the dual operators. This follows because of their $SO(4) \times SO(4)$ invariance, i.e. it is possible to write 
$V_{-8}(n)$ with all the indices $\alpha, \dot{\alpha}, i, \bar{i}$ contracted with $\epsilon$'s tensors.
Alternatively, one can show this explicitly by computing the global 
$PSU(2,2|4)$
transformations of the wordsheet variables.
Recall that under a global $PSU(2,2|4)$ transformation with an element $g_p$ the coset representative $g$ of  
(\ref{eq:cosetparametrization}) transforms as 
\begin{equation}
g_p \, g(z,w,\theta,x,y)  = g(z',w',\theta',x',y') h \, ,
\label{finitetransformations}
\end{equation}
where the variables with primes are transformed variables and $h$ is a gauge transformation. 
For the case of an infinitesimal transformation with parameter $\epsilon^A$  the formula above takes the form 
($T_A$ are generators of the supergroup and $X^M$ are the collection of all worldsheet variables 
with $X^{'M} = X^M + \delta X^M$)  
\begin{equation}
(1 + \epsilon^A T_A) g(X) = (g(X) + \delta X^M \frac{\partial g(X)}{\partial X^M})(1+ \delta h) \, ,   
\end{equation} 
equivalently 
\begin{equation}
g(X)^{-1} \, (\epsilon^A T_A) \, g(X)= \delta X^M \hat{e}_M^A T_A + \delta h \, , \quad {\rm{with}} \quad 
\hat{e}_M^A T_A = g(X)^{-1} \frac{\partial g(X)}{\partial X^M} \, . 
\label{transformationsinfinitesimal}
\end{equation}
Notice that for a general $PSU(2,2|4)$ transformation 
there is a compensating gauge transformation (the factor $\delta h$ above) 
and the pure spinor variables rotate accordingly. 
This is similar to the BRST transformations of the pure spinors given in (\ref{brstlambda}). 
By computing the transformations of the worldsheet variables using (\ref{transformationsinfinitesimal}) for different generators, it is possible 
to show that our vertex operators $V_{-8}(n)$ given in (\ref{eq:minus8vertex}) are annihilated by all susy generators with 
$J-\Delta \geq 0$ and by both $K^{\dot{\alpha}}_{\alpha}$ and $K^{i}_{\bar{i}}$. In fact, the combination $z+w$ appearing 
in the exponential has the only nontrivial transformations   
\begin{equation}
\delta z + \delta w =  - \epsilon^i_{\alpha} \theta^{\alpha}_i + \epsilon^{\dot{\alpha}}_{\bar{i}} \theta^{\bar{i}}_{\dot{\alpha}} 
- y^{\bar{i}}_i \epsilon^i_{\bar{i}} + x^{\alpha}_{\dot{\alpha}} \epsilon^{\dot{\alpha}}_{\alpha} \, . 
\end{equation}
Note that the parameters $\epsilon^i_{\alpha}$ and $\epsilon^{\dot{\alpha}}_{\bar{i}}$ are 
associated with the supersymmetry generators $q^a_-$ which all have 
$J-\Delta = -1$ and both $\epsilon^i_{\bar{i}}$ and $\epsilon^{\dot{\alpha}}_{\alpha}$ are associated 
with the translation generators $P^{\bar{i}}_i$ and $P^{\alpha}_{\dot\alpha}$. Moreover, not considering transformations 
generated by the $P$'s and $q^a_-$, we have 
\begin{equation}
\begin{aligned}
&\delta \theta^i_{\alpha} = - \frac{1}{2}  \theta^i_{\alpha} (\epsilon^j_{\dot\alpha}\theta^{\dot\alpha}_j - \epsilon^{\beta}_{\bar{i}} \theta^{\bar{i}}_{\beta} )
- \epsilon^{i}_{\dot\alpha} \theta^j_{\alpha} \theta^{\dot\alpha}_j + \epsilon^{\beta}_{\bar{i}} \theta^i_{\beta} \theta^{\bar{i}}_{\alpha} \, , \\
& \delta \theta^{\dot\alpha}_{\bar{i}} =  \frac{1}{2}  \theta^{\dot\alpha}_{\bar{i}} 
(\epsilon^{\alpha}_{\bar{j}} \theta^{\bar{j}}_{\alpha} - \epsilon^i_{\dot\beta} \theta^{\dot\beta}_i)
- \epsilon^i_{\dot\beta} \theta^{\dot\beta}_{\bar{i}} \theta^{\dot\alpha}_i 
- \epsilon^{\alpha}_{\bar{i}} \theta^{\bar{j}}_{\alpha} \theta^{\dot\alpha}_{\bar{j}}  \, . 
\end{aligned} 
\end{equation}
Despite the fact that $ \theta^i_{\alpha}$ and  $\theta^{\dot\alpha}_{\bar{i}}$ have nontrivial transformations,
the combination of eight thetas appearing in the vertex $V_{-8}(n)$ is invariant due to cancellations. Finally,  
again setting $\epsilon^i_{\alpha}, \epsilon^{\dot{\alpha}}_{\bar{i}} , \epsilon^i_{\bar{i}}, \epsilon^{\dot{\alpha}}_{\alpha} 
\rightarrow 0$, the compensating gauge transformation $\delta h$ is given in terms 
of the following rotation parameters (up to factors involving traces), see 
\eqref{gaugetransformation},
\begin{equation}
\begin{aligned}
\Sigma^{\prime \beta}_{\alpha}  = - \epsilon^{\beta}_{\bar{i}} \theta^{\bar{i}}_{\alpha} \, , \quad
\Sigma^{\prime \dot\beta}_{\dot\alpha} = - \epsilon^i_{\dot\alpha} \theta^{\dot\beta}_i  \, , \quad 
 \Sigma^{\prime j}_{i} = - \epsilon^j_{\dot{\alpha}} \theta^{\dot\alpha}_i \, , \quad  
 \Sigma^{\prime \bar{j}}_{\bar{i}} = - \epsilon^{\alpha}_{\bar{i}} \theta^{\bar{j}}_{\alpha} \, , \quad
 \Sigma_{\alpha}^{\prime \dot\alpha}  =0 \, , \quad  \Sigma^{\prime i}_{\bar{i}} =0  \, . 
\end{aligned} 
\end{equation}
It is easy to show that the prefactor $(\bar{\lambda}_+ \bar{\lambda}_+)$ and the product of eight 
deltas are invariant under the above gauge transformations. These arguments prove that $V_{-8}(n)$ are annihilated
by both twenty four supersymmetries and the $K$'s as required. 

To show that the vertices $V_{-8}(n)$ are BRST invariant one uses the explicit 
transformations given in the Appendix \ref{appendixcoset}. 
First, all the terms in $\delta z$ + $\delta w$ contains a $\theta^a_+$. 
Second, all the terms in the sigmas $\Sigma^{\dot{\alpha}}_{\alpha}$ and $\Sigma^i_{\bar{i}}$ of (\ref{eq:BRST}) contain 
also at least one $\theta^a_+$.  
The remaining $\Sigma$'s are related with transformations generated 
by the $\hat{M}$'s and these transformations annihilated the vertex because 
both the prefactor   $(\bar{\lambda}_+ \bar{\lambda}_+)$ and the product of eight delta functions
are $SO(4) \times SO(4)$ invariant. Finally, $\delta \theta^a_+$ has the term $\lambda^a_+$ which is killed by the delta functions
and the remaining terms cancel between the transformations of the eight thetas.  
Alternatively, one can think that the arguments of the delta functions are $Q \cdot \lambda^a_+$ as was done in 
(\ref{generalminuseight}), so 
the invariance of the vertex operators under the BRST transformations of the $\theta^a_+$ is manifest. 
Then expanding the delta functions,
all terms proportional to derivatives contain at least one 
$\theta^a_+$ and are killed, thus we return to the expression for $V_{-8}(n)$ in 
(\ref{eq:minus8vertex}). 
During our computation we have checked that all the $V_{-n^{\prime}}(n)$ are BRST invariant for any $n^{\prime}$ between zero and eight. This is expected since the 
BRST operator is nilpotent. 
In the subsection below, we also argue that the picture raising procedure is well defined in $AdS_5 \times S^5$. 

The picture zero vertex operators $V_{0}(n)$ that are obtained by the picture raising procedure starting with    
the picture minus eight vertices $V_{-8}(n)$ are 
 \begin{equation}
e^{-n(z+w)} V_{0}(n) = V_0^{0}+ V_0^{2} +  V_0^{4} +  V_0^{6} +  V_0^{8}  \, , 
\label{vertexoperatorso4}
\end{equation} 
and in the notation above the superscript indicates the number of thetas. We have 
\begin{equation}
\begin{aligned}
V^0_{0} = (\bar{\lambda}_+ \bar{\lambda}_+) \, ,  
\end{aligned} 
\end{equation}
where $(\bar{\lambda}_+ \bar{\lambda}_+)$ was defined in (\ref{eq:lambdabarplus}). We will also need the definitions 
\begin{equation}
\begin{aligned}
& \bar{\lambda}^2_{+1} \equiv \lambda^{\alpha}_{\bar{i}} \lambda^{\beta}_{\bar{j}} \epsilon^{\bar{i} \bar{j}} \epsilon_{\alpha \beta} \, , \; \quad   \bar{\lambda}^2_{+2} \equiv \lambda^{i}_{\dot{\alpha}} \lambda^{j}_{\dot{\beta}}  \epsilon^{\dot{\alpha} \dot{\beta}} \epsilon_{i j} \, , \\
& \bar{\lambda}^2_{-1} \equiv \lambda_{\alpha}^{\bar{i}} \lambda_{\beta}^{\bar{j}} \epsilon_{\bar{i} \bar{j}} \epsilon^{\alpha \beta} \, , \; \quad \bar{\lambda}^2_{-2}  \equiv \lambda^{\dot\alpha}_i \lambda^{\dot\beta}_j \epsilon_{\dot\alpha \dot\beta} \epsilon^{i j}  \, . \; \quad
 \end{aligned}
\end{equation}
Note that the terms above 
are not all independent due to the
pure spinor cointraints
\begin{equation}
\bar{\lambda}^2_{+1}  
- \bar{\lambda}^2_{+ 2}
+ \bar{\lambda}^2_{-1}
- \bar{\lambda}^2_{-2} =0 \, . 
\end{equation}
The term $V_0^2$ is given by
\begin{equation}
\begin{aligned}
&V_0^2 =  -(1+n) (\lambda^{\alpha}_{\bar{i}} \lambda^{\bar{i}}_{\gamma} \theta^i_{\alpha} \theta^{j}_{\beta} \epsilon_{i  j} \epsilon^{\beta \gamma}+\lambda^i_{\dot{\alpha}} \lambda^{\dot{\beta}}_i \theta^{\dot{\alpha}}_{\bar{i}} \theta^{\dot{\gamma}}_{\bar{j}} \epsilon_{\dot{\gamma} \dot{\beta}} \epsilon^{\bar{i} \bar{j}}) \\
-(1-n) ( \lambda^{\alpha}_{\bar{i}} \lambda_{\alpha}^{\bar{j}} \theta^{\dot{\alpha}}_{\bar{k}} \theta^{\dot{\beta}}_{\bar{j}} & \epsilon_{\dot{\alpha} \dot{\beta}}
\epsilon^{\bar{i} \bar{k}} +\lambda^i_{\dot{\alpha}} \lambda^{\dot{\alpha}}_j \theta^j_{\beta} \theta^k_{\gamma} \epsilon^{\beta \gamma} \epsilon_{i k} 
)+ 2 n (\lambda^{\alpha}_{\bar{i}} \lambda^{\dot{\alpha}}_{j} \theta^{\dot{\beta}}_{\bar{j}} \theta^{j}_{\alpha} \epsilon^{\bar{i} \bar{j}} \epsilon_{\dot{\beta}
 \dot{\alpha}}-\lambda^{i}_{\dot{\alpha}} \lambda^{\bar{i}}_{\alpha} \theta^{\dot{\alpha}}_{\bar{i}} \theta^{j}_{\beta} \epsilon_{i j} \epsilon^{ \beta \alpha}) \, . 
\end{aligned} 
\end{equation}
Moreover, 
\begin{equation}
V^4_0 = V^4_{0,1} + V^4_{0,2} \, ,     
\end{equation}
with 
\begin{equation}
\begin{aligned}
V^4_{0,1} &= (2+n^2)(\bar{\lambda}^2_{+1}  - \bar{\lambda}^2_{-2})(
\theta^{1}_{1} \theta^1_{2} \theta^2_1 \theta^2_2+\theta^{\dot{1}}_{\bar{1}} \theta^{\dot{1}}_{\bar{2}} \theta^{\dot{2}}_{\bar{1}} \theta^{\dot{2}}_{\bar{2}})
\\
& + 3 n \, 
(\bar{\lambda}^2_{+1}  + \bar{\lambda}^2_{-2})
\theta^{1}_{1} \theta^1_{2} \theta^2_1 \theta^2_2 + 3 n \, 
( \bar{\lambda}^2_{+2} + \bar{\lambda}^2_{-1})
\theta^{\dot{1}}_{\bar{1}} \theta^{\dot{1}}_{\bar{2}} \theta^{\dot{2}}_{\bar{1}} \theta^{\dot{2}}_{\bar{2}} \, ,
\end{aligned}
\end{equation}
and 
\begin{equation}
\begin{aligned}
V^4_{0,2} &=\left(1-\frac{3}{2}n - \frac{1}{2} n^2 \right)
(\lambda^{\alpha}_{\bar{i}} \lambda^{\beta}_{\bar{j}} \theta^i_{\alpha} \theta^j_{\beta} \theta^{\dot{\alpha}}_{\bar{k}} \theta^{\dot{\beta}}_{\bar{l}} \epsilon^{\bar{i} \bar{k}} \epsilon^{\bar{j} \bar{l}}
\epsilon_{ij} \epsilon_{\dot{\alpha} \dot{\beta}}-\lambda^i_{\dot{\alpha}} \lambda^j_{\dot{\beta}} \theta^k_{\alpha} \theta^p_{\beta} \theta^{\dot{\alpha}}_{\bar{i}} \theta^{\dot{\beta}}_{\bar{j}}  \epsilon_{i k} \epsilon_{j p} \epsilon^{\bar{i} \bar{j}} \epsilon^{\alpha \beta}) \\
&+\left(- \frac{2}{3} + 2 n + \frac{2}{3} n^2\right)(\lambda^i_{\dot{\alpha}} \lambda^{\alpha}_{\bar{i}} \theta^{\dot{\alpha}}_{\bar{j}}  \theta^j_{\beta} \theta^k_{\alpha} \theta^l_{\gamma}  \epsilon_{i j} \epsilon^{\bar{i} \bar{j}} \epsilon^{\beta \gamma}  \epsilon_{k l}+\lambda^i_{\dot{\alpha}} \lambda^{\alpha}_{\bar{i}} \theta^{j}_{\alpha} \theta^{\dot{\alpha}}_{\bar{k}} \theta^{\dot{\beta}}_{\bar{j}} \theta^{\dot{\gamma}}_{\bar{l}}  \epsilon_{i j}  \epsilon^{\bar{i} \bar{j}} \epsilon^{\bar{k} \bar{l}} \epsilon_{\dot{\beta} \dot{\gamma}}) \\
&+\left(1+\frac{3}{2} n - \frac{1}{2} n^2 \right)
(\lambda^{\dot{\alpha}}_i \lambda^{\dot{\beta}}_j \theta^i_{\alpha} \theta^j_{\beta}  \theta^{\dot{\gamma}}_{\bar{i}} \theta^{\dot{\delta}}_{\bar{j}} \epsilon^{\bar{i} \bar{j}} \epsilon^{\alpha \beta} \epsilon_{\dot{\alpha} \dot{\gamma}} \epsilon_{\dot{\beta} \dot{\delta}}-
\lambda^{\bar{i}}_{\alpha} \lambda^{\bar{j}}_{\beta} \theta^i_{\gamma} \theta^j_{\delta} \theta^{\dot{\alpha}}_{\bar{i}} \theta^{\dot{\beta}}_{\bar{j}} \epsilon^{\alpha \gamma} \epsilon^{\beta \delta} \epsilon_{ij} \epsilon_{\dot{\alpha} \dot{\beta}}) \\
&+\left(\frac{2}{3} + 2 n - \frac{2}{3} n^2\right)(
\lambda^{\dot{\alpha}}_i \lambda^{\bar{i}}_{\alpha} \theta^{\dot{\beta}}_{\bar{i}} \theta^i_{\gamma} \theta^j_{\beta}  \theta^k_{\delta} \epsilon_{\dot{\alpha} \dot{\beta}} \epsilon^{\alpha \beta} \epsilon^{\gamma \delta} \epsilon_{j k}-\lambda^{\dot\alpha}_i \lambda^{\bar{i}}_{\alpha} \theta^i_{\beta} \theta^{\dot\beta}_{\bar{j}} \theta^{\dot\gamma}_{\bar{i}} \theta^{\dot\delta}_{\bar{k}} \epsilon^{\beta \alpha} \epsilon_{\dot\alpha \dot\beta} \epsilon_{\dot\gamma \dot\delta} \epsilon^{\bar{j} \bar{k}}) \, . 
\end{aligned}
\end{equation}
In addition, the term with six
$\theta_+$'s is 
\begin{equation}
\begin{aligned}
V_0^6 & = (-8 + 19 n + 2 n^2 - n^3) 
(\lambda^i_{\dot\alpha} \lambda^{\dot\beta}_i \theta^{\dot\alpha}_{\bar{i}} \theta^{\dot\gamma}_{\bar{j}} \epsilon^{\bar{i} \bar{j}}   \epsilon_{\dot\beta \dot\gamma} 
\theta^{1}_{1} \theta^1_{2} \theta^2_1 \theta^2_2-
\lambda^{\alpha}_{\bar{i}} \lambda^{\bar{i}}_{\beta} \theta^i_{\alpha} \theta^j_{\gamma} \epsilon_{i j} \epsilon^{\beta \gamma}  
\theta^{\dot{1}}_{\bar{1}} \theta^{\dot{1}}_{\bar{2}} \theta^{\dot{2}}_{\bar{1}} \theta^{\dot{2}}_{\bar{2}} )\\
&+ (8 + 19 n - 2 n^2 - n^3)
(\lambda^i_{\dot{\alpha}} \lambda^{\dot\alpha}_j \theta^j_{\alpha} \theta^k_{\beta} \epsilon^{\alpha \beta} \epsilon_{i k} 
\theta^{\dot{1}}_{\bar{1}} \theta^{\dot{1}}_{\bar{2}} \theta^{\dot{2}}_{\bar{1}} \theta^{\dot{2}}_{\bar{2}}-
\lambda^{\alpha}_{\bar{j}} \lambda^{\bar{i}}_{\alpha} \theta^{\dot\alpha}_{\bar{i}} \theta^{\dot\beta}_{\bar{k}} \epsilon^{\bar{j} \bar{k}} \epsilon_{\dot\alpha \dot\beta}
\theta^{1}_{1} \theta^1_{2} \theta^2_1 \theta^2_2) \\
+&\left( \frac{38}{9}n- \frac{2}{9} n^3 \right) 
(\lambda^{\alpha}_{\bar{i}} \lambda^{\dot{\alpha}}_i \theta^i_{\beta} \theta^j_{\alpha} \theta^k_{\gamma} 
\theta^{\dot\beta}_{\bar{k}} \theta^{\dot\gamma}_{\bar{j}} \theta^{\dot\delta}_{\bar{l}} \epsilon^{\bar{i} \bar{j}} 
\epsilon_{\dot{\alpha} \dot{\beta}} \epsilon_{j k} \epsilon^{\beta \gamma} \epsilon^{\bar{k} \bar{l}} \epsilon_{\dot\gamma \dot\delta}-
\lambda^i_{\dot{\alpha}}  \lambda^{\bar{i}}_{\alpha}  \theta_{\gamma}^j  \theta^k_{\beta} \theta^l_{\delta}  \theta_{\bar{j}}^{\dot{\alpha}} \theta_{\bar{i}}^{\dot{\beta}} \theta_{\bar{k}}^{\dot{\gamma}}
\epsilon_{i j} \epsilon^{\alpha \beta} \epsilon^{\gamma \delta} \epsilon_{k l} \epsilon_{\dot{\beta} \dot{\gamma}} \epsilon^{\bar{j} \bar{k}} ) \, .
\end{aligned}
\end{equation}
Finally, 
\begin{equation}
V_0^8 =(\bar{\lambda}_+ \bar{\lambda}_+) \theta^8_+ (72 -37 n^2 + n^4) \, ,
\end{equation}
where we have used the definition
\begin{equation}
\theta^8_+ = 
\theta^1_{1} \theta^1_2 \theta^2_1
\theta^2_ 2 \theta^{\dot{1}}_{\bar{1}}
\theta^{\dot{1}}_{\bar{2}}
\theta^{\dot{2}}_{\bar{1}}
\theta^{\dot{2}}_{\bar{2}} \, .
\label{theta8def}
\end{equation}

The vertex operators 
(\ref{vertexoperatorso4}) 
are the main results of this paper. Notice that they are in the gauge where only 
the component $A_{\dot{a} \dot{b}}$ of the gauge superfield (\ref{vertextypeII}) 
is nonzero.
However, it is possible to write  
the vertices in the more usual form 
(\ref{vertextypeII}) by adding to it BRST trivial term and making appropriated $USp(2,2) \times USp(4)$ gauge transformations.
This will be explained below. 
These vertices correspond to a particular dual operator as 
explained in the beginning of the section. 
However, it is possible to translate the vertices to different spacetime positions and arbitrary $R$-charge orientations. 
This can be done by computing 
the finite $PSU(2,2|4)$ 
transformations of the worldsheet variables by solving
(\ref{finitetransformations}). Notice that for computing amplitudes and getting non-trivial results, it will be important to have vertices with different $R$-charge orientations. 
It is also possible to get the other states of the supermultiplet by applying supersymmetry transformations.
The result of this procedure 
is equivalent to write the vertex in terms of the chiral superfield $\Phi$ described in the section 
\ref{flatspace}.

\subsection{Arguing 
that the $AdS$ picture raising procedure is well defined}

In this subsection, we are going to argue that the procedure of changing pictures used above is justified in $AdS_5 \times S^5$.  
In other words that all the denominators are canceled at every step 
and globally defined states are mapped into globally defined states.  
In $AdS$, the pure spinors are not BRST invariant and the same is true for the delta functions. In our conventions
\begin{equation}
    Q \cdot \delta^{(n)}(\lambda) = (Q \cdot \lambda) \,  
    \delta^{(n+1)} (\lambda) \, ,  
    \label{eq:derivativeofdelta}
\end{equation}
where the superscripts denote the number of derivatives.  
Thus a vertex operator $\tilde{V}_{-n}$ in a picture $-n$ has the following expansion 
\begin{equation}
\tilde{V}_{-n} =
A_0 \, \delta(\lambda^i) 
+
A_1 \, \delta^{(1)}(\lambda^i) 
+
A_2 \, \delta^{(2)}(\lambda^i) + \ldots \, ,
\end{equation}
where we have selected a particular pure spinor component $\lambda^i$.
Note that if $n \neq 1$, the other delta functions are inside the  
$A_i$'s. Since the vertex 
operator 
$\tilde{V}_{-n}$ for a particular state
has a definite grading all the $A_i$'s have necessarily the same grading $a$ (the delta functions and their derivatives are fermionic). 
By construction the vertex is BRST closed. Using the notation (one can check that terms with $(\lambda^a_+)^2$ are only generated by the BRST transformations $Q \cdot \lambda$ for $\lambda^a_+$ and if one replaces $\lambda^a_-$ by \eqref{eq:minusintermsofplus}, see the explicit formulas in the Appendix 
\ref{appendixcoset})  
\begin{equation}
Q \cdot A_j = X^0_j + \lambda^i X^1_j  + (\lambda^i)^2 X^2_j  \, , 
\quad Q \cdot \lambda^i = Z_0 + \lambda^i  Z_1 + (\lambda^i)^2 Z_2 .  
\label{eq:QA} 
\end{equation}
The condition 
$Q \cdot \tilde{V}_{-n} =0 $
implies by reading the coefficients of both the delta function and its derivatives that
\begin{equation} 
\begin{aligned}
& 0= X^0_0 - X^1_1+ 2 X^2_2 - (-1)^a(A_0 Z_1 - 2 A_1 Z_2) \, , \\
&0= X^0_1 - 2 X^1_2 + 6 X^2_3 + (-1)^a (A_0 Z_0 - 2 A_1 Z_1 + 6 A_2 Z_2) \, , \\
& \ldots \\
& 0= X^0_{i-1} - i X^1_{i} + i (i+1)X^2_{i+1} + 
(-1)^a (A_{i-2} Z_0 - i A_{i-1} Z_i + i(i+1) A_{i} Z_2)  \, . 
\label{sistemofequations}
\end{aligned}
\end{equation}
To derive the formulas above, we have used that 
\begin{equation}
\lambda^i \,  \delta^{(n)}(\lambda^i) = - n \, \delta^{(n-1)}(\lambda^i) \, ,  \quad {\rm{and}} \quad 
(\lambda^i)^2 \,  \delta^{(n)}(\lambda^i) =  n (n+1) \, \delta^{(n-2)}(\lambda^i) \, . 
\label{lambdadelta}
\end{equation}
and these formula will be justified below. 
Now using 
\begin{equation}
\xi_{(i)} \cdot \delta^{(n)} (\lambda^i) = \left( \frac{d}{d \lambda^i} \right)^n \frac{1}{\lambda^i} \, ,
\label{derivativedelta}
\end{equation}
we have 
\begin{equation}
\xi_{(i)} \cdot \tilde{V}_{-n}  = 
(-1)^a A_0 \frac{1}{\lambda^i}
+ (-1)^a A_1 \frac{-1}{(\lambda^i)^2} +
(-1)^a A_2 \frac{2}{(\lambda^i)^3}
+ \ldots \, . 
\end{equation}
It is not difficult to see that the conditions  (\ref{sistemofequations}) imply
\begin{equation}
\tilde{V}_{-(n-1)}=Q \cdot \xi_{(i)} \cdot \tilde{V}_{-n} = (-1)^a X^1_0-
 (-1)^a X^2_1 - A_0 Z_2 + (-1)^a \lambda^{i} X^2_0  \, . 
\label{thesurvivorofX}
\end{equation}
The conclusion is that all terms with the pure spinor $\lambda^i$  in the denominator have cancelled.
This is necessary for the picture raising procedure to be well defined in $AdS$.
In addition the term $\lambda^{i} X^2_0$
necessarily have a derivative of delta inside it.

Let us work out an example to see this cancellation in practice. Suppose we have 
the following minus two 
picture
vertex operator ($\tilde{A}$ 
can in principle depends on all the worldsheet variables) 
\begin{equation}
\begin{aligned}
V_{-2} &= \tilde{A}_0 \, \delta(\lambda^{2}) \,  \delta(\lambda^{1}) +  \tilde{A}_1 \,  \delta^{(3)}(\lambda^{2}) \,  \delta^{(1)}(\lambda^{1}) + \tilde{A}_2  \, \delta^{(1)}(\lambda^{2}) 
\, \delta^{(2)}(\lambda^{1}) \,   \\
& 
= A_0 \, \delta(\lambda^{1}) +  A_1 \,   \delta^{(1)}(\lambda^{1}) + A_2  \, 
 \delta^{(2)}(\lambda^{1}) \, . 
 \label{TheVertexV2}
\end{aligned} 
\end{equation}
By construction the vertex is BRST invariant and this implies (We are going to suppose  
that $V_{-2}$ is bosonic and accordingly all the $A_i$'s are fermionic) \begin{equation}
\begin{aligned} 
& Q\cdot V_{-2} =0 
= 
(Q \cdot A_0) \, \delta(\lambda^{1})
- A_0 \, (Q \cdot \delta(\lambda^{1})) \\
& +
(Q \cdot A_1) \, \delta^{(1)} (\lambda^{1})
- A_1 \, (Q \cdot \delta^{(1)} (\lambda^{1}))
+ (Q \cdot A_2) \, \delta^{(2)}(\lambda^{1})
- A_2 \, (Q \cdot \delta^{(2)} (\lambda^{1})) \, . 
\end{aligned} 
\end{equation}
Now using both \eqref{eq:QA} with $i=1$ and  \eqref{eq:derivativeofdelta}, we have that the formula above
is equal to 
\begin{equation}
\begin{aligned}
0 & =   
\, (\lambda^1 X^1_0 + X^0_0 + (\lambda^1)^2 X^2_0) \, \delta(\lambda^{1})
- A_0 \, (Z_0 + \lambda^1 Z_1 + (\lambda^1)^2 Z_2) \, \delta^{(1)}(\lambda^{1}) 
 \\ 
 & +
(X^0_1 + \lambda^1 X^1_1+ (\lambda^1)^2 X^2_1 ) \, \delta^{(1)} (\lambda^{1}) 
 - A_1 \,  (Z_0 + \lambda^1 Z_1 + (\lambda^1)^2 Z_2) \, \delta^{(2)}(\lambda^{1}) \\
& + (X^0_2 + \lambda^1 X^1_2+ (\lambda^1)^2 X^2_2 ) \, \delta^{(2)}(\lambda^{1})
- A_2 \,(Z_0 + \lambda^1 Z_1 + (\lambda^1)^2 Z_2) \, \delta^{(3)}(\lambda^{1}) \, . 
\end{aligned}    
\end{equation}
Finally, by using  \eqref{lambdadelta} and collecting the terms proportional to each  $\delta^{(i)}(\lambda^{1})$, we arrive at the following conditions (to be compared with \eqref{sistemofequations})
\begin{equation}
\begin{aligned}
&0 = X^0_0 - X^1_1 + 2 X^2_2 + A_0 Z_1 - 2 A_1 Z_2 \, , \\
&0= X^0_1 - 2 X^1_2 -A_0 Z_0 + 2 A_1 Z_1 - 6 A_ 2 Z_ 2 \, , \\
&0 = X^0_2 -A_1 Z_0 + 3 A_2 Z_1 \, , \\
&0 = A_2  Z_0 \, . 
\label{setexample} 
\end{aligned}
\end{equation}
The next step is to apply 
a picture raising operator 
in the vertex $V_{-2}$ of 
\eqref{TheVertexV2}. 
We have 
\begin{equation}
\begin{aligned}
 V_{-1} & = Q\cdot\xi_{1}\cdot V_{-2}= - Q\cdot \left(  A_0 \frac{1}{\lambda^1}
+  A_1 \frac{-1}{(\lambda^1)^2} +
 A_2 \frac{2}{(\lambda^1)^3} \right) \\
& =- (Q \cdot A_0) \frac{1}{\lambda^1} + A_0 \left(Q \cdot \frac{1}{\lambda^1} \right)
 + (Q \cdot A_1) \frac{1}{(\lambda^1)^2} - A_1 \left(Q \cdot \frac{1}{(\lambda^1)^2} \right) \\
& - 2 (Q \cdot A_2) 
\frac{1}{(\lambda^1)^3} + 
2 A_ 2
\left(Q \cdot \frac{1}{(\lambda^1)^3} \right)
\\
& = - X^1_0+ X^2_1 - A_0 Z_ 2 - \lambda^1 X^2_0 \, . 
\label{exemplofinal}
 \end{aligned} 
\end{equation}
where we have used 
\eqref{eq:QA}, 
\eqref{setexample} and
\begin{equation}
\left( Q \cdot \frac{1}{(\lambda^1)^i} \right) = - i \,  \frac{1}{(\lambda^1)^{i+1} } \, 
(Q \cdot \lambda^1) \, . 
\end{equation}
We see from 
\eqref{exemplofinal} that 
only (\ref{thesurvivorofX}) survives as
claimed. Since we have replaced all $\lambda_-$ by $\eqref{eq:minusintermsofplus}$ (we 
are working with delta functions), one question is if all the $1/(\bar\lambda_+ \bar\lambda_+)$ disappears. For our coset, we have checked in a massive number of cases that this denominator either cancels or it recombined back when derivatives of deltas are presented. So the vertex in any picture does not have 
any pure spinor in the denominator. For our case, we believe that this can be proved 
because the $\lambda_-$ only appears in the rotation of the pure spinors and in a particular combination in $Q \cdot \theta_+$.   
However, the proof of this cancellation in general deserves further investigation. 
    
We will now explain the formulas 
(\ref{lambdadelta}) and 
(\ref{derivativedelta}). 
There are three ways of deducing them. The first way uses the theory of distribution where we view $\delta(\lambda_+^{a})$ and $\xi_{(a)}=\Theta(w_{a}^{+})$ (Heaviside step function) as being integrated against test functions $f(\lambda_{+}^{a})$ and $\tilde f(w_{a}^{+})$, which are related by Fourier transform
\begin{equation}
f(\lambda_{+}^{a})=\int dw_{a}^{+}e^{i w_{(a)}^{+}\lambda^{a}_{+}}\cdot \tilde  f (w_{a}^{+}) \, , \quad \quad 
\delta (\lambda^a_+) 
= \int \frac{dw_{a}^{+}}{2\pi}
\, e^{i w_{(a)}^{+} \lambda^a_+} \cdot 1\, . 
\end{equation}
Thus
\begin{equation}
\xi_{(a)}\cdot \delta^{(n)}(\lambda_{+}^{a})= \left(\frac{d}{d\lambda_{+}^{a}}\right)^{n}\int \frac{dw_{a}^{+}}{2\pi} \Theta(w_{a}^{+})
\, e^{i w_{(a)}^{+} \lambda^a_+} =  \left(\frac{d}{d\lambda_{+}^{a}}\right)^{n} \frac{1}{\lambda_{+}^{a}} \, ,
\end{equation}
where we have performed the contour integral on $w_{a}^{+}$ from $0$ to $i\infty$ along the imaginary line. Now, to obtain rules for the multiplication between $\delta^{(n)}(\lambda_+^{a})$ and $\lambda_{+}^{a}$ we view this product as a new distribution being integrated against a test function $f(\lambda_{+}^{a})$:
\begin{equation}
\int d\lambda_{+}^{a} \left(\lambda_{+}^{a}\cdot \delta^{(n)}(\lambda_+^{a})\right) f(\lambda_{+}^{a})=\int d\lambda_{+}^{a}  \delta^{(n-1)}(\lambda_+^{a}) \left(\frac{d}{d\lambda_{+}^{a}}\right)\left(-\lambda_{+}^{a}f(\lambda_{+}^{a})\right)=   
\end{equation}
$$
=\int d\lambda_{+}^{a}  \delta(\lambda_+^{a})\left(\frac{d}{d\lambda_{+}^{a}}\right)^{n} \left((-1)^{n}\lambda_{+}^{a}f(\lambda_{+}^{a})\right)=\int d\lambda_{+}^{a}  \delta(\lambda_+^{a})\left(\frac{d}{d\lambda_{+}^{a}}\right)^{n-1} \left((-1)^{n}nf(\lambda_{+}^{a})\right)
$$
$$
=\int d\lambda_{+}^{a} \left(- n \delta^{(n-1)}(\lambda_+^{a})\right) f(\lambda_{+}^{a}) \, , 
$$
where we have used integration by parts, the commutator $[\frac{d}{d\lambda_{+}^{a}},\lambda_{+}^{a}]=1$ and $\delta(\lambda_{+}^{a})\lambda_{+}^{a}=0$.

The second way
uses the formal expressions for $\xi_{(a)}$ appearing in  \cite{LucasFlat}
\begin{equation}
\xi_{(a)} = \int \frac{1}{\tau} 
 \, e^{- \tau \frac{\partial}{\partial \lambda^a_+}} \, d \tau 
\end{equation}
and the Fourier transform of $\delta(\lambda_{+}^{a})$ such that 
\begin{equation}
\xi_{(a)} \cdot   
\delta^{(n)} (\lambda^a_+) 
=
\left( \frac{d}{d \lambda^a_+} \right)^n 
\int d \tau  \int \frac{d w}{2 \pi}
\frac{1}{\tau} e^{i w (\lambda^a_+ - \tau)} = 
\left( \frac{d}{d \lambda^a_+} \right)^n  \frac{1}{\lambda^a_+} \, , 
\end{equation}
and 
\begin{equation}
\lambda^a_+ \cdot   
\delta^{(n)} (\lambda^a_+)
= \int \frac{d w}{2 \pi}
(i w)^n \lambda^a_+ 
e^{i w \lambda^a_+} =
\int \frac{d w}{2 \pi}
(i w)^n  
\left( \frac{d \, e^{i w \lambda^a_+}}{i d w}
\right) = - n 
\, \delta^{(n-1)} (\lambda^a_+) \, . 
\end{equation}

The third way makes use of  the
bosonized formulas given in (\ref{bosonization}) and written again below for the readers convenience   
\begin{equation}
\lambda^a_+ \cong \eta^{(a)} e^{\phi^{(a)}} \, , \quad w^+_{a} \cong - e^{- \phi^{(a)}} \partial\xi_{(a)}  \, , \quad \delta(\lambda^a_+) \cong - e^{- \phi^{(a)}} \, . 
\end{equation}
The nontrivial OPE's of the fields are \begin{equation}
\begin{aligned}
\phi^{(a)}(z) \phi^{(a)}(0) \sim 
- \, {\rm{log}}(z) \, , \quad \quad
\eta^{(a)}(z) \xi^{(a)}(0) \sim \frac{1}{z} \, , \quad \quad
 \xi^{(a)}(z) \eta^{(a)}(0) \sim \frac{1}{z} \, . 
\end{aligned}
\end{equation}
and using the above expressions, we have for example (omitting the indices $(a)$ and the $+$)
\begin{equation}
\oint w (z) \lambda (0) = 1 \, ,  \end{equation}
or more generally 
\begin{equation}
\oint w(z) f(\lambda (0)) = f^{\prime}(\lambda(0)) \, .  \end{equation}

Using the rule above
it is not difficult to see that 
\begin{equation}
\delta^{(1)}(\lambda) \cong \oint :  \partial \xi 
e^{- \phi} (z) : :  - e^{-\phi}(0) : = - \partial  \xi e^{-2 \phi}
(0) \, ,
\end{equation}
and the $n$-th derivative 
case is 
\begin{equation}
\delta^{(n)}(\lambda) 
\cong
f(n) \prod_{i=1}^n \frac{\partial^i \xi}{(i-1)!} e^{-(n+1) \phi} (0) \, ,
\end{equation}
where $f(n)$ is a sign alternating function $\pm 1$ for every two terms\footnote{One example being $f(n) = -\sqrt{2} \, {\rm{sin}} \left( \frac{(2 n +1) \pi}{4} \right)$.}. 
From the expression above, one can easily derive \eqref{lambdadelta}:
\begin{equation}
 \lambda \cdot \delta^{(n)}(\lambda) = \underset{z\rightarrow 0}{{\rm{lim}}} 
\;  : \eta e^{\phi}(z): : f(n) \prod_{i=1}^n \frac{\partial^i \xi}{(i-1)!} e^{-(n+1) \phi} (0): = - n \,  
\delta^{(n-1)}(\lambda) \, . \end{equation}
Moreover, we have 
\begin{equation}
     \xi \cdot \delta^{(n)}(\lambda) 
     = f(n) \, \xi \prod_{i=1}^n \frac{\partial^i \xi}{(i-1)!} e^{-(n+1) \phi} (0) \, .
\label{epsilondelta}     
\end{equation}
The final step consists 
in rewriting the expression above in terms of the $\lambda$'s. 
The easiest way of doing this is by acting on
\eqref{epsilondelta}
with $n+1$ $\lambda$'s successively. 
For just one $\lambda$, we have, for example
\begin{equation}
\begin{aligned}
     \lambda \cdot \xi \cdot \delta^{(n)}(\lambda) \Big|_{n \neq 0} & =
    \underset{z\rightarrow 0}{{\rm{lim}}} 
\;  : \eta e^{\phi}(z):
: f(n) \, \xi \prod_{i=1}^n \frac{\partial^i \xi}{(i-1)!} e^{-(n+1) \phi} (0): \\
& = 
-n \, f(n) (-1)^{(n+1)}  \, \xi \prod_{i=1}^{n-1} \frac{\partial^i \xi}{(i-1)!} e^{-n \, \phi} (0) \, .  
\end{aligned} 
\end{equation}
Proceeding in this way, one eventually concludes that
\begin{equation}
 \xi \cdot \delta^{(n)}(\lambda) =
 \left( \frac{d}{d \lambda^a_+} \right)^n  \frac{1}{\lambda^a_+} \, . 
\end{equation}
Note that 
there are various deltas involved in our computation and we always have considered them as fermions. In the bosonized language we need to work with cocycles to ensure that the deltas are fermions, see \cite{Kostelecky:1986xg} for example.

We end this section by commenting on the fact that our vertices 
(\ref{vertexoperatorso4}) are in the gauge where only the component 
$A_{\dot{a} \dot{b}}$ of the superfield appearing in (\ref{vertextypeII}) is nonzero. 
We proved in the Introduction that this gauge can be reached in a flat background 
by starting with (\ref{vertextypeII}) and by adding BRST trivial quantities
. 
Certainly, a similar proof can be given in $AdS$ but there are many new details due to the BRST transformations of the $\lambda$'s and complicated covariant derivatives. 
We are not going to try to write this proof here but only argue why the  existence of this gauge is justified. Yet a second argument will be given in the last subsection of this section. 
Notice that only a few terms survive in \eqref{thesurvivorofX}, 
in particular, this implies that if the vertex operator $V_{-8}$ is in this gauge then the term with no derivatives of the delta functions in all the vertices $V_{-n}$ are going to have only $\bar{\lambda}$'s. 
This follows
because for our variables 
the covariant derivatives, see Appendix \ref{appendixcoset}, which contains $USp(2,2) \times USp(4)$ gauge transformations that mix $\bar\lambda^{\dot{a}}$ with $\lambda^a$
are the ones in (\ref{complicatecovariant}). These covariant derivatives appear in the BRST operator always multiplied by a $\bar\lambda^{\dot{a}}$ and this implies that the number of $\bar{\lambda}$ is conserved after canceling 
a possible $\lambda_+$ in the denominator.

\subsection{The Dilaton Vertex Operator} 

One important test of our vertex operators 
(\ref{vertexoperatorso4}) is that for $n=0$
it reduces to the known dilaton vertex 
operator of \cite{BerkovitsPrescription}. 
The dilaton vertex operator is manifestly 
$PSU(2,2|4)$ invariant and it 
is given by 
\begin{equation}
V_{{\rm{dilaton}}} = \eta_{\underline\alpha \hat{\underline\alpha}} \lambda_L^{\underline\alpha} \lambda_R^{\hat{\underline\alpha}} \, ,  
\label{dilaton}
\end{equation}
where $\eta_{\underline\alpha \hat{\underline\alpha}}$ is the background value of the component 
$B_{\underline\alpha \hat{\underline\alpha}}$ of the 
$B$ field.  Its value is numerically 
given by $\eta_{\underline\alpha \hat{\underline\alpha}} = \left( \gamma^{01234} \right)_{\underline\alpha \hat{\underline\alpha}}$.
In \cite{BerkovitsPrescription}, 
it was argued that this state 
is physical in $AdS$ despite the 
fact of being trivial in a flat background 
($\eta_{\underline\alpha \hat{\underline\alpha}} \lambda_L^{\underline\alpha} \lambda_R^{\hat{\underline\alpha}} = Q|_{\rm{flat}}  \cdot
\eta_{\underline\alpha \hat{\underline\alpha}} \theta_L^{\underline\alpha} \lambda_R^{\hat{\underline\alpha}}$). 
A direct proof that 
(\ref{dilaton}) is not
BRST exact is complicated.
The easiest way 
to argue this is by verifying 
that the integrated form of this vertex operator is the Lagrangian so it must be in the cohomology.  

For $n=0$, our vertex 
operators (\ref{vertexoperatorso4})  still have a very nontrivial dependency on
the $\theta$'s. However, we can add to it the following BRST exact quantity 
\begin{equation}
V^{\prime}_{n=0}=V_{n=0} + Q 
\cdot (\lambda_+^a \theta_+^a)+ 
\frac{1}{12} \, Q 
\cdot (
\lambda^{\alpha}_{i} \theta^i_{\beta}   
\theta_{\alpha}^j \theta^k_{\gamma} 
\epsilon^{\beta \gamma} \epsilon_{j k}
-
\lambda^{\bar{i}}_{\dot\alpha} \theta^{\dot\alpha}_{\bar{j}}   
\theta_{\bar{i}}^{\dot{\beta}} \theta^{\dot{\gamma}}_{\bar{k}} 
\epsilon^{\bar{j} \bar{k}} \epsilon_{\dot{\beta} \dot{\gamma}}) \, , 
\label{exactBRST}
\end{equation}
and we have 
\begin{equation}
V^{\prime}_{n=0}= (\bar{\lambda}_+ \bar{\lambda}_+) + (\lambda_+ \lambda_+) \, ,
\label{dilaton2}
\end{equation}
with
\begin{equation}
(\lambda_+ \lambda_+)=    
\lambda^i_{\alpha} \lambda^j_{\beta} 
\epsilon^{\alpha \beta} \epsilon_{i j}
- \lambda^{\dot\alpha}_{\bar{i}} \lambda^{\dot \beta}_{\bar{j}} \epsilon_{\dot\alpha \dot\beta} \epsilon^{\bar{i} \bar{j}} \, , 
\end{equation}
and $(\bar{\lambda}_+ \bar{\lambda}_+)$ was defined in 
(\ref{eq:lambdabarplus}).
Note that all the dependency on the 
$\theta$'s have disappeared and we have recovered the dilaton vertex operator of 
(\ref{dilaton}) in our notation. 
It is possible to write 
(\ref{dilaton2}) more symmetrically  including 
the dependency on
the $\lambda^a_-$ and $\bar\lambda^{\dot{a}}_-$ by using the pure spinor constraints (\ref{eq:so8pure}). 

Using the same reasoning, it is possible to remove all the $\theta$'s dependency of the term 
with $n^0$ of our vertices  
by adding the exponential factors to (\ref{exactBRST}) and we are left with 
\begin{equation}
V_2^{\prime}(n) =
V^{\prime}_{n=0} e^{n(z+w)}
+ {\rm{terms \; with \; }}n^i \; {\rm{and \;}}i>0 \, .
\end{equation}
Since $V^{\prime}_{n=0}$ is a state in the cohomology, it is possible 
to argue that $V_2^{\prime}(n)$ is also in the cohomology. 
This is one way of seeing that our vertex operators are not 
BRST exact. 
First, let us focus in the term
of order $n^0$ and consider $n$ generic.
The 
only possibility 
for removing this term 
is by adding to the vertices the following schematically trivial quantity $Q \cdot (\lambda \theta e^{n(z+w)} + \ldots)$. 
We known that for $n=0$ 
the vertex is not exact, but when the $Q$ acts 
on the exponential it drops a factor of $n$ and it does not help with possible
cancellations. In addition, adding terms 
with factors of $n$ in the denominator does not help as well (the possible remaining 
terms with $n$ in the denominator have to
be also exact by themselves)
because such terms 
could also be constructed in the case $n=0$ by 
multiplying any such term by $z+w$ (any 
remaining terms with $z+w$ have to be exact by themselves as well, because no BRST transformation produce a $z+w$, see Appendix \ref{appendixcoset}). 
It is also possible to see that the vertices 
are not trivial by starting with
their original expressions (\ref{vertexoperatorso4}). 
The only way to cancel $V^0_0 =(\bar{\lambda}_+ \bar{\lambda}_+) e^{n(z+w)}$ is by
adding minus the term $Q \cdot ((\bar{\lambda}_+ \bar{\theta}_+) e^{n(z+w)})$ to the vertices. 
Many terms will survive after this cancellation, 
in particular, a term $B \propto
(\bar{\lambda}_+ \bar{\theta}_+)
(\lambda_- {\theta}_+)$. 
This term has multiple origins,
one is when $Q$ acts on $e^{n(z+w)}$ and 
as a consequence its numerical prefactor 
is linear in $n$. There are other independent terms with a $\bar{\theta}_+$ and a $\theta_+$ as well. It is not difficult 
to see 
that there no $C$ with three $\theta$'s 
such that  
$Q\cdot C$ will remove it.
Yet another way to see 
that the vertices are not BRST exact is by analysing the flat space limit and verifying that the vertices reduce correctly to the flat space ones which we know are not BRST trivial. This limit, interesting by its own, is considered below.  

\subsection{The Flat Space Limit}

Another important test 
of our vertex operators
(\ref{vertexoperatorso4})
is that they reduce correctly in the flat space limit. 
This limit corresponds 
to take the radius of $AdS_5\times S^{5}$ very large such that the superspace approach 
to the type IIB flat superspace \cite{Berkovits:2007zk}. Here we perform this limit by splitting the $PSU(2,2|4)$ generators into three groups that we will call rotations $(M)$, translations $(P)$ and supercharges $(q)$:
\begin{equation}
\begin{aligned}
& M=\{ \hat{M}^{\alpha}_{\beta} , \, \hat{M}^{\dot\alpha}_{\dot\beta}, \, (P^{\alpha}_{\dot\alpha} +\epsilon_{\dot\alpha \dot\beta} \epsilon^{\alpha \beta} 
K_{\beta}^{\dot\beta}),\, \hat{M}^{i}_{j} , \, \hat{M}^{\bar{i}}_{\bar{j}}, \, (P^{\bar{i}}_{i} +\epsilon_{i j} \epsilon^{\bar{i} \bar{j}} 
K_{\bar{j}}^{j})  \}  \, ,  \\ 
& P=\{
K_{\beta}^{\dot\beta},
K_{\bar{j}}^{j} , \Delta, J \} \, ,\quad 
q =\{ q^{\alpha}_{\bar{i}}, q^i_{\dot{\alpha}} , q^{\bar{i}}_{\alpha}, q^{\dot\alpha}_i , q^i_{\alpha}, q^{\dot\alpha}_{\bar{i}} ,q^{\bar{i}}_{\dot\alpha}, q^{\alpha}_i \} \, . 
\label{generatorsgroup}
\end{aligned}
\end{equation}
Note that this procedure gauge fix the $USp(2,2) \times USp(4)$ gauge symmetry by picking what are the generators of $SU(2,2)\times SU(4)$ that will be interpreted as generators of translations. 
It could have been any linear combination of the bosonic generators that is linearly independent from the generators in $M$.
The commutation relations 
are schematically of the form
\begin{equation}
\begin{aligned}
&[M,M] \sim M,\quad [P,P]\sim M+P,\quad [M,P]\sim P +M, \\
& [P,q]\sim q,\quad [M, q]\sim q,\quad [q,q]\sim P + M \, . 
\end{aligned}
\end{equation}
In terms 
of these generators, the flat space limit is defined by the re-scaling
\begin{equation}
M\rightarrow M, \quad P\rightarrow \Lambda^{-1} P,\quad q\rightarrow \Lambda^{-\frac{1}{2}}q \, . 
\end{equation}
with $\Lambda\rightarrow 0$. Note that for $\Lambda\neq 1$ we lose $PSU(2,2|4)$ if we do not 
re-scale some of the structure constants as well. This will give us a continuously deformation from the $PSU(2,2|4)$ algebra to the super-Poincaré algebra.
Actually after taking 
this limit we obtain just a subalgebra of super-Poincaré, where the missing generators are actually outer automorphisms of this subalgebra. Once we promote this outer automorphisms to generators, we get the full Poincaré algebra. Notice that this limit is singular since we have inverse powers of $\Lambda$. Physically this limit is given by looking at small distances in $AdS_5\times S^{5}$. It describes a small neighborhood around the origin where the coset acts. Since we are 
rescaling all the momenta, in order to get well defined expressions we rescale the worldsheet coordinates as well 
\begin{equation}
z \rightarrow \Lambda z \, , \quad 
w \rightarrow \Lambda w \, , \quad 
x \rightarrow \Lambda x \, , \quad y \rightarrow \Lambda y \, , \quad
\theta \rightarrow \Lambda^{\frac{1}{2}} \theta \, . 
\label{rescalingvariables}
\end{equation}

The relevant $AdS_5 \times S^5$ BRST transformations, 
see the Appendix \ref{appendixcoset},  
have the following form 
in the $\Lambda\rightarrow 0$ limit 
\begin{equation}
Q(z+w)=\lambda^{\alpha}_i \theta^i_{\alpha} - \lambda^{\bar{i}}_{\dot{\alpha}} \theta^{\dot{\alpha}}_{\bar{i}},\qquad Q(\theta^a_+)=\lambda^a_+ \, , 
\label{flatBRST}
\end{equation}
and the transformations above are the usual flat space ones of \eqref{flatspaceBRST2}. 
To do the flat space limit of our half-BPS vertex operators with charge $n$ we also need to maintain the combination $n(z+w)$ appearing in the exponentials fixed, so $n\rightarrow \Lambda^{-1} n$. 
In addition, to compare the flat space limit of our vertices with the vertex operators obtained in \cite{LucasFlat} 
and revised in the section \ref{flatspace}, 
it is also necessary 
to make the replacements 
$n \leftrightarrow ik_+$ 
and $(z+w) \leftrightarrow y $. 

Due to 
(\ref{rescalingvariables}), the terms surviving the flat space limit will be the ones of the form $n^{k}\theta^{2k}$. 
Looking 
at the Appendix \ref{so8vertexap}, we see that the terms with this structure preserve $SO(8)$ and there is just one term for 
every $n^k$. This means that our $AdS$ vertices have the correct flat space limit.
A further verification of this fact is that the vertex operators after the limit are BRST closed considering the flat BRST transformations of 
(\ref{flatBRST}).
Note that the $n=0$ vertex operator becomes BRST exact in the flat space limit, see the discusion in previous subsection. This is expected since the $n=0$ vertex operator correspond to the moduli of type IIB $AdS_5\times S^{5}$ background which is absent in the flat background.

Using the fact 
that 
the vertex operators must have the correct flat space limit, it is possible to give an alternative argument for the existence of the gauge where only the component $A_{\dot{a} \dot{b}}$ is nonzero for the picture zero vertex.
Notice that all the terms of the form $n^k \theta^{2k}$ can be put in this gauge because the $\lambda$'s do not scale with $\Lambda$ only the coordinates, see
(\ref{rescalingvariables}). 
In addition, again due to the flat space limit, it is not possible to have terms of the type $n^k \theta^{k^{\prime}}$ with $k^{\prime} < 2k$. However, it is possible to have terms with $k^{\prime}> 2k$. 
As mentioned before, the covariant derivatives  
that mix $\bar\lambda^{\dot{a}}$ with $\lambda^a$ appear in 
the BRST operator 
always multiplied by a $\bar\lambda^{\dot{a}}$.
The conclusion, due to the fact that the vertices are BRST closed and only depends on $\theta_+^a$ by supersymmetry, is that this show recursively that it is possible to put the full vertex in the gauge where only $A_{\dot{a} \dot{b}}$ is nonzero.


\section{The $AdS_5\times S^{5}$ boundary}\label{theboundary}

In this section, we are going to locate the $AdS_5$ boundary in
our coordinates appearing in the supercoset (\ref{eq:cosetparametrization}).
This is important because we would like to apply translations to our vertex operators and find their expressions at different points.  In order to do this 
we start by setting all the fermionic coordinates to zero.
The supercoset (\ref{eq:cosetparametrization}) reduces to 
\begin{equation}
g^{\prime}=uv,\quad
{\rm{with}} \quad 
u \, \in \, 
\frac{SU(2,2)}{USp(2,2)}
 \, , \quad v \, \in \, 
\frac{SU(4)}{USp(4)} \, . 
\end{equation}
Note 
that $u$ parametrizes the $AdS_5$ space and $v$ parametrizes the $S^{5}$ space. Explicitly 
\begin{equation}
u=(e^{x^{\alpha}_{\dot\alpha}K^{\dot\alpha}_{\alpha}}e^{z\Delta}) \, , \quad v=(e^{y^{\bar i}_{i}K^{i}_{\bar i}}e^{w J} ) \, , 
\label{BosonicCosets}
\end{equation}
and we have used that the 
the $SU(2,2)$ generators commute with the $SU(4)$ generators. The generators that appear in the exponent are given in terms of matrices by (see (\ref{MatrixMAB}))
\begin{equation}
[x^{\alpha}_{\dot\beta}K^{\dot \beta}_{\alpha}]= 
\left( 
\begin{tabular}{cc}
$0$&$0$ \\ $x^{\alpha}_{\dot\beta}$&$ 0$
\end{tabular}\right) \, , \quad 
[y^{\bar i}_{j}K^{j}_{\bar i}]=
\left( 
\begin{tabular}{cc}
$0$&$y^{\bar i}_{j}$ \\ $0$&$ 0$
\end{tabular}\right) ,
\end{equation}
which means that the exponentiation stops at first order in $(xK)$ and $(yK)$. 
The coset elements (\ref{BosonicCosets}) are given by 
\begin{equation}
\centering
[u_{R}^{\tilde R}]= 
\left( 
\begin{tabular}{cc}
$e^{\frac{z}{2}}\delta^{\alpha}_{\beta}$&$0$ \\ $e^{\frac{z}{2}}x^{\alpha}_{\dot\beta}$&$ e^{-\frac{z}{2}}\delta^{\dot\alpha}_{\dot\beta}$
\end{tabular}\right) \, , \quad  [v_{I}^{\tilde J}]= 
\left( 
\begin{tabular}{cc}
$e^{\frac{w}{2}}\delta^{i}_{j}$&$e^{\frac{-w}{2}}y^{\bar i}_{j}$ \\ $0$&$ e^{-\frac{w}{2}}\delta^{\bar i}_{\bar j}$
\end{tabular}\right) \, . 
\label{matrixelements} 
\end{equation}
It is possible to construct an embedding of both $AdS_{5}$ and $S^{5}$ into $\mathbb{R}^{4,2}$ and $\mathbb{R}^{6}$ respectively by using the coset elements above.
We define 
\begin{equation}
X_{RS}=u_{R}^{\tilde R}u_{S}^{\tilde S}(\sigma_{-1})_{\tilde R\tilde S} \, , \quad \quad 
Y_{IJ}=v_{I}^{\tilde I}v_{J}^{\tilde J}(\sigma_{6})_{\tilde I\tilde J},
\label{eq:Xcoset}
\end{equation}
where the sigma matrix $(\sigma_6)_{\tilde{I} \tilde{J}}$ can be obtained from the one in  
(\ref{sigma6}) by using
(\ref{loweringindices})
and 
the sigma matrix  
$(\sigma_{-1})_{\tilde{I} \tilde{J}}$ is similar.
The variables defined above satisfy
\begin{equation}
\frac{1}{8}\varepsilon^{RSTU}X_{RS}X_{TU}=\frac{1}{8}\varepsilon^{IJKL}Y_{IJ}Y_{KL}=-1 \, .     
\label{AdSequations}
\end{equation}
which are the $SU(2,2)$ and $SU(4)$ notation for the usual 
$AdS_5$ and $S^5$ equations with embedding coordinates
\begin{equation}
\begin{aligned}
-(X_{-1})^{2}-(X_{0})^{2}+(X_{1})^{2}+(X_{2})^{2}+(X_{3})^{2}+(X_{4})^{2}=-1 \, , \\ 
(Y_{1})^{2}+(Y_{2})^{2}+(Y_{3})^{2}+(Y_{4})^{2}+(Y_{5})^{2}+(Y_{6})^{2}=1 \, , 
\end{aligned}
\end{equation}
where $X_{RS}=X^{m}(\sigma_{m})_{RS}$ and $Y_{IJ}=Y^{m'}(\sigma_{m'})_{IJ}$. 
The $X_{RS}$ and $Y_{IJ}$ of (\ref{eq:Xcoset}) have the following representation in our conventions
\begin{equation}
\centering
[X_{RS}] = 
\left( 
\begin{tabular}{cc}
$e^{z}\varepsilon_{\alpha\beta}$&$ e^{z} \epsilon_{\alpha \beta} x^{\beta}_{\dot\beta}$ \\ $ e^{z}x^{\gamma}_{\dot\alpha} \epsilon_{\gamma \beta}$&
$-(e^z x^2
+ e^{-z}) \epsilon_{\dot\alpha \dot \beta}$
\end{tabular}
\right) , \; \,  [Y_{IJ}] = 
\left( 
\begin{tabular}{cc}
$(e^w + e^{-w} y^2)\epsilon_{ij}$&$ -e^{-w}y^{\bar k}_{i} \epsilon_{\bar k \bar j}$\\$- e^{-w}y^{\bar k}_j \epsilon_{\bar i \bar k}$&$ -e^{-w}\epsilon_{\bar i\bar j}$ 
\end{tabular}
\right) \, ,  
\label{eq:Xmatrix}
\end{equation}
where $x^2 = -(1/2) \epsilon_{\alpha \beta} \epsilon^{\dot\alpha \dot\beta}
x^{\alpha}_{\dot\alpha} x^{\beta}_{\dot\beta}$ 
and $y^2 = -(1/2) \epsilon^{ij} 
\epsilon_{\bar{i} \bar{j}} 
y^{\bar{i}}_i y^{\bar{j}}_j $. 

Let's focus in the $AdS_5$ part, we will make further comments about the variables $Y_{IJ}$ and $S^5$ in the next subsection. 
From the expression above, we can see that $X_{12}>0$, which means that our coordinate system covers only a patch of $AdS_5$.
Recall that the boundary of $AdS_5$ in global coordinates 
is the intersection of the embedding with the $\mathbb{R}^{4,2}$ infinity. 
If we scale $X_{RS}\rightarrow \bar X_{RS}= \Lambda X_{RS}$ in order to keep the embedding coordinates finite as we move towards the $\mathbb{R}^{4,2}$ infinity, the embedding equation becomes
\begin{equation}
\varepsilon^{RSTU}\bar X_{RS}\bar X_{TU}=0 \, ,    
\end{equation}
since $1/\Lambda^2 \rightarrow 0$. This means that the boundary can be defined by the equation above under the scaling equivalence $\bar X_{RS}\cong \Lambda \bar X_{RS}$. The result is the well known conformal compactification of Minkowski space \cite{Witten:1998qj}.

Since in our coordinates we have $\bar X_{12}\neq 0$ we can parametrize the boundary of our patch by $x_{\alpha\dot\beta}=
\bar X_{\alpha\dot\beta} /\bar X_{12}$. This means that in our coordinates the boundary lies at $z=\infty$ where $X_{12}\rightarrow \infty$.
Recall that the points $X_{12}=0$ are not covered and these points are the horizon of our coordinate system. 
Notice that these points belong to the bulk of $AdS_5$ in global coordinates. 
Another standard way of computing 
the $AdS_5$ metric and locating 
the boundary is by using the vielbeins 
$\hat{e}_M^A$ defined 
in 
(\ref{transformationsinfinitesimal}). Setting all the fermionic coordinates to zero, we have 
($\eta_{A B}$ is the flat metric)
\begin{equation}
d s^2_{{\rm{AdS}}} = 
\eta_{A B} \, \hat{e}_M^A \, \hat{e}_N^B \, dx^M dx^N
= dz^2 + e^{2 z} 
\epsilon_{\alpha \beta} \epsilon^{\dot \alpha \dot \beta}d x^{\alpha}_{\dot \alpha}
\, d x^{\beta}_{\dot \beta} \, , \end{equation}
and once again we see that the boundary is at $z \rightarrow \infty$. 

It is instructive to compare our coordinates with the usual Poincaré coordinates $(\tilde z,\tilde x)$ where 
(these coordinates are defined 
by using the coset 
$\tilde{u}=(e^{\tilde{x}_{\alpha}^{\dot\alpha}P_{\dot\alpha}^{\alpha}}e^{\tilde{z}\Delta})$ 
instead of the coset $u$ given in 
(\ref{BosonicCosets})).
\begin{equation}
[\tilde{X}_{RS}] = 
\left( 
\begin{tabular}{cc}
$(e^{\tilde{z}} - e^{-\tilde{z}} \tilde{x}^2)\epsilon_{\alpha \beta}$&$ -e^{-\tilde{z}}\tilde{x}^{\dot\gamma}_{\alpha} \epsilon_{\dot\gamma \dot\beta}$\\$- e^{-\tilde{z}}x^{\dot\gamma}_{\beta} \epsilon_{\dot\alpha \dot\gamma}$&$ -e^{-\tilde{z}}\epsilon_{\dot\alpha \dot\beta}$ 
\end{tabular}
\right) \, , 
\end{equation}
Note that these coordinates cover the region where $\tilde{X}_{43}>0$ and  
the intersection of this coordinates with the boundary can be parametrized by $\tilde x_{\alpha\dot\beta}=(\tilde{X}_{\alpha\dot\beta}/\tilde{X}_{34})$. 

We can relate our coordinates
$(z, x)$ with the Poincaré coordinates $(\tilde{z},\tilde{x})$, when both $X_{12}$ and $\tilde{X}_{43}$ are non-zero. The relation is given by
\begin{equation}
\tilde x_{\alpha\dot\beta}= \frac{x_{\alpha\dot\beta}}{(e^{-2z}+x^{2})},\qquad e^{-\tilde z}=(e^{-z}+e^{z}x^{2}) \, .    
\end{equation}
and at the boundary $z\rightarrow \infty$ ($\tilde z\rightarrow -\infty$) it becomes
\begin{equation}
\tilde x_{\alpha\dot\beta}= \frac{x_{\alpha\dot\beta}}{x^{2}} \, . 
\label{eq:nspole}
\end{equation}
This means that at the boundary the origin of the Poincaré patch is a point at infinity of our coordinates and vice-versa. This is indeed expected since the action of conformal boost in $x_{\alpha\dot\beta}$ is simply a translation while in $\tilde x_{\alpha\dot\beta}$ is the sequence of an inversion, a translation and an inversion again.

One last comment 
is that it is also possible to choose the coset 
$\hat{u}=(e^{\hat{x}_{\alpha}^{\dot\alpha}(P^{\alpha}_{\dot\alpha} -\epsilon_{\dot\alpha \dot\beta} \epsilon^{\alpha \beta} 
K_{\beta}^{\dot\beta})}e^{\hat{z}\Delta})$. 
In fact, this would be the most natural choice, since this combination of generators with a relative plus sign is the one appearing in the isotropy group, see (\ref{generatorsgroup}).  
The coordinates $\hat{x}$ are interesting as they cover $AdS_5$ completely. 
Our choice to work with the coset $u$ of  (\ref{BosonicCosets}) 
instead of $\hat{u}$
is only technical. The calculations greatly simplify
if the generators appearing in the coset are the ones that annihilate the vertex operators.  

We have considered up to now the case of an $AdS_5$ space with Lorentz signature. 
If we perform a Wick rotation of the time coordinate $X_{0}$ we obtain the Euclidean $AdS_5$. The points $X_{12}^{E}=0$ does not belong to the Euclidean $AdS_5$ anymore since $ (x_E^2) \ge 0$ and 
\begin{equation}
X_{12}^{E}X_{34}^{E}+(x_E^2)= -1 \, .     
\end{equation} 
This happens because the Wick rotation maps the horizon of our coordinates to a single point at the boundary, so our coordinates cover all the bulk of Euclidean $AdS_5$ 
\cite{BragaBoundary}. The same is true for the Poincaré coordinates.

\begin{figure}[h]
    \centering
    \includegraphics[scale=0.28]{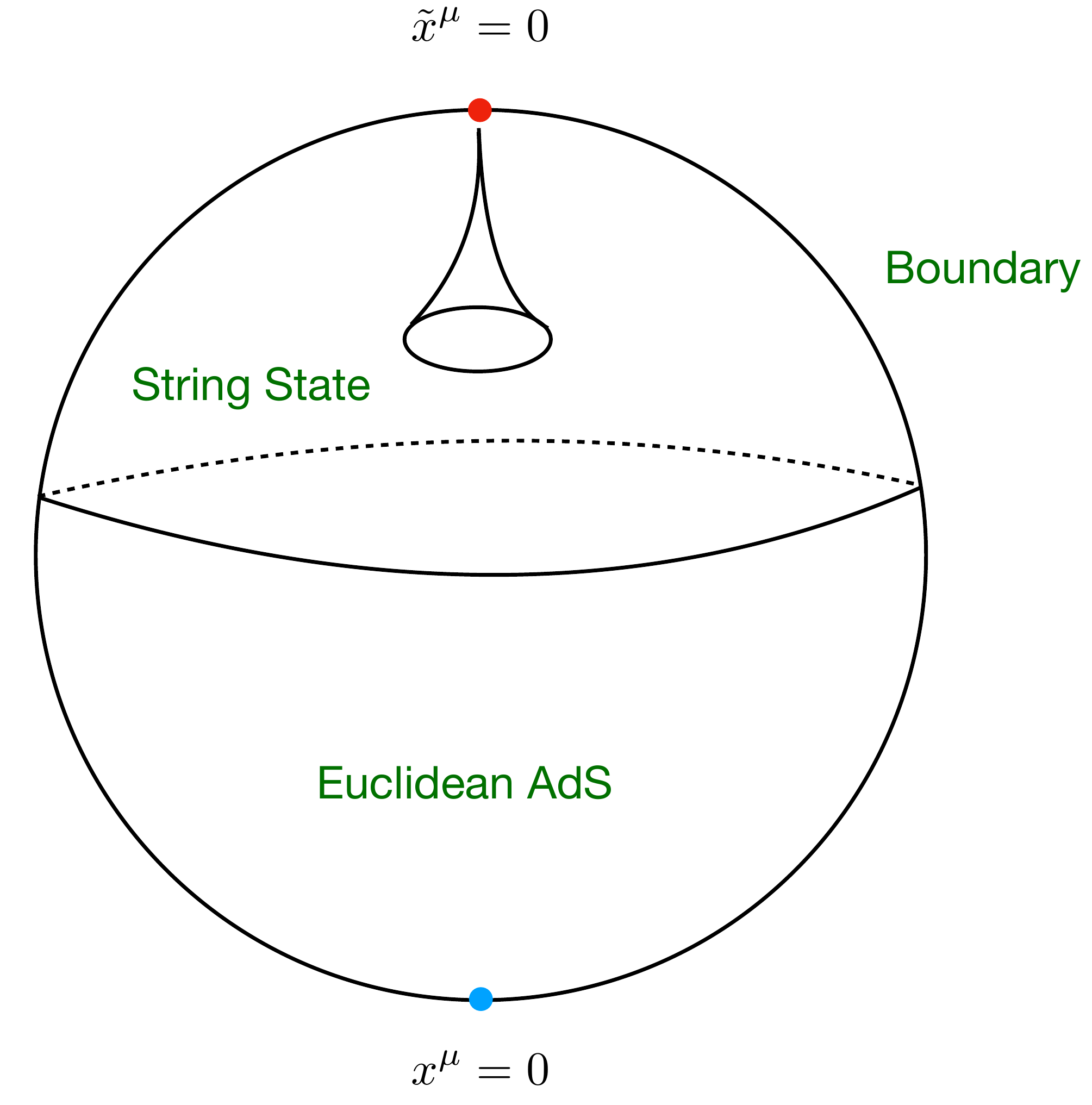}
    \caption{Euclidean $AdS_5$ as a unit ball: the boundary is $S^{4}$ and our half-BPS state is coming from the north pole $x=\infty$ ($\tilde x=0$).}
    \label{fig:boundary}
\end{figure}

The boundary of Euclidean $AdS_5$ have the topology of $S^{4}$ where the Poincaré coordinates and ours are complementary, as indicated in equation \eqref{eq:nspole} and Figure \ref{fig:boundary}. Our half-BPS vertex operators describe the state in which the string is coming from the point $x=\infty$ ($\tilde x=0$) at the boundary. This string state have $n$ units of angular momenta along the sixth direction  of $S^{5}$ and also $n$ units of $z$ momenta in $AdS_5$.

\subsection{The $S^{5}$ parametrization and harmonic variables}

In this subsection, we 
define the so called harmonic variables and we related them to our variables. 
As mentioned in the Introduction, previously 
the supergravity vertex operators were only known close to the boundary and the harmonic superspace tecniques were used in that construction.
It will be nice two compare the two results in the region of common validity and for this we will need to understand better the harmonic variables.  
The $S^{5}$ in this paper is viewed as the following coset
\begin{equation}
S^5=\left(\frac{SO(6)}{SO(5)}\right) \, .  
\end{equation}
The interpretation of this expression is that we can generate the whole $S^{5}$ by acting with $SO(6)$ on 
a point called 
the origin. 
The points are vectors in a six dimensional Euclidean space. 
The $SO(5)$ is the subgroup that preserves the origin. We can take, for example, the origin to be an unit vector in the sixth direction. 
In this case, 
the $SO(5)$ subgroup that preserve this unit vector are rotations generated by the ten generators $(M_{12},\dots,M_{45})$ of $SO(6)$. 
Thus, 
an arbitrary point of $S^{5}$ can be obtained by acting 
on the origin with the five remaining generators $(M_{16},\dots,M_{56})$, or any combinations of them 
that are linearly independent from the chosen $\mathfrak{so}(5)$ generators. In terms of spin groups, we have 
\begin{equation}
\left(\frac{SO(6)}{SO(5)}\right) \cong
\left(\frac{SU(4)}{USp(4)}\right) \, ,  
\label{spingroup}
\end{equation} 
where $SO(6)$ unit vectors are described now by the variables $Y_{IJ}=-Y_{JI}$ 
introduced previosuly. 
In our conventions, the unit vector in the sixth direction
corresponds to 
$Y_{12}^0=Y_{34}^0=1$ and $Y_{13}^0=Y_{14}^0=Y_{23}^0=Y_{24}^0=0$, see 
(\ref{sigma6}). 
By breaking the 
$SU(4)$ indices $I,J$
into $I=(i,\bar i)$ and $J=(j,\bar j)$, we have compactly 
$Y_{ij}^0=\epsilon_{ij}$, 
$Y_{\bar i\bar j}^0=\epsilon_{\bar i\bar j}$ and $Y_{i\bar j}^0=0$.
We can now act 
with $SU(4)$ elements $u^{I}_{J}$ in 
$Y^0_{IJ}$ and obtain the other points, we have 
\begin{equation}
Y_{IJ}=u_{I}^{i}u_{J}^{j}\varepsilon_{ij}+ u_{I}^{\bar i}u_{J}^{\bar j}\varepsilon_{\bar i\bar j} \, .    
\end{equation}
The $USp(4)$ subgroup, 
i.e. the isotropy group in 
(\ref{spingroup}), 
constitute of the $(\tilde u_{I}^{i},\tilde u_{I}^{\bar i})$ variables that satisfy
\begin{equation}
\tilde u_{i}^{k}\tilde u_{j}^{l}\epsilon_{kl}+ \tilde u_{i}^{\bar k}\tilde u_{j}^{\bar l}\epsilon_{\bar k\bar l}=\epsilon_{ij},\quad  \tilde u_{\bar i}^{k}\tilde u_{\bar j}^{l}\epsilon_{kl}+ \tilde u_{\bar i}^{\bar k}\tilde u_{\bar j}^{\bar l}\epsilon_{\bar k\bar l}=\epsilon_{\bar i\bar j},\quad \tilde u_{i}^{k}\tilde u_{\bar j}^{l}\epsilon_{kl}+ \tilde u_{i}^{\bar k}\tilde u_{\bar j}^{\bar l}\epsilon_{\bar k\bar l}=0\,.     
\end{equation}

The $S^{5}$ can then be parametrized by the $SU(4)$ elements $u_{I}^{\tilde I}$ with the identification
\begin{equation}
u_{I}^{\tilde I}\cong u_{I}^{\tilde J}\tilde u_{\tilde J}^{\tilde I} \, . \end{equation}
This provides a global parametrization of $S^{5}$ which maintain the compactness of the space manifest. However, the parametrization we used in the paper, see 
(\ref{BosonicCosets}), is a different one. We have complefixied the $S^{5}$ and gauge fixed the $USp(4)_{\mathbb{C}}$ subgroup such that we end up with $u_{I}^{\tilde J}$
given in 
(\ref{matrixelements}) 
where both $w$ and $y^{i}_{\bar i}$ are now complex variables. In order to obtain the real $S^{5}$ we need to impose a reality condition on the variables 
which guarantee that 
\begin{equation}
(Y_{IJ})^{*}=\frac{1}{2}\epsilon^{IJKL}Y_{KL} \, , 
\end{equation} 
with $Y_{KL}$ given in (\ref{eq:Xmatrix}).
These parametrization does not make the compactness of $S^{5}$ manifest since it covers just a patch defined by $Y_{43}\neq 0$.

Now we are going to relate our $S^{5}$ variables with the harmonic variables, which parametrize projective six-dimensional complex null vectors
\begin{equation}
\bar Y_{IJ}\cong \Lambda \bar Y_{IJ},\qquad \epsilon^{IJKL}\bar Y_{IJ}\bar Y_{KL}=0\, ,    
\end{equation}
with $\Lambda$ a constant. 
Picking the null vector 
$\bar{Y}_{ij}^0\cong \varepsilon_{ij}$ and $Y_{\bar i\bar j}^0=Y_{i\bar j}^0=0$ to be the origin in this case, we can obtain all the other projective null vectors by acting on $Y_{\bar i\bar j}^0$ with $SO(6)$ rotations. The subgroup that preserves the origin is given by $SO(2)\times SO(4)$ so we have a coset
\begin{equation}
\bar{Y}=\left(\frac{SO(6)}{SO(2)\times SO(4)}\right) \, . \end{equation}
In terms of spin groups we have
\begin{equation}
\left(\frac{SO(6)}{SO(2)\times SO(4)}\right)\cong \left(\frac{SU(4)}{S(U(2)\times U(2))} \right) \, ,     
\end{equation}
and this is the usual way in which harmonic variables are usually presented. 
Acting with $SU(4)$ elements $u^{I}_{J}$ in $\bar{Y}_{IJ}^0$ will give
\begin{equation}
\bar{Y}_{IJ}\cong u_{I}^{i}u_{J}^{j}\epsilon_{ij} \, .     
\end{equation}
and the subgroup $S(U(2)\times U(2))$ will constitute the $\hat u_{I}^{J}$ variables that satisfy
\begin{equation}
\hat u_{i}^{k}\hat u_{j}^{l}\epsilon_{kl}=\epsilon_{ij},\quad \hat u_{\bar i}^{k}\hat u_{\bar j}^{l}\epsilon_{kl}=0 ,\quad \hat u_{\bar i}^{k}\hat u_{j}^{l}\epsilon_{kl}=0\,.    
\end{equation}
Note that the $\hat u^J_I $ are also $SU(4)$ variables so they satisfy additional constraints. The harmonic variables will be given by the identification
\begin{equation}
u^{\prime \tilde I}_{I} \cong u_{I}^{\tilde J}\tilde u_{\tilde J}^{\tilde I}\,.   \end{equation}
Another parametrization of the projective six-dimensional null vectors is given by complexifying $SU(4)$ such that we can gauge fix 
\cite{Harmonics,Harmonics2}
\begin{equation}
[u^{\prime \tilde I}_{I}]=\left( 
\begin{tabular}{cc}
$\delta^{i}_{j}$&$(y^{\prime})^{\bar i}_{j}$ \\ $0$&$ \delta^{\bar i}_{\bar j}$
\end{tabular}\right) \, .   
\end{equation}
which is similar to our $S^{5}$ variables but here $(y^{\prime})^{\bar i}_{i}$ does not satisfy any reality condition. It is interesting to recover the radius dependence $R$ for both the $AdS_5$ and the $S^{5}$ space which gives
\begin{equation}
\varepsilon^{RSTU} X_{RS}X_{TU}=\varepsilon^{IJKL} Y_{IJ}Y_{KL} = -8R^{2}.   
\end{equation}
Now, both the boundary and the harmonic variables can be obtained by simply taking the limit $R\rightarrow 0$ together with the appropriate identifications $X^{RS}\cong \Lambda X^{RS}$ and $Y^{IJ}\cong \Lambda' Y^{IJ}$.

\subsection{Boundary superspace}

As showed above the 
boundary is located 
at $z \rightarrow \infty$.
At the boundary we expect to recover a superspace resembling the $d=4$ $\mathcal{N}=4$ 
Minkowski superspace with sixteen odd dimensions. 
We will define the 
boundary superspace 
to be the superspace that represents the scaling preserving transformations of the boundary in a given patch. Recall that the boundary can constitute in many patches.  
This happens for example in our coordinates $x$ and the Poincaré coordinates $\tilde{x}$ introduced previously.
They together cover the boundary and are glued by an inversion transformation. Notice that scale preserving transformation of one patch is not in general a scale preserving transformation of another patch. 
In the patch 
covered by $x$ 
the scaling preserving transformations are the ones in which the generators have negative charge under $\Delta$. 
The generators satisfying this property are $\{ K,
q^a_+, \bar{q}^{\dot{a}}_-\}$
which implies that the superspace is parametrized by $\{x,\theta_-^a,\bar{\theta}^{\dot{a}}_+ \}$. 
In the Poincaré patch the scaling preserving transformations are the ones respective to generators with positive charge under $\Delta$. It is also possible to glue these two superspaces by a super-diffeomorphism and obtain a global superspace for the boundary. This super-diffeomorphism is nothing more than the inversion acted on both the bosonic and fermionic coordinates. Here the inversion of fermionic coordinates is defined by
\begin{equation}
\theta^{I}_{\alpha}\leftrightarrow \theta^{I}_{\dot\alpha},\qquad  \theta_{I}^{\alpha}\leftrightarrow \theta_{I}^{\dot\alpha} \, . 
\end{equation}

It is interesting to study the $PSU(2,2|4)$ transformations of the worldsheet variables when they approach the boundary.
We compute them in infinitesimal form by using 
the formula 
(\ref{transformationsinfinitesimal}) and taking the limit 
$z \rightarrow \infty$. 
For some cosets it is possible 
to consistently set sixteen 
of the thirty two $\theta$'s
at the boundary to zero 
because the transformations of these variables under 
$PSU(2,2|4)$ tend to zero as 
$z \rightarrow \infty$. 
For other cosets all the transformations are nonvanishing at the boundary and one has instead to verify  
that the dependence of the bulk fields
tends to only sixteen $\theta$'s at the boundary, see \cite{ChiralSuperfields}. 
Our coset belongs to the second class because it is inconsistent to set 
$\theta^a_+=\bar{\theta}^{\dot{a}}_{-}=0$ at the boundary and indeed 
our vertex operators only depend on eight $\theta$'s 
instead of all the thirty two. 
For example, we have 
under the infinitesimal transformation parametrized by
$\epsilon^{\dot\alpha}_{\alpha}$ that
\begin{equation}
\delta \theta^i_{\alpha} \Big|_{\rm{boundary}}
= \epsilon^{\dot\alpha}_{\alpha} \theta^i_{\dot\alpha} \, , \qquad 
\delta \theta^{\dot\alpha}_{\bar{i}} 
\Big|_{\rm{boundary}} =- \epsilon^{\dot\alpha}_{\alpha}
\theta^{\alpha}_{\bar{i}} \, . 
\end{equation}

Apparently, it seems inconsistent 
that our vertex operators depend on $\theta^a_+$ which are not the $\theta$'s in the superspace covered by our variables $x$ at the boundary. However, recall that our vertex operators are at the point 
$x=\infty$ which is only covered by the Poincaré path
whose superspace contains 
$\theta^a_+$.

 

\section{Conclusions} 
\label{Conclusions}

In this paper, we have found 
for the first time
explicit expressions for the gauge superfields appearing in the superstring
vertex operators 
\cite{BerkovitsChandia} in the pure spinor formalism 
in an $AdS_5 \times S^5$ background. 
This generalizes the results of 
\cite{BerkovitsFleury}, where 
the vertices were computed only close to the boundary. 
The vertex operators are labelled by an integer $n$ 
which corresponds to the dimension and $R$-charge of the states. 
For $n=0$, it reduces correctly
to the well known dilaton
vertex operator which in integrated form corresponds 
to the Lagrangian. 
We have also taken the flat space limit of our results.
This amounts  
to deform the $\mathfrak{psu}(2,2|4)$ algebra continuously to the 
ten dimensional super-Poincaré algebra. 
In the limit, our vertices coincide with the flat space ones as expected.  
Moreover, we have both located 
the boundary of $AdS_5$ in our coordinates and described a 
way of finding the vertex operators at different positions. 

One immediate application
of our results would be the calculation of string amplitudes \cite{FleuryMartinsII}. 
Many amplitudes have been 
computed using the pure spinor formalism in a flat space background.
It is likely that we can 
use in $AdS_5$ some of the techniques and the computer packages used in those calculations such as   
\cite{MafraPackage}. 
Technically, the main difference between the calculations 
is that the OPE's of the worldsheet variables receive
$\alpha^{\prime}$ corrections in $AdS$ and they have to be computed order by order, or eventually bootstraped. 
There were many progresses and new ideas 
for computing these amplitudes using bootstrap, localization and integrability
techniques and  several amplitude results already exist in the literature for four-point and five-point functions 
\cite{AmplitudeI,AmplitudeII,AmplitudeIII,AmplitudeIV,AmplitudeV,AmplitudeVI,AmplitudeVII,newAlday,AmplitudeVIII,AmplitudeIX,AmplitudeX,AmplitudeXI,AmplitudeXII,AmplitudeXIII,AmplitudeXIV,AmplitudeXV,AmplitudeXVI,AmplitudeXVII,AmplitudeXVIII,Bissi,Bissi2,AmplitudeXIX,AmplitudeXX,AmplitudeXXI,IntegrabilidadeI,IntegrabilidadeII,IntegrabilidadeIII,IntegrabilidadeIV,IntegrabilidadeV,IntegrabilidadeVI,IntegrabilidadeVII,IntegrabilidadeVIII,IntegrabilidadeIX,IntegrabilidadeX,IntegrabilidadeXI,IntegrabilidadeXII}.
It will be great if we could rederive any of these results using 
the pure spinor language. 


Another research direction is the construction of the massive vertex operators.
They are known 
in a flat background \cite{Massive,NewMassiveI,NewMassiveII,NewMassiveIII}, but these tecniques
have not been used in $AdS$ yet. 
One difficult is that 
the masses of the states are very complicated functions of $\alpha^{\prime}$ as the worldsheet theory is interacting. The spectrum can be obtained by  
using string and integrability tecniques 
\cite{Konishi,ReviewIntegrability,Exact,QSC}.
In this direction, it will be very interesting to take the OPE's 
of our vertices after translations. 
The brute force calculation can be done 
by reading the interaction terms from
the Lagragian and using standard methods for
computing OPE's in interacting 
quantum field theories, see \cite{WeinbergvolII} 
for example. Another way of trying 
to construct the massive vertex operators is by generalizing the techniques developed in  \cite{RenannSpectrum}.
It will be also very interesting to better understand the current algebra of the model at the quantum level, see 
\cite{Current3,Currents} for progresses. 
Notice that the BRST operator is 
a pure spinor times a 
fermionic current.
In addition, 
the current algebra 
was already used to compute part of the spectrum in 
\cite{Ysystem} by deriving the so called $Y$-system.
The current algebra is very constrained and 
maybe one can try to set a bootstrap program for it.

\section*{Acknowledgement}

We would like to thank Thiago Araujo 
for collaboration during the initial 
stages of this work.
We acknowledge useful discussions with 
Thiago Araujo, Nathan Berkovits, Luis Ypanaque and 
Dennis Zavaleta. 
This work was supported by
the Serrapilheira Institute (grant number Serra-1812-26900). L. M. would like
to thank FAPESP grant 18/07834-8 for partial financial support.
 


\appendix


\section{The $\mathfrak{psu}(2,2|4)$ algebra}\label{SectionAlgebra}

We start by describing the superalgebra $\mathfrak{pu}(2,2|4)$. The $\mathfrak{p}$ is due to the fact that we always work with representations 
with vanishing central charge. The generators of this superalgebra can be organized as follows
\begin{equation}
M^A_B = 
\left[ 
\begin{tabular}{cccc}
$M^{\alpha}_{\beta}$ & $P^{\alpha}_{\dot\beta}$ & $q^{\alpha}_{j}$ & $q^{\alpha}_{\bar{j}}$ \\
$K^{\dot\alpha}_{\beta}$ & $M^{\dot{\alpha}}_{\dot{\beta}}$ & $q^{\dot\alpha}_{j}$ & $q^{\dot\alpha}_{\bar{j}}$ \\
$q^{i}_{\beta}$ & $q^{i}_{\dot\beta}$ & $M^{i}_j$ & $K^{i}_{\bar{j}}$ \\
$q^{\bar{i}}_{\beta}$ & $q^{\bar{i}}_{\dot\beta}$ & $P^{\bar{i}}_{j}$ & $M^{\bar{i}}_{\bar{j}}$  
\end{tabular}
\right] \, ,
\label{MatrixMAB}
\end{equation}
where $\alpha, \dot{\alpha}, i, \bar{i} = 1, 2$ and the $q$'s are the thirty two fermionic generators.   
Note that the position of the indices in the $q$'s are used to distinguish the different generators so we cannot raise and lower the indices with the $\epsilon$ tensors while preserving the symbol $q$. The $P$'s are the translation generators and the $K$'s are the special translation generators of the bosonic subgroups.  
The diagonal generators are further decomposed
into  
\begin{equation} 
\begin{aligned}
& M^{\alpha}_{\beta} = \hat{M}^{\alpha}_{\beta} + \frac{1}{2} \delta^{\alpha}_{\beta} \Delta
+ \frac{1}{2} \delta^{\alpha}_{\beta} \mathcal{B} \, , \quad \quad M^{\dot\alpha}_{\dot\beta}  = \hat{M}^{\dot\alpha}_{\dot\beta} - \frac{1}{2} \delta^{\dot\alpha}_{\dot\beta} \Delta
+ \frac{1}{2} \delta^{\dot\alpha}_{\dot\beta} \mathcal{B} \, , \\
& M^i_j =\hat{M}^{i}_{j} + \frac{1}{2} \delta^{i}_{j} J
- \frac{1}{2} \delta^{i}_{j} \mathcal{B} \, , \quad \quad \quad 
M^{\bar{i}}_{\bar{j}} =\hat{M}^{\bar{i}}_{\bar{j}} - \frac{1}{2} \delta^{\bar{i}}_{\bar{j}} J
- \frac{1}{2} \delta^{\bar{i}}_{\bar{j}} \mathcal{B} \, , 
\end{aligned} 
\label{TheHattedGenerators} 
\end{equation} 
where the $\hat{M}$'s are the rotation generators, $\Delta$ is the dilatation, $J$ is the diagonal $R$-charge generator 
and $\mathcal{B}$ is the so called hypercharge generator. The commutation relations with the last three generators are diagonal
and takes the form
\begin{equation}
[ \Delta, M^A_B ] = C_{\Delta}(M^A_B) M^A_B \, , \quad [ J, M^A_B ] = C_{J}(M^A_B) M^A_B \, , 
\quad [ \mathcal{B}, M^A_B ] = C_{\mathcal{B}}(M^A_B) M^A_B \, , 
\end{equation}
and the nonzero numerical values of the $C$'s can be read from the tables \ref{tab:ChargesDJ} and \ref{tab:ChargesB} below. It will be convenient to group
the fermionic generators in groups with definite $\Delta-J$ charge using $SO(8)$ notation as follows 
\begin{table}[h]
\centering
\begin{tabular}{ccccccc}
 & $K^{\dot{\alpha}}_{\alpha}$ & $P^{\alpha}_{\dot{\alpha}}$ & $K^i_{\bar{i}}$ & $P^{\bar{i}}_i$ & $q^a_{\pm}$ & $\bar{q}^{\dot{a}}_{\pm}$ \\ 
 \hline 
 $C_{J}:$ &  0 & 0 & 1 & -1 & $\pm 1/2$ & $\pm 1/2$  \\
 $C_{\Delta}:$ &  -1 & 1 & 0 & 0 & $\mp 1/2$ & $\pm 1/2$  
\end{tabular}
\caption{The charge of the generators under $\Delta$ and $J$. The values of $C_J$ and $C_{\Delta}$ of the generators not 
appearing in the table are zero. Both the $q$'s and the $\bar{q}$'s are defined in (\ref{eq:qinso8}).} 
\label{tab:ChargesDJ}
\end{table}
\begin{table}[h]
\centering
\begin{tabular}{ccccc}
 & $(q^{\alpha}_{i}, q^{\alpha}_{\bar{i}})$ & $(q_{\alpha}^{i}, q_{\alpha}^{\bar{i}})$ & $(q^{\dot\alpha}_{i}, q^{\dot\alpha}_{\bar{i}})$ & $(q_{\dot\alpha}^{i}, q_{\dot\alpha}^{\bar{i}})$ \\ \hline
$C_{\mathcal{B}}: $ & 1/2  & -1/2 & 1/2 & -1/2 
\end{tabular}
\caption{The hypercharge of the generators. The values of $C_{\mathcal{B}}$ for the generators not 
appearing in the table are zero.} 
\label{tab:ChargesB}
\end{table}
\begin{equation}
\bar{q}^{\dot{a}}_+ = (q^{\alpha}_{\bar{i}}, q^i_{\dot{\alpha}} ) \, , \quad \bar{q}^{\dot{a}}_- = (q^{\bar{i}}_{\alpha}, q^{\dot\alpha}_i) \, , \quad
q^a_+ = (q^i_{\alpha}, q^{\dot\alpha}_{\bar{i}}) \, , \quad q^a_- =(q^{\bar{i}}_{\dot\alpha}, q^{\alpha}_i) \, . 
\label{eq:qinso8} 
\end{equation} 
In the definition above, the subscript indicates the charge under $J$ and $a, \dot{a}$ are chiral and antichiral $SO(8)$ spinor indices.

The remaining commutators of the algebra are of the form 
\begin{equation}
[ M^B_A , M^D_C ] = \delta_A^D M_C^B - (-1)^{([A]+[B])([C]+[D])} \delta^B_C M^A_D \, .   
\label{BasicCommuta} 
\end{equation}
where $[A]=1$ if $A=i$ or $A=\bar{i}$ and zero otherwise. Notice that it is possible to deduce the commutators involving the $\hat{M}$ generators
from the expression above and the definitions in (\ref{TheHattedGenerators}). One has,
\begin{equation}
[ \hat{M}^{\alpha}_{\beta}, \hat{M}^{\gamma}_{\delta}] =[ M^{\alpha}_{\beta}, M^{\gamma}_{\delta}] \, ,
\end{equation}
and 
\begin{equation}
[ \hat{M}^{\alpha}_{\beta}, M^A_B ] = [ M^{\alpha}_{\beta}, M^A_B ]- \frac{1}{2} \delta^{\alpha}_{\beta} [\Delta, M^A_B ] - \frac{1}{2}   
 \delta^{\alpha}_{\beta} [\mathcal{B}, M^A_B ] \, . 
\end{equation} 
which gives, for example,
\begin{equation}
[ \hat{M}^{\alpha}_{\beta}, q^{\gamma}_i ] = \delta^{\gamma}_{\beta} q^{\alpha}_i - \frac{1}{2} \delta^{\alpha}_{\beta} q^{\gamma}_i\, . 
\end{equation} 
	
One important property of the commutation relations is that the hypercharge generator $\mathcal{B}$ never appears
on the right hand side of (\ref{BasicCommuta}). Keeping the commutations relations and dropping $\mathcal{B}$ 
one arrives at the $\mathfrak{psu}(2,2|4)$ algebra. An important subalgebra (isotropy group) is $\mathfrak{usp}(2,2) \times \mathfrak{usp}(4)$ which are generated by the following 
generators 
\begin{equation}
\{ \hat{M}^{\alpha}_{\beta} , \, \hat{M}^{\dot\alpha}_{\dot\beta}, \, (P^{\alpha}_{\dot\alpha} +\epsilon_{\dot\alpha \dot\beta} \epsilon^{\alpha \beta} 
K_{\beta}^{\dot\beta})  \}  \; \times \; \{ \hat{M}^{i}_{j} , \, \hat{M}^{\bar{i}}_{\bar{j}}, \, (P^{\bar{i}}_{i} +\epsilon_{i j} \epsilon^{\bar{i} \bar{j}} 
K_{\bar{j}}^{j})  \} \, , 
\end{equation} 
and in our conventions $\epsilon_{12}=\epsilon^{12}=1$.

 \section{BRST transformations} \label{appendixcoset}

In this Appendix, we provide all the BRST transformations of the worldsheet variables for the coset given in (\ref{eq:cosetparametrization}).  
In addition, we also consider a second coset ($SU(2,2)\times SU(4)$ covariant parametrization) where it is easier to define gauge invariant pure spinors.  

\subsection{BRST transformations for the coset (\ref{eq:cosetparametrization})}

In order to find the BRST transformations for the worldsheet variables parametrizing the coset (\ref{eq:cosetparametrization}),
one has to solve the equation (\ref{eq:BRST}). In the process, one also gets the compensating gauge transformation parameters
$\Sigma$'s.  The manipulations are standard but tedious and the result is given below in terms of commutators. 
We will use the definitions
\begin{equation}
\tilde{K}^i_{\bar{i}} = e^w K^i_{\bar{i}} \, , \quad \tilde{K}^{\dot\alpha}_{\alpha} = e^{-z} K^{\dot\alpha}_{\alpha} \, , \quad 
\end{equation}
and the BRST transformations are ($\Sigma^{\bar i}_{i}$ 
and $\Sigma^{\alpha}_{\dot\alpha}$ will be given below and we are using a compact notation where the contractions of the indices are the obvious ones) 
\begin{equation}
(Qz) \Delta + (Qw) J - \Sigma \hat{M} = \left[ \theta_+ q_-, e^{\frac{w-z}{2}} \lambda_- q_+ \right] +
\left[ \bar{\theta}_+ \bar{q}_-, e^{\frac{w+z}{2}} \bar{\lambda}_- \bar{q}_+ \right] + 
\left[ \bar{\theta}_+ \bar{q}_-, \left[ \theta_+ q_-, \Sigma \tilde{K} \right] \right] \, , 
\end{equation} 
and
\begin{equation}
\begin{aligned}
(Q \theta_+) q_- &= e^{\frac{z-w}{2}} \lambda_+ q_- - \frac{1}{2} \left[ \theta_+ q_-, \left[ \theta_+ q_-, e^{\frac{w-z}{2}} \lambda_- q_+ \right] \right] - \left[ \theta_+ q_-, \left[ \bar{\theta}_+ \bar{q}_-, e^{\frac{w+z}{2}} \bar{\lambda}_- \bar{q}_+ \right] \right] \\
&- \left[ \theta_+ q_- , \left[ \bar{\theta}_+ \bar{q}_-, \left[ \theta_+ q_-, \Sigma \tilde{K} \right] \right] \right] \, , \\
(Q \bar{\theta}_+) \bar{q}_- & = e^{- \frac{w+z}{2}} \bar{\lambda}_+ \bar{q}_- - \frac{1}{2} \left[ \bar{\theta}_+ \bar{q}_-, \left[ \bar{\theta}_+ \bar{q}_-, \left[ \theta_+ q_-, \Sigma \tilde{K} \right] \right] \right] \\
&- \frac{1}{2} \left[ \bar{\theta}_+ \bar{q}_- , \left[ \bar{\theta}_+ \bar{q}_-, e^{\frac{w+z}{2}} \bar{\lambda}_- \bar{q}_+ \right] \right] +
\left[ \theta_+ q_-, \Sigma \tilde{K} \right]_{\bar{q}_-} \, , \\
(Q \bar{\theta}_-) \bar{q}_+ & = e^{\frac{w+z}{2}} \bar{\lambda}_- \bar{q}_+ + \left[ \theta_+ q_-, \Sigma \tilde{K} \right]_{\bar{q}_+} \, , 
\end{aligned}
\end{equation}
finally, 
\begin{equation}
\begin{aligned}
(Q \theta_-) q_+ &= e^{\frac{w-z}{2}} \lambda_- q_+ + \left[ \bar{\theta}_- \bar{q}_+ , \left[ \bar{\theta}_+ \bar{q}_-, e^{\frac{w-z}{2}}\lambda_- q_+ \right] \right]
+ \left[ \bar{\theta}_- \bar{q}_+, \Sigma \tilde{K} \right]  \\
&+ \left[\bar{\theta}_+ \bar{q}_-, \Sigma \tilde{K} \right]+ 
\frac{1}{2} \left[ \bar{\theta}_- \bar{q}_+,
\left[ \bar{\theta}_+ \bar{q}_-, \left[ \bar{\theta}_+ \bar{q}_-, \, \Sigma \tilde{K} \right]\right] \right] \, , 
\end{aligned} 
\end{equation} 
and 
\begin{equation}
\begin{aligned}
 (Qx)K + (Qy)K & = \Sigma \tilde{K}+ \left[ \bar{\theta}_+ \bar{q}_-, e^{\frac{w-z}{2}} \lambda_- q_+ \right] 
+ \left[ \bar{\theta}_- \bar{q}_+, e^{\frac{w-z}{2}} \lambda_- q_+ \right] + \\
&+ \frac{1}{2} \left[ \bar{\theta}_- \bar{q}_+, \left[ \bar{\theta}_- \bar{q}_+, \left[ \bar{\theta}_+ \bar{q}_-, e^{\frac{w-z}{2}} \lambda_- q_+ \right]\right]\right] 
+ \frac{1}{2} \left[ \bar{\theta}_- \bar{q}_+ \left[ \bar{\theta}_- \bar{q}_+, \Sigma \tilde{K} \right]\right] \\
&+ \frac{1}{2} \left[ \bar{\theta}_+ \bar{q}_-, \left[ \bar{\theta}_+ \bar{q}_-, \Sigma \tilde{K} \right]\right]
+ \left[ \bar{\theta}_- \bar{q}_+, \left[ \bar{\theta}_+ \bar{q}_-, \Sigma \tilde{K} \right]\right] \\
&+ \frac{1}{4} \left[ \bar{\theta}_- \bar{q}_+, \left[ \bar{\theta}_- \bar{q}_+, \left[ \bar{\theta}_+ \bar{q}_-, \left[ \bar{\theta}_+ \bar{q}_-, \Sigma \tilde{K} \right]\right]\right]\right]
\, . 
\end{aligned}
\end{equation}

The remaining $\Sigma$'s can be obtained by solving the following equations
\begin{equation}
\begin{aligned}
\Theta^{\dot\alpha}_{\alpha} + \Sigma^{\dot\alpha}_{\alpha}
&+ \Sigma^{\bar{i}}_i (\tilde\theta^i_{\alpha} \tilde\theta^{\dot{\alpha}}_{\bar{i}}) =0 \, , \quad 
 \Theta^i_{\bar{i}} + \Sigma^{i}_{\bar{i}}
+ \Sigma^{\alpha}_{\dot\alpha} (\tilde\theta^{\dot\alpha}_{\bar{i}} \tilde\theta^{i}_{\alpha}) =0 \, , \\
 & \Sigma^{\alpha}_{\dot\alpha} = \epsilon^{\alpha \beta} \epsilon_{\dot\alpha \dot\beta} \Sigma^{\dot\beta}_{\beta} \, , \quad
\Sigma^i_{\bar{i}} = \epsilon^{i j} \epsilon_{\bar{i} \bar{j}} \Sigma^{\bar{j}}_{j} \, , 
\end{aligned}
\end{equation}
where 
\begin{equation}
\Theta^{\dot\alpha}_{\alpha} \equiv \tilde\theta^i_{\alpha} \lambda^{\dot\alpha}_i + \tilde\theta^{\dot\alpha}_{\bar{i}} \lambda^{\bar{i}}_{\alpha} \, , \quad 
\Theta^i_{\bar{i}} \equiv \tilde\theta^i_{\alpha} \lambda^{\alpha}_{\bar{i}} + \tilde\theta^{\dot{\alpha}}_{\bar{i}} \lambda^i_{\dot\alpha} \, ,
\end{equation}  
and the $\tilde{\theta}$'s were defined in  
(\ref{eq:thetatilde}).  
The solution to these equations is ($\Theta_{\dot\alpha}^{\alpha}=\epsilon^{\alpha \beta} \epsilon_{\dot\alpha \dot\beta}
\Theta^{\dot\beta}_{\beta}$ and $\Theta_i^{\bar{i}}= \epsilon_{i j} \epsilon^{\bar{i} \bar{j}}\Theta_{\bar{j}}^j$)
\begin{equation}
\begin{aligned}
\Sigma^{\dot\alpha}_{\alpha} & = -\Theta^{\dot\alpha}_{\alpha} + (\tilde\theta^{\dot\alpha}_{\bar{i}} \Theta^{\bar{i}}_{i} \tilde\theta^i_{\alpha}) -  
(\epsilon^{\bar{i} \bar{j}} \tilde\theta^{\dot\alpha}_{\bar{i}} \tilde\theta^{\dot\beta}_{\bar{j}})
(\epsilon_{i j} \tilde\theta^i_{\alpha} \tilde\theta^j_{\beta}) \Theta^{\beta}_{\dot\beta} 
 +
 (\epsilon^{\bar{i} \bar{j}} \tilde\theta^{\dot\alpha}_{\bar{i}} \tilde\theta^{\dot\beta}_{\bar{j}})
 (\epsilon_{i j} \tilde\theta^i_{\alpha} \tilde\theta^j_{\beta})\epsilon_{\dot\gamma \dot\beta} \epsilon^{\gamma \beta} 
 (\tilde\theta^{\dot\gamma}_{\bar{k}} \Theta^{\bar{k}}_i \tilde\theta^i_{\gamma}) \, , \\
 \Sigma^i_{\bar{i}} & = -\Theta^i_{\bar{i}} + (\tilde\theta^i_{\alpha} \Theta^{\alpha}_{\dot\alpha} \tilde\theta^{\dot\alpha}_{\bar{i}})  - 
 (\epsilon^{\alpha \beta} \tilde\theta^i_{\alpha} \tilde\theta^j_{\beta})(\epsilon_{\dot\alpha \dot\beta} \tilde\theta^{\dot\alpha}_{\bar{i}} \tilde\theta^{\dot\beta}_{\bar{j}}) \Theta^{\bar{j}}_j 
 + (\epsilon^{\alpha \beta} \tilde\theta^i_{\alpha} \tilde\theta^j_{\beta})
(\epsilon_{\dot\alpha \dot\beta} \tilde\theta^{\dot\alpha}_{\bar{i}} \tilde\theta^{\dot\beta}_{\bar{j}} ) \epsilon_{j k} \epsilon^{\bar{j} \bar{k}} 
(\tilde\theta^k_{\gamma} \Theta^{\gamma}_{\dot\gamma} \tilde\theta^{\dot\gamma}_{\bar{k}}) \, . 
 \end{aligned}
\end{equation} 
 
\subsection{
Covariant Derivatives 
for the coset (\ref{eq:cosetparametrization})} 

In this subsection, we will write 
down the covariant derivatives 
$\nabla$'s relatively to 
the coset 
(\ref{eq:cosetparametrization}).
It is possible to read 
the fermionic derivatives 
from 
the BRST transformations given previously, since the 
BRST charge takes the following 
form in our conventions
\begin{equation}
Q = \lambda^i_{\alpha} \nabla^{\alpha}_i +
\lambda^i_{\dot\alpha} \nabla^{\dot\alpha}_i
+\lambda^{\bar{i}}_{\alpha} \nabla^{\alpha}_{\bar{i}} +
\lambda^{\bar{i}}_{\dot\alpha} \nabla^{\dot\alpha}_{\bar{i}}+
\lambda_i^{\alpha} \nabla_{\alpha}^i +
\lambda_i^{\dot\alpha} \nabla_{\dot\alpha}^i
+\lambda_{\bar{i}}^{\alpha} \nabla_{\alpha}^{\bar{i}} +
\lambda_{\bar{i}}^{\dot\alpha} \nabla_{\dot\alpha}^{\bar{i}} \, . 
\label{ourBRST}
\end{equation}
An alternative procedure 
\cite{Galperin} that 
does not involve the knowledge 
of the BRST transformations  
and enables the computation of all
covariant derivatives including 
the bosonic ones 
uses the Cartan form 
($Y_a$ are the generators appearing in the
coset and $X_b$ are the generators 
of the isotropy group) 
\begin{equation}
g^{-1} d g 
= \omega_Y^a Y_a + \omega_X^b X_b \ , 
\end{equation}
and consists in solving the system of equations below
\begin{equation}
(d + \omega_X^b X_b) \psi_k = 
\omega_Y^a \nabla_a \psi_k \, ,
\end{equation}
where $\psi_k$ is any field in the $k$
representation of the isotropy group.
The covariant derivatives are complicated objects and some of them have a very high order expansion in the $\theta$'s. We have
\begin{equation}
\begin{aligned}
& \hspace{30mm} \nabla^{\alpha}_i =  e^{\frac{1}{2}(z-w)} \frac{\partial}{\partial \theta^i_{\alpha}} 
\, , \quad \quad 
\nabla^{\bar{i}}_{\dot\alpha} =
e^{\frac{1}{2}(z-w)} 
\frac{\partial}{\partial \theta^{\dot{\alpha}}_{\bar{i}}} \, , \\
\nabla^i_{\alpha}
&=e^{\frac{1}{2}(w-z)} 
\frac{\partial}{\partial \theta^{\alpha}_i}+
e^{\frac{1}{2}(z-w)} \tilde\theta^i_{\beta} \, \tilde\theta^j_{\alpha} \, 
\frac{\partial}{\partial \theta^j_{\beta}} 
+e^{\frac{1}{2}(w-z)} \tilde\theta^{i}_{\dot\alpha}
\tilde\theta^{\dot\alpha}_j
\frac{\partial}{\partial \theta^{\alpha}_j} 
- e^{\frac{1}{2}(w-z)} \tilde{\theta}^i_{\dot\alpha} \tilde{\theta}^{\bar{i}}_{\alpha} \frac{\partial}{\partial \theta^{\bar{i}}_{\dot\alpha}}
- \tilde\theta^i_{\beta} \hat{M}^{\beta}_{\alpha} 
- \tilde\theta^j_{\alpha} \hat{M}^i_j \\
&+\frac{1}{2} \tilde\theta^i_{\alpha} \left(
\frac{\partial}{\partial z}+
\frac{\partial}{\partial w} \right) 
+ e^{-z} \tilde\theta^i_{\dot\alpha} \frac{\partial}{\partial x^{\alpha}_{\dot\alpha}} 
+ e^w \tilde\theta^{\bar{i}}_{\alpha} 
\frac{\partial}{\partial y^{\bar{i}}_i} 
- e^{w} \tilde{\theta}^i_{\dot\alpha} 
\tilde{\theta}^{\bar{i}}_{\alpha} 
\tilde{\theta}^{\dot\alpha}_j \frac{\partial}{\partial y^{\bar{i}}_j} \, , 
\end{aligned}
\end{equation}
and 
\begin{equation}
\begin{aligned}
\nabla^{\dot\alpha}_{\bar{i}} &= 
e^{\frac{1}{2}(w-z)} \frac{\partial}{\partial \theta^{\bar{i}}_{\dot\alpha}}
+e^{\frac{1}{2}(w-z)} \tilde{\theta}^{\alpha}_{\bar{i}} 
\tilde{\theta}^{\dot\alpha}_i \frac{\partial}{\partial \theta^{\alpha}_i}+
e^{\frac{1}{2}(z-w)}
\, \tilde\theta^{\dot\alpha}_{\bar{j}}
\, \tilde\theta^{\dot\beta}_{\bar{i}} \,
\frac{\partial}{\partial \theta^{\dot\beta}_{\bar{j}}} 
+ e^{\frac{1}{2} (w-z)} 
\tilde{\theta}^{\bar{j}}_{\alpha} \tilde{\theta}^{\alpha}_{\bar{i}}
\frac{\partial}{\partial \theta^{\bar{j}}_{\dot\alpha}}
- \tilde{\theta}^{\dot\alpha}_{\bar{j}} 
\hat{M}^{\bar{j}}_{\bar{i}}
- \tilde{\theta}^{\dot\beta}_{\bar{i}} \hat{M}^{\dot\alpha}_{\dot{\beta}}
\\
&- \frac{1}{2} \tilde\theta^{\dot\alpha}_{\bar{i}} \left(\frac{\partial}{\partial z}+
\frac{\partial}{\partial w}\right)
+e^{-z} \tilde{\theta}^{\alpha}_{\bar{i}} \frac{\partial}{\partial x^{\alpha}_{\dot\alpha}}
+e^w 
\tilde{\theta}^{\dot\alpha}_i \frac{\partial}{\partial y^{\bar{i}}_{i}}+
e^w \tilde{\theta}^{\bar{j}}_{\beta} \tilde{\theta}^{\beta}_{\bar{i}} 
\tilde{\theta}^{\dot\alpha}_j 
\frac{\partial}{\partial y^{\bar{j}}_j} \, . 
\end{aligned} 
\end{equation}
In the expressions above the $\tilde{\theta}$'s were defined in 
(\ref{eq:thetatilde}) and the $\hat{M}$'s 
only act on the $\lambda$'s. The action is canonical and given by 
\begin{equation}
\hat{M}^A_B \cdot \lambda^C = 
\delta^C_B \, \lambda^A 
- \frac{1}{2} \delta^A_B \lambda^C \, , \qquad  
\hat{M}^A_B \cdot \lambda_C = 
- \delta^A_C \, \lambda_B 
+ \frac{1}{2} \delta^A_B \lambda_C \, . 
\end{equation}
The remaining covariant derivatives are the most complicated ones. 
Due to the fact that our vertex operators
only depend on $z+w$, $\lambda$'s and $\tilde{\theta}^a_+$, we 
are going to write down only the terms
that act non-trivially on the vertices.
We have, schematically,
\begin{equation}
\begin{aligned}
&\nabla^{\dot\alpha, \bar{i}}_{i,\alpha} \Big|^{\rm{no \, gauge}}_{\rm{vertex}} = 
e^{\frac{1}{2} (z-w)} 
\left[ \tilde\theta^2_+ \tilde{\bar{\theta}}_+ \frac{\partial}{\partial \theta_+} +
\tilde\theta^7_+ \tilde{\bar{\theta}}_+ \frac{\partial}{\partial \theta_+}  \right]
^{\dot\alpha, \bar{i}}_{i,\alpha} + \frac{1}{2}
\left[ \tilde\theta_+^2 \tilde{\bar{\theta}}_+
+ \tilde\theta_+^6 \tilde{\bar{\theta}}_+ \right]
^{\dot\alpha, \bar{i}}_{i,\alpha} \left(\frac{\partial}{\partial z} - \frac{\partial}{\partial w} \right) \, , \\
& \nabla^{i,\alpha}_{\dot\alpha, \bar{i}} \Big|^{\rm{no \, gauge}}_{\rm{vertex}}= 
e^{\frac{1}{2} (z-w)} \left[ 
\tilde\theta_+ \tilde{\bar\theta}_+ 
\frac{\partial}{\partial \theta_+}
+ 
\tilde\theta^5_+ \tilde{\bar\theta}_+ 
\frac{\partial}{\partial \theta_+}\right]^{i,\alpha}_{\dot\alpha, \bar{i}}
+ \frac{1}{2}
\left[ \tilde\theta_+^4 \tilde{\bar{\theta}}_+
+ \tilde\theta_+^8 \tilde{\bar{\theta}}_+ \right]
_{\dot\alpha,\bar{i}}^{i,\alpha} \left(\frac{\partial}{\partial z} - \frac{\partial}{\partial w} \right) \, ,
\label{complicatecovariant}
\end{aligned}
\end{equation}
where {\it{no gauge}} means that we have omitted terms acting on the $\lambda$'s which are quite complicated. In any case, these terms can be deduced from the BRST transformations. 

The BRST operator given in (\ref{BRSToperator}) and written in terms of $\lambda_L^{\underline\alpha}$ and $\lambda_R^{\hat{\underline\alpha}}$ 
can be obtained from (\ref{ourBRST}) by reorganising the terms and
by using the definitions of
both the left and right pure spinors of 
(\ref{definitionLR})
and (\ref{Definitionsoftildespinors}). We have 
\begin{equation}
Q = \frac{1}{2} 
\left( (\lambda_L)^{\alpha}_i (\nabla_L)^i_{\alpha} +(\lambda_L)^{\dot\alpha}_{\bar{i}}
(\nabla_L)^{\bar{i}}_{\dot\alpha}
+ (\lambda_L)^{\alpha}_{\bar{i}}
(\nabla_L)^{\bar{i}}_{\alpha} 
+
(\lambda_L)^{\dot\alpha}_{i}
(\nabla_L)^{i}_{\dot\alpha} + 
L \rightarrow R 
\right) \, , 
\end{equation}
where 
\begin{equation}
\begin{aligned}
(\nabla_L)^i_{\alpha} 
=\nabla^i_{\alpha} + \epsilon^{i j} \epsilon_{\alpha \beta} \nabla^{\beta}_j 
\, , \quad \quad 
(\nabla_L)^{\bar{i}}_{\dot\alpha} 
=\nabla^{\bar{i}}_{\dot\alpha} + \epsilon^{\bar{i} \bar{j}} \epsilon_{\dot\alpha \dot\beta} \nabla^{\dot\beta}_{\bar{j}} \, , \\
(\nabla_L)^i_{\dot\alpha} 
=\nabla^i_{\dot\alpha} - \epsilon^{i j} \epsilon_{\dot\alpha \dot\beta} \nabla^{\dot\beta}_j 
\, , \quad \quad 
(\nabla_L)^{\bar{i}}_{\alpha} 
=\nabla^{\bar{i}}_{\alpha} - \epsilon^{\bar{i} \bar{j}} \epsilon_{\alpha \beta} \nabla^{\beta}_{\bar{j}} \, ,
\end{aligned}
\end{equation}
and the covariant derivatives with $R$ are similar and 
they are obtained by flipping
the sign before the second $\nabla$ on the right hand side of the expressions above.
In the Introduction, we proved that the supergravity vertex operators in a flat background can be put in 
the gauge $\bar\lambda^{\dot a}_{L}\bar\lambda^{\dot b}_{R}A_{\dot a\dot b}$. The proof was based on the commutation algebra 
of the flat space covariant derivatives.
We believe that a similar proof is
possible for $AdS_5$ and we compute 
some of the commutation relations of the covariant 
derivatives below. 
In the expressions, we only show 
the potentially nonzero terms when 
acting on our vertices, we have
\begin{equation}
\begin{aligned}
\left[(\nabla_L)^i_{\alpha}, 
(\nabla_L)^{i^{\prime}}_{\alpha^{\prime}}\right] \Bigg|_{\rm{vertex}} &=
\epsilon^{i^{\prime} i}
\epsilon_{\alpha^{\prime} \alpha } 
 \left(\frac{\partial}{\partial z} + \frac{\partial}{\partial w} \right) \\
&+ \epsilon^{i^{\prime} i}
\epsilon_{\alpha^{\prime} \beta}
\theta^j_{\alpha} 
\frac{\partial}{\partial \theta^j_{\beta}}
- \epsilon^{i^{\prime} j}
\epsilon_{\alpha^{\prime} \alpha} \theta^i_{\beta} \frac{\partial}{\partial \theta^j_{\beta}}
+ \epsilon^{i i^{\prime}}
\epsilon_{\alpha \beta}
\theta^j_{\alpha^{\prime}}
\frac{\partial}{\partial \theta^j_{\beta}} 
-\epsilon^{ij}\epsilon_{\alpha \alpha^{\prime}} \theta^{i^{\prime}}_{\beta} 
\frac{\partial}{\partial \theta^j_{\beta}} \\
&- \epsilon^{i^{\prime} i}
\epsilon_{\alpha^{\prime} \beta}
\hat{M}^{\beta}_{\alpha}
- \epsilon^{i^{\prime} j}
\epsilon_{\alpha^{\prime}
\alpha}
\hat{M}^i_j 
-\epsilon^{i i^{\prime}}
\epsilon_{\alpha \beta} \hat{M}^{\beta}_{\alpha^{\prime}}
-\epsilon^{ij} \epsilon_{\alpha \alpha^{\prime}} \hat{M}^{i^{\prime}}_{j} \, .
\end{aligned} 
\end{equation}
In fact, due to the fact 
that our vertex operators 
preserve a $SO(4) \times SO(4)$ symmetry
and only depends on $\bar\lambda^{\dot{a}}$, the terms 
with $\hat{M}$'s and derivatives of $\theta$'s
cancel among themselves as they correspond to $SO(4) \times SO(4)$ rotations. So, in fact, the expressions simplify to 
\begin{equation}
\begin{aligned}
&\left[(\nabla_L)^i_{\alpha}, 
(\nabla_L)^{i^{\prime}}_{\alpha^{\prime}}\right] \Bigg|_{\rm{vertex}}
=\epsilon^{i^{\prime} i}
\epsilon_{\alpha^{\prime} \alpha } 
 \left(\frac{\partial}{\partial z} + \frac{\partial}{\partial w} \right) \, , \quad 
 \left[(\nabla_L)^i_{\alpha}, 
(\nabla_R)^{i^{\prime}}_{\alpha^{\prime}}\right] \Bigg|_{\rm{vertex}}
=0 \, ,\\
& \left[(\nabla_R)^i_{\alpha}, 
(\nabla_R)^{i^{\prime}}_{\alpha^{\prime}}\right] \Bigg|_{\rm{vertex}}
=-\epsilon^{i^{\prime} i}
\epsilon_{\alpha^{\prime} \alpha } 
 \left(\frac{\partial}{\partial z} + \frac{\partial}{\partial w} \right) \, ,
\end{aligned}
\end{equation}
and the commutators involving  $(\nabla_L)^{\bar{i}}_{\dot\alpha}$ and
$(\nabla_R)^{\bar{i}}_{\dot\alpha}$
are similar to the ones above. 
\subsection{The $SU(2,2)\times SU(4)$ covariant parametrization}

In \cite{BerkovitsMaldacenaConjecture}, gauge invariant pure spinors were defined and used. 
We give here the explicit BRST transformations for a coset similar to the one used in \cite{BerkovitsMaldacenaConjecture}.
One of the difficulties in working with gauge invariant pure spinors is that the pure spinor constraints become 
more complicated. In the main text, we decide to work with the coset 
(\ref{eq:cosetparametrization}) and pure spinors that transform under gauge transformations because
it was easier to work out the simplifications for moving from the picture minus eight 
to the picture zero. 
In any case, we hope that the explicit BRST transformations of this alternative coset might be useful for some readers
and future applications. The alternative coset is 
\begin{equation}
g = e^{\bar{\theta}_- \bar{q}_+}e^{\bar{\theta}_+ \bar{q}_-}e^{\theta_- q_+}e^{\theta_+ q_-}G(X) H(Y) \, , 
\end{equation}
where now $G(X)$ parametrizes the coset $\frac{SO(2,4)}{SO(1,4)}$ and $H(Y)$ parametrizes $\frac{SO(6)}{SO(5)}$. 
We are going to denote the elements of the coset by both $G(X)_R^{\tilde{R}}$ and $H(Y)^{\tilde{I}}_I$ 
where the indices with a tilde are spinor indices of the isotropy group. 
The BRST transformations of the variables can be computed as usual from the coset transformation
\begin{equation}
\begin{aligned}
Q \, \cdot \, g & = g (\lambda q + \Sigma) \\
& = e^{\bar{\theta}_- \bar{q}_+}e^{\bar{\theta}_+ \bar{q}_-}e^{\theta_- q_+}e^{\theta_+ q_-} 
(\tilde{\lambda} q) G(X) H(Y) \\
& + e^{\bar{\theta}_- \bar{q}_+}e^{\bar{\theta}_+ \bar{q}_-}e^{\theta_- q_+}e^{\theta_+ q_-} (G(X) \Sigma_1) H(Y) +
e^{\bar{\theta}_- \bar{q}_+}e^{\bar{\theta}_+ \bar{q}_-}e^{\theta_- q_+}e^{\theta_+ q_-} G(X) (H(Y) \Sigma_2) \, , 
\label{qalternative}
\end{aligned}
\end{equation}
where $\tilde\lambda$ are gauge invariant pure spinors given by
\begin{equation}
\tilde\lambda^R_{I} = G^{-1}(X)^R_{\tilde{R}} H(Y)^{\tilde{I}}_I \lambda^{\tilde{R}}_{\tilde{I}} \, , \quad 
\tilde\lambda_R^{I} = G(X)_R^{\tilde{R}} H^{-1}(Y)_{\tilde{I}}^I \lambda_{\tilde{R}}^{\tilde{I}} \, . 
\label{gaugeinvariant}
\end{equation}
The $AdS_5$ and $S^5$ coordinates are given as
\begin{equation}
X_{RS} \equiv G(X)_R^{\tilde{R}} \sigma^{-1}_{\tilde{R} \tilde{S}} G(X)^{\tilde{S}}_S \, , \quad Y_{IJ} \equiv H(Y)^{\tilde{I}}_I \sigma^6_{\tilde{I}\tilde{J}} H(Y)^{\tilde{J}}_J 
\, , 
\end{equation}
where the direction, i.e. indices -1 and 6  in the sigma matrices $\sigma^n_{\tilde{I} \tilde{J}}$ is the one invariant under the isotropy groups.  
The variables above are constrained as follows
\begin{equation}
\frac{1}{8}\epsilon^{RSTU} X_{RS} X_{TU} = \frac{1}{8}\epsilon^{IJKL} Y_{IJ} Y_{KL}=-1 \, . 
\end{equation}

In the formulas below, the pure spinor variables are always the gauge invariant ones and we are going to suppress  the tildes to avoid cluttering. 
To write down the formulas, we will use the following definitions
\begin{equation}
\begin{aligned}
\Sigma^{(0,--)} = - a_0 + a_1-a_2+a_3-a_4+a_5 \, , 
\end{aligned} 
\end{equation}
where $a_0=\left[ \theta_- q_+ , \bar{\lambda}_+ \bar{q}_- \right]$ and $a_n = \left[\theta_- q_+, \left[ \theta_+ q_-, a_{n-1} \right] \right]$ and 
\begin{equation}
\begin{aligned}
\Sigma^{(++,0)} = b_0 + b_1-b_2+b_3-b_4+b_5 \, , 
\end{aligned} 
\end{equation}
where $b_0=\left[ \theta_- q_+ , \bar{\lambda}_- \bar{q}_+ \right]$ and $b_n = \left[\theta_- q_+, \left[ \theta_+ q_-, b_{n-1} \right] \right]$.
Moreover
\begin{equation}
\Sigma^{(0,++)} = - \frac{1}{2} \left[ \theta_+ q_-, \left[ \theta_+ q_-, \Sigma^{(++,0)} \right]\right] \, , \quad 
\Sigma^{(--,0)} = - \frac{1}{2} \left[ \theta_+ q_-, \left[ \theta_+ q_-, \Sigma^{(0,--)} \right]\right] \, . 
\end{equation} 
In terms of the $\Sigma$'s above, the BRST transformations of the fermionic variables are 
\begin{equation}
\begin{aligned}
&(Q \theta_+) q_- = \lambda_+ q_- + \frac{1}{2} \left[ \left[ (\theta_+ q_-), (\lambda_- q_+) \right], (\theta_+ q_-) \right] +
\left[ \left[ \bar{\theta}_+ \bar{q}_-, ( \bar{\lambda}_- \bar{q}_+ - \left[ \theta_+ q_-, \Sigma^{(++,0)} \right])\right], \theta_+ q_- \right] \, , \\
& (Q \theta_-) q_+ = \lambda_- q_+ + \left[ \left[ \bar{\theta}_+ \bar{q}_-, (\bar{\lambda}_- \bar{q}_+ + \left[ \theta_+ q_-, \Sigma^{(++,0)} \right])\right], 
\theta_- q_+ \right] \, , \\
&(Q \bar{\theta}_+ ) \bar{q}_- = \bar{\lambda}_+ \bar{q}_- + \left[ \theta_+ q_- , \Sigma^{(0,--)} \right] +
\frac{1}{2} \left[ \left[ \bar{\theta}_+ \bar{q}_- , (\bar{\lambda}_- \bar{q}_+ + \left[ \theta_+ q_- , \Sigma^{(++,0)} \right]) \right], \bar{\theta}_+ \bar{q}_- \right] \, , \\
&(Q \bar{\theta}_-) \bar{q}_+ = \bar{\lambda}_- \bar{q}_+ + \left[ \theta_+ q_-, \Sigma^{(++,0)} \right] \, . 
\end{aligned} 
\end{equation}
Note that the indices $R$ and $I$ appearing in $G$ and $H$ are not BRST invariant and they rotate. 
The rotation parameters    
\begin{equation}
\Sigma M = \Sigma^I_J M^J_I + \Sigma^R_S M^S_R \, ,
\end{equation}
are given by 
\begin{equation}
\begin{aligned}
&\Sigma M  = \left[ (\theta_+ q_-) , (\lambda_- q_+) \right] + \left[ (\theta_+ q_-) , (\bar{\lambda}_+ \bar{q}_-) \right] + \left[\theta_+ q_-, \bar{\lambda}_- \bar{q}_+ \right] + \\
&\left[ \bar{\theta}_+ \bar{q}_- , (\bar{\lambda}_- \bar{q}_+ + \left[ \theta_+ q_-, \Sigma^{(++,0)} \right])\right] -
(\Sigma^{(--,0)} + \Sigma^{(0,--)} + \Sigma^{(++,0)} + \Sigma^{(0,++)} ) \, . 
\end{aligned} 
\end{equation} 
So, for example, 
\begin{equation}
Q \, \cdot \, X_{RS}  = \Sigma_R^T \, X_{TS} + \Sigma_S^{T} \, X_{R T} \, .  
\end{equation}
The explicit form of $\Sigma_1$ and $\Sigma_2$ appearing 
in (\ref{qalternative}) is not necessary in general because one only works 
with gauge invariant quantities and these matrices rotates the indices
belonging to the isotropy groups. 

\subsection{The gauge invariant pure spinor constraints}

In order to perform computations using 
the $SU(2,2) \times SU(4)$ 
covariant parametrization just described, we have to write the pure spinor constraints in terms of the gauge invariant pure spinors
$\tilde\lambda$ defined in (\ref{gaugeinvariant}) and solve them. 
One way of doing it is to relate the $\tilde\lambda$ 
with the $\lambda$'s used in the main text, see (\ref{eq:BRST})
and (\ref{eq:so8lambda}), and  we already know the solution of the constraints in terms of the $\lambda$'s. The relation is as follows ($u$ and $v$ were 
defined in (\ref{BosonicCosets}))
\begin{equation}
\tilde{\lambda} q = (uv)\lambda q (uv)^{-1}= e^{yK}e^{(z\Delta+wJ)}(\lambda q)e^{-(z\Delta+wJ)}e^{-yK} = e^{yK}(\lambda^{\prime} q)e^{-yK} \, , 
\end{equation}
and we have used the shorthand notation 
$yK=x^{\alpha}_{\dot\alpha}K^{\dot\alpha}_{\alpha} + y^{\bar i}_{i}K^{i}_{\bar i}$. Since $K$ have charge $+1$ under $J-\Delta$ the series truncate
\begin{equation}
\tilde\lambda q = \lambda^{\prime} q+[yK, \lambda^{\prime} q]+\frac{1}{2}[yK,[yK, \lambda^{\prime} q]] \, .     
\end{equation}
In components, we have
\begin{equation}
\begin{aligned}
& \tilde \lambda_{+}q_-=\lambda^{\prime}_{+}q_{-},\quad \tilde{\bar{\lambda}}_{+}\bar q_{-}+\tilde{\bar\lambda}_{-} q_{+}= {\bar\lambda}^{\prime}_{+}\bar q_{-}+\bar\lambda^{\prime}_{-} q_{+} + [yK,\lambda^{\prime}_+ q_-] \, , \\
& \tilde\lambda_{-}q_+=\lambda^{\prime}_{-}q_{+}+[yK,(\bar\lambda^{\prime}_{+}\bar q_{-}+\bar\lambda^{\prime}_{-} q_{+})]+\frac{1}{2}[yK,[yK,\lambda^{\prime}_+ q_-]] \, . 
\label{eqap1}
\end{aligned}
\end{equation}
Moreover, we can easily invert the expressions above and get
\begin{equation}
\begin{aligned}
& \lambda^{\prime}_{+}q_-=\tilde\lambda_{+}q_{-},\quad \bar\lambda^{\prime}_{+}\bar q_{-}+\bar\lambda^{\prime}_{-} \bar q_{+} =\tilde{\bar\lambda}_{+}\bar q_{-}+\tilde{\bar\lambda}_{-} \bar q_{+} - [yK,\tilde\lambda_+ q_-] \, ,\\
& \lambda^{\prime}_{-}q_+=\tilde\lambda_{-}q_{+}-[yK,(\tilde{\bar\lambda}_{+}\bar q_{-}+\tilde{\bar\lambda}_{-} \bar q_{+})]+\frac{1}{2}[yK,[yK,\tilde\lambda_+ q_-]] \, .
\label{eqap2}
\end{aligned}
\end{equation}
It is easy to evaluate 
both (\ref{eqap1}) and
\eqref{eqap2} given above. 
The result can be written 
in terms of the gauge invariant variables $X_{RS}$ and $Y_{IJ}$ defined in (\ref{eq:Xmatrix}) as follows 
\begin{equation}
\begin{aligned}
& \lambda_{\dot\alpha}^{i}=\left(\frac{Y_{12}}{X_{34}}\right)^{-\frac{1}{2}}\left(\tilde{\lambda}^{i}_{\dot\alpha}+X^{\alpha}_{\dot\alpha}\tilde\lambda_{\alpha}^{i}\right) \, ,\quad  \lambda_{\bar i}^{\alpha}=\left(\frac{Y_{12}}{X_{34}}\right)^{-\frac{1}{2}}\left(\tilde{\lambda}_{\bar i}^{\alpha}-X^{\alpha}_{\dot\alpha}\tilde\lambda^{\dot\alpha}_{\bar i}\right) \, ,    \\
& \lambda^{\dot \alpha}_{ i}=\left(\frac{Y_{12}}{X_{34}}\right)^{+\frac{1}{2}}\left(\tilde{\lambda}^{\dot\alpha}_{i}+Y^{\bar i}_{i}\tilde\lambda_{\bar i}^{\dot\alpha}\right) \, ,\qquad  \lambda_{\alpha}^{\bar i}=\left(\frac{Y_{12}}{X_{34}}\right)^{+\frac{1}{2}}\left(\tilde{\lambda}_{\alpha}^{\bar i}-Y^{\bar i}_{i}\tilde\lambda^{i}_{\alpha}\right) \, , \\
& \lambda^{\bar i}_{\dot\alpha}=\left(Y_{12}X_{34}\right)^{+\frac{1}{2}}\left(\tilde\lambda^{\bar i}_{\dot\alpha}+X_{\dot\alpha}^{\alpha}\tilde\lambda_{\alpha}^{\bar i}-Y^{\bar i}_{i}\tilde\lambda^{i}_{\dot\alpha}-\frac{1}{2}X^{\alpha}_{\dot\alpha}Y^{\bar i}_{i}\tilde\lambda^{i}_{\alpha}\right) \, ,\\
& \lambda^{\alpha}_{i}=\left(Y_{12}X_{34}\right)^{+\frac{1}{2}}\left(\tilde\lambda_{ i}^{\alpha}-X_{\dot\alpha}^{\alpha}\tilde\lambda^{\dot\alpha}_{i}+Y^{\bar i}_{i}\tilde\lambda_{\bar i}^{\alpha}-\frac{1}{2}X^{\alpha}_{\dot\alpha}Y^{\bar i}_{i}\tilde\lambda_{\bar i}^{\dot\alpha}\right) \, ,
\\
& \lambda^{i}_{\alpha}=\left(Y_{12}X_{34}\right)^{-\frac{1}{2}}\tilde\lambda^{i}_{\alpha} \, ,\qquad \lambda_{\bar i}^{\dot\alpha}=\left(Y_{12}X_{34}\right)^{-\frac{1}{2}}\tilde\lambda_{\bar i}^{\dot\alpha} \, , 
\end{aligned}
\end{equation}
equivalently, 
\begin{equation}
\begin{aligned}
& \tilde\lambda_{\dot\alpha}^{i}=e^{+\frac{(w+z)}{2}}\left(\lambda^{i}_{\dot\alpha}-X^{\alpha}_{\dot\alpha}\lambda_{\alpha}^{i}\right),\quad  \tilde\lambda_{\bar i}^{\alpha}=e^{+\frac{(w+z)}{2}}\left(\lambda_{\bar i}^{\alpha}+X_{\dot\alpha}^{\alpha}\lambda^{\dot\alpha}_{\bar i}\right) \, , \\
& \tilde\lambda^{\dot\alpha}_{ i}=e^{-\frac{(w+z)}{2}}\left(\lambda^{\dot\alpha}_{ i}-Y^{\bar i}_{i}\lambda_{\bar i}^{\dot\alpha}\right),\qquad  \tilde\lambda_{\alpha}^{\bar i}=e^{-\frac{(w+z)}{2}}\left(\lambda_{\alpha}^{\bar i}+Y^{\bar i}_{i}\lambda^{i}_{\alpha}\right) \, , \\
& \tilde\lambda^{\bar i}_{\dot\alpha}=e^{-\frac{(w-z)}{2}}\left(\lambda^{\bar i}_{\dot\alpha}-X_{\dot\alpha}^{\alpha}\lambda_{\alpha}^{\bar i}+Y^{\bar i}_{i}\lambda^{i}_{\dot\alpha}-\frac{1}{2}X^{\alpha}_{\dot\alpha}Y^{\bar i}_{i}\lambda^{i}_{\alpha}\right) \, ,\\
& \tilde\lambda^{\alpha}_{i}=e^{-\frac{(w-z)}{2}}\left(\lambda_{ i}^{\alpha}+X_{\dot\alpha}^{\alpha}\lambda^{\dot\alpha}_{i}-Y^{\bar i}_{i}\lambda_{\bar i}^{\alpha}-\frac{1}{2}X^{\alpha}_{\dot\alpha}Y^{\bar i}_{i}\lambda_{\bar i}^{\dot\alpha}\right) \, , \\
& \tilde\lambda^{i}_{\alpha}=e^{+\frac{(w-z)}{2}}\lambda^{i}_{\alpha} \, ,\qquad \tilde\lambda_{\bar i}^{\dot\alpha}=e^{+\frac{(w-z)}{2}}\lambda_{\bar i}^{\dot\alpha} \, . 
\end{aligned}
\end{equation}
In the main text, we 
demonstrated that it is possible to express 
$\lambda^a_-$ as 
function  of the other $\lambda$'s and this 
implies that 
$\lambda^{i}_{\alpha}$ and $\lambda^{\dot\alpha}_{\bar i}$ can be taken as 
unconstrained variables.
The same is true for 
$\tilde\lambda^{i}_{\alpha}$ and $\tilde\lambda^{\dot\alpha}_{\bar i}$ 
and this follows from the equations  
above. In fact, using
\begin{equation}
\lambda_-^{a}=\frac{(\bar\lambda_+\sigma^{mn}\bar\lambda_{-})}{4(\bar\lambda_{+}\bar\lambda_+)}(\sigma^{mn}\lambda_+)^{a},
\end{equation}
it is not difficult to see that $\tilde\lambda^a_-$
can be written in terms of the others $\tilde\lambda$'s.
The conclusion is that 
$\delta(\tilde\lambda_{+}^{a})$ makes sense and it is possible to define 
minus eight picture vertex operators
for the $SU(2,2) \times SU(4)$ covariant
parametrization. 

\section{
Half-BPS vertex operators in $SO(8)$ notation}
\label{so8vertexap}

In this Appendix, we rewrite the vertex operators given in  (\ref{vertexoperatorso4}) in a more compact 
$SO(8)$ notation, see (\ref{eq:so8lambda}) and (\ref{eq:deftheta}). We will use the $SO(8)$ Pauli Matrices
\begin{equation}
\sigma^{\underline{m}}_{a \dot{a}} \, , \quad
{\rm{and}} \quad \sigma^{\underline{m}}_{\dot{a} a} \, , 
\end{equation}
with $\underline{m},a,\dot{a}=1,\ldots,8.$ 
In what follows we will use the notation
\begin{equation}
\hat\sigma_{a b} =(\sigma_{\underline{1234}})_{a b} \, , 
\end{equation}
where $\sigma_{\underline{1} \underline{2} 
\underline{3} \underline{4}}$ is totally anti-symmetric in the
indices $\{\underline{1},\underline{2},\underline{3},\underline{4}\}$.
The vertices take the form 
\begin{equation}
e^{-n(z+w)} V_{\mathfrak{so}(8)}(n) = V_{\mathfrak{so}(8)}^{0}+ V_{\mathfrak{so}(8)}^{2} +  V_{\mathfrak{so}(8)}^{4} +  V_{\mathfrak{so}(8)}^{6} +  V_{\mathfrak{so}(8)}^{8}  \, , 
\label{so8vertex}
\end{equation} 
and as usual the superscripts 
indicate the number of $\theta_+$'s. 
We have 
\begin{equation}
\begin{aligned}
V^0_{\mathfrak{so}(8)} = (\bar{\lambda}_+ \bar{\lambda}_+) \, , \quad  \quad V^8_{\mathfrak{so}(8)}
= (\bar{\lambda}_+ \bar{\lambda}_+) \theta^8_+ (72 -37 n^2 + n^4) \, , 
\end{aligned} 
\end{equation}
where $(\bar{\lambda}_+ \bar{\lambda}_+)$ and 
$\theta^8_+$
were defined in (\ref{eq:lambdabarplus}) and 
(\ref{theta8def}) respectively. 
For the remaining terms, we are going to use the notation
\begin{equation}
\begin{aligned}
& (\theta^2_{+})_{ij} = 
\theta_+^a (\sigma_{\underline{ij}})_{a b} 
\theta_+^b \, , \quad \quad 
(\hat\theta^2_{+})_{ij} = 
\theta_+^a (\hat\sigma \sigma_{\underline{ij}})_{a b} 
\theta_+^b \, , \quad  \\
& (\bar\lambda_x \bar\lambda_y)^{i_1 j_1,\ldots,i_m j_m} =
\bar\lambda_x^{\dot{a}} 
(\sigma^{\underline{i_1 j_1}} 
\ldots \sigma^{\underline{i_m j_m}})_{\dot{a} \dot{b}} \bar\lambda_y^{\dot{b}} \, , \quad \\
& (\bar\lambda_x  \bar\lambda_y)_{\rm{hat}}^{i_1 j_1,\ldots,i_m j_m}=\bar\lambda_x^{[ \dot{a}} (\hat\sigma \sigma^{\underline{i_1 j_1}} 
\ldots \sigma^{\underline{i_m j_m}})_{\dot{a}
\dot{b}} \bar\lambda_y^{\dot{b} ]} \, , 
\end{aligned} 
\end{equation}
where $[ \dot{a} \dot{b} ] =
\dot{a} \dot{b} - \dot{b} \dot{a}$
without a factor of two and
\begin{equation}
\sigma^{\underline{i j}}_{a b} = \frac{1}{2}
\left( \sigma^{\underline{i}} \sigma^{\underline{j}} -
\sigma^{\underline{j}} \sigma^
{\underline{i}} \right)_{a b} \, . 
\end{equation}
The remaining terms in (\ref{so8vertex}) are given by
\begin{equation}
V_{\mathfrak{so}(8)}^{2} =
- \frac{1}{16} 
(\bar\lambda_+  \bar\lambda_-)_{\rm{hat}}^{i j}
(\hat\theta^2_{+})_{ij} 
- \frac{n}{8} 
(\bar\lambda_+ \bar\lambda_-)^{i j}
(\theta^2_{+})_{ij} \, , 
\end{equation}
and 
\begin{equation} 
\begin{aligned}
V_{\mathfrak{so}(8)}^{6} & =  
- (\frac{1}{2304} - \frac{n^2}{9216})
(\bar\lambda_+  \bar\lambda_-)_{\rm{hat}}^{i j 
k l m n}
(\hat\theta^2_{+})_{ij}
(\hat\theta^2_{+})_{kl}
(\hat\theta^2_{+})_{mn} \\
&+ ( \frac{19 \, n}{46080} - \frac{n^3}{46080})  
(\bar\lambda_+ \bar\lambda_-)^{i j k l m n}
(\theta^2_{+})_{ij}
(\theta^2_{+})_{kl}
(\theta^2_{+})_{mn} \, .
\end{aligned} 
\end{equation}
Finally, 
\begin{equation}
\begin{aligned}
V_{\mathfrak{so}(8)}^{4} 
& =
\frac{1}{768} 
(1 - n^2)
(\bar\lambda^2_+ -\bar\lambda^2_-)^{ij kl}
(\theta^2_{+})_{ij} 
(\theta^2_{+})_{kl} 
-\frac{3 n}{768} 
 (\bar\lambda^2_+ + \bar\lambda^2_-)^{ij kl} 
 (\theta^2_{+})_{ij} 
(\theta^2_{+})_{kl} \\
& - \frac{1}{256} 
(\bar\lambda^2_+  
- \bar\lambda^2_-)^{i j 
k l} 
(\hat\theta^2_{+})_{ij}
(\hat\theta^2_{+})_{kl}
- \frac{n}{128}
(\bar\lambda^2_+)_{\rm{hat}}^{i j 
k l} 
(\theta^2_{+})_{ij} 
(\hat\theta^2_{+})_{kl}  \, . 
\end{aligned}
\end{equation}

\end{document}